\documentclass{aa}  

\usepackage{graphicx}
\usepackage[section]{placeins} 
\usepackage{txfonts}
\usepackage{comment}
\usepackage{xcolor}

\def\msun{M$_{\odot}$}
\def\omast{$\omega^*$ }
\def\iso#1#2{$^{#2}$#1}

\usepackage[normalem]{ulem}
\usepackage{lscape}

\begin{document}

   \title{Barium stars as tracers of \textit{s}-process nucleosynthesis in AGB stars}

   \subtitle{III. Systematic deviations from the AGB models}

   \author{B. Világos
          \inst{1, 2, 3, 4}\fnmsep\thanks{vilagos.blanka@csfk.hun-ren.hu}
          \and
          B. Cseh\inst{1, 2, 3}
          \and
          A. Yagüe López\inst{5}
          \and M. Joyce \inst{1,2}
          \and A. Karakas\inst{6,7} 
          \and G. Tagliente\inst{8} 
          \and M. Lugaro\inst{1,2,4,6}
          }

   \institute{
   Konkoly Observatory, HUN-REN Research Centre for Astronomy and Earth Sciences, Konkoly Thege Mikl\'os út 15-17., H-1121, Hungary 
   \and
   CSFK, MTA Centre of Excellence, Budapest, Konkoly Thege Mikl\'os út 15-17., H-1121, Hungary 
   \and
   MTA--ELTE Lend{\"u}let `Momentum' Milky Way Research Group, Hungary 
   \and 
    ELTE E\"{o}tv\"{o}s Lor\'and University, Institute of Physics and Astronomy, Budapest 1117, P\'azm\'any P\'eter s\'et\'any 1/A, Hungary 
    \and
    Computer, Computational and Statistical Sciences (CCS) Division, Center for Theoretical Astrophysics, Los Alamos National Laboratory, Los Alamos, NM 87545, USA 
    \and
    School of Physics and Astronomy, Monash University, VIC 3800, Australia 
    \and ARC Centre of Excellence for All Sky Astrophysics in 3 Dimensions (ASTRO 3D) 
    \and
    Istituto Nazionale di Fisica Nucleare (INFN), Bari, Italy 
    }

   \date{Received Mar 22, 2024; accepted May 23, 2024}

 
  \abstract
   {Barium (Ba) stars help to verify asymptotic giant branch (AGB) star nucleosynthesis models since they experienced pollution from an AGB binary companion and thus their spectra carry the signatures of the slow neutron capture process ($s$ process).}
   {For a large number (180) of Ba stars, we searched for AGB stellar models that match the observed abundance patterns. We aim to uncover any systematic deviations of the sample abundances from the predictions of the nucleosynthesis models.
   }
   {We employed three machine learning algorithms as classifiers: a Random Forest method, developed for this work, and the two classifiers used in our previous study. Compared to that work, we also expanded our observational sample with 11 Ba stars available in the supersolar metallicity range.
   We studied the statistical behaviour of the different $s$-process elements in the observational sample to investigate if the AGB models systematically under- or overpredict the abundances observed in the Ba stars and show the results in the form of violin plots of the residuals between spectroscopic abundances and model predictions.
   We inspected the correlations between the observed [Fe/H], the $s$-process elemental abundances, and the residuals. 
   We employed the [Zr/Fe] and [Nb/Fe] abundances as a thermometer to constrain the operational temperature that rules the production of these elements in the sample stars, assuming a steady-state $s$ process.
   We also investigated the mass distribution of the identified polluter AGB stars and the behaviour of the $\delta$ parameter, which describes the fraction of accreted AGB material relative to the Ba star envelope.}
   {We find a significant trend in the residuals that implies an underproduction of the elements just after the first $s$-process peak (Nb, Mo, and Ru) in the models relative to the observations. This may originate from a neutron-capture process (e.g. the intermediate neutron-capture process, $i$ process) not yet included in the AGB models of metallicity from solar to roughly 1/5 solar, corresponding to the range of the Ba stars.   
    Correlations are found between the residuals of these peculiar elements, suggesting a common origin for the deviations from the models. In addition, there is a weak metallicity dependence of the residuals of these elements.  
    The $s$-process temperatures derived with the [Zr/Fe] -- [Nb/Fe] thermometer have an unrealistic value for the majority of our stars. The most likely explanation is that at least a fraction of these elements are not produced in a steady-state $s$ process, and instead may be due to processes not included in the AGB models.
   The mass distribution of the identified models confirms that our sample of Ba stars was polluted by low-mass AGB stars (< 4 \msun). Most of the matching AGB models require low accreted mass, but a few systems with high accreted mass are needed to explain the observations.}
   {}


   \maketitle
%

\section{Introduction}
The peculiar Barium (Ba) stars are particularly interesting for the study of the slow neutron-capture process ($s$ process), which is responsible for the production of about half of the elements heavier than iron in the Universe (\citealt{bbfh, busso99}; for a recent review see \citealt{Lugaro-review}). 
Asymptotic giant branch (AGB) stars are the main astrophysical site of the $s$ process, providing neutrons through the $^{13}$C($\alpha$,n)$^{16}$O reaction in the so-called $^{13}$C pocket (in AGB stars with masses below 4\, \msun) and through the $^{22}$Ne($\alpha$,n)$^{25}$Mg reaction when $T \gtrsim$ 300 MK in AGBs with masses above 4\,\msun~ \citep{lugaro03, cristallo09, kaeppeler11, Karakas-Lattanzio-review, Karakas-Lugaro-22Ne}.
The abundances of the $s$ elements peak at the magic neutron numbers: the first peak corresponds to neutron number 50 (e.g. Sr, Y, and Zr), the second to 82 (e.g. Ba, La, and Ce), and the third to 126 (Pb). 

Although Ba stars have not yet undergone the AGB phase, they are enriched in the material produced by the $s$ process. After their discovery by \cite{Bafirst}, it was concluded that Ba stars are in binary systems, where the companion star has already completed the AGB phase and is now a white dwarf (e.g. \citealt{RV1-McCFN, RV2-McC}). Hence, the enrichment of $s$ process material is caused by accretion from the companion during its previous AGB phase onto the Ba star (e.g. \citealt{accretion-Han, accretion-Jori1, jorissen19}).
AGB stars are cooler than Ba stars; therefore, many molecular bands appear in their spectra and their atmospheres are highly dynamical, complicating the derivation of precise abundances.
Abundances derived from Ba stars are therefore useful for testing nucleosynthesis models of $s$-process production, since their direct spectroscopic study carries much smaller uncertainties. 

The difficulty in finding the best-fitting AGB models\footnote{By AGB models, we mean AGB nucleosynthesis models throughout the paper.} to have produced an observed Ba star abundance pattern arises from the fact that we have limited or no information on the initial mass of the AGB star and the mass of the transferred AGB material ($M\mathrm{_{AGB, trans}}$).
Therefore, a parameter $\delta$ is usually introduced, which represents the fraction of the transferred AGB material to the total envelope mass of the Ba star ($M\mathrm{_{tot}} = M\mathrm{_{Ba, env}} + M\mathrm{_{AGB, trans}}$, where $M\mathrm{_{Ba, env}}$ is the mass of the Ba star envelope before accretion, see \citealt{paper1-Ce}):
\begin{equation}
    \delta = \frac{M\mathrm{_{AGB, trans}}}{M\mathrm{_{tot}}} = \frac{M\mathrm{_{AGB, trans}}}{M\mathrm{_{Ba, env}} + M\mathrm{_{AGB, trans}}}.
\end{equation}
Since in this work we only consider Ba giants with convective envelopes, $\delta$ is effectively a measure of the dilution of the AGB transferred material in the Ba star envelope. 
The value of $\delta$ lies between 0 and 1, where $\delta = 1$ represents undiluted, pure material from the AGB star in the envelope, while $\delta = 0$ corresponds to no $s$-process enhanced material.
The real value of $\delta$ is in between these extremes, but is unknown and different for each system.

In the previous papers of this series (\citealt{paper1-Ce} and \citealt{paper2-ML}, hereafter Paper I and Paper II, respectively) we individually analysed 28 Ba stars (Paper I) with the masses of the Ba star and the initial mass of the AGB star derived from the binary orbital parameters \citep{jorissen19}; and the sample of 169 Ba stars (Paper II) from \citet{deC} with additional elemental abundances from \citet{roriz21a} and \citet{roriz21b}.
In Paper I we calculated the $\delta$ parameter by constraining the AGB model to match the [Ce/Fe]\footnote{Throughout the paper we use the spectroscopic notation: for elements $A$ and $B$, $[A/B] = \log_{10}(A/B)_\mathrm{star}-\log_{10}(A/B)_{\odot}$.} value of the Ba star. Using this $\delta$ value, we transformed every other elemental abundance predicted by the AGB model into a diluted envelope and compared it with the Ba star abundance pattern.  By finding the best-fitting model, we estimated the mass of the polluter AGB and compared it with that derived by \cite{jorissen19}.
In Paper II we introduced an improved classification\footnote{By classification we mean the process of identifying the best-fitting polluter AGB models for the abundance pattern of each Ba star, since classification ML algorithms were used for this task. A specific algorithm for this purpose is therefore referred to as a `classifier'.} by incorporating machine learning (ML) techniques with the aim of identifying the parameters of the polluter AGB model for all systems, even those without prior information on the initial AGB mass, and leaving $\delta$ and [Fe/H] as free parameters. The first classifier used a nearest-neighbour algorithm, here called the closest method (classifier CM), while the other used a set of neural networks (classifier NN).
In Papers I and II, we introduced plots for the individual stars showing the abundances of the Ba stars and the identified AGB models on the same figure. These can be used to inspect the individual fit for each star and to categorise them. 

In this study we extended the previous ML work described in Paper II in three ways: (i) we used it to statistically study the Ba stars as a population instead of individually; (ii) we developed a new classifier using the Random Forest ML technique (RF classifier); and (iii) we added 11 stars with supersolar metallicity. We applied the three classifiers to all the Ba stars studied and made statistical inferences for the entire sample. For the description of our methods, including the observational sample, the AGB nucleosynthesis models and our algorithms, readers can refer to Sect. \ref{sec:methods}.

We first investigate how the model predictions systematically differ from the Ba star abundance features using violin- and boxplots of their residuals (Subsect. \ref{sec:boxplot}). This indicates an underproduction of the elements Nb, Mo, and Ru in the AGB models relative to the Ba star abundances, which may be due to an unknown nucleosynthesis process in the models, possibly the intermediate neutron capture process ($i$ process). 
We then examine the underlying relationships between different elements using several types of correlation analysis (Subsect. \ref{sec:correl}). This concludes in that Nb, Mo and Ru behave similarly and our ability to fit them slightly depend on the metallicity of the star. 
We then show the mass and $\delta$ value distribution of the identified models (Subsect. \ref{sec:distributions}). This points to AGB polluters of low mass and systems of low $\delta$ with some exceptions.
We further investigate the operational temperature of the neutron capture process assuming steady-state $s$ process with the help of the [Zr/Fe] and [Nb/Fe] abundances (Sect. \ref{sec:NbZr}). We find that the majority of our stars cannot be explained with reasonable $s$-process temperatures, most probably because \iso{Zr}{93} is not solely produced in steady-state $s$ process.
For our conclusions and discussions, we refer to Sect. \ref{sec:conclusions}.
 

\section{Methods}
\label{sec:methods}
\subsection{Ba star abundances and AGB models} 
Here we extended our data set, as compared to Paper II. The elemental abundances included in Paper II were the 169 Ba stars homogeneously analysed by \citet{deC}, \citet{roriz21a} and \citet{roriz21b}. We added here 11 supersolar Ba stars from \citet{Pereira11}, for which additional elemental abundances were also published in \citet{roriz21a} and \citet{roriz21b}. These authors selected the Ba stars as those with a significant $s$-process enrichment of [$s$/Fe] $>$ 0.25 dex, where [$s$/Fe] is the average of the abundances of Y, Zr, La, Ce and Nd. 
The abundances were derived assuming local thermal equilibrium (LTE) in the stellar atmosphere.

As in Paper II, we used two families of AGB nucleosynthesis models, Monash \citep{2012lugaro,2014fishlock,2016karakas,2018karakas} and FRUITY\footnote{\url{http://fruity.oa-teramo.inaf.it/}} \citep{2009cristallo,2011cristallo,2013piersanti,2015cristallo}.
We adopted 69 Monash and 83 FRUITY AGB model predictions, on a grid of masses between 1.5 and 8 \msun~and [Fe/H] values between $-$1.67 and 0.33. The grid of masses and metallicities of the models used is presented in Paper II, Tables C.1 and D.1.
We applied the three classifiers (CM, NN, and RF) to the entire sample of 180 Ba star abundances. 
We note that the FRUITY models do not originally take into account the decay of \iso{Zr}{93} to \iso{Nb}{93}. However, this process is the main source of \iso{Nb}{93}, and without its inclusion the models lack most of the Nb. This decay has a half-life of 1.5 Myrs, so it is reasonable to assume that all the \iso{Zr}{93} has decayed to \iso{Nb}{93}. In this work, to account for this decay, we used the isotopic tables provided on the FRUITY website and added all the \iso{Zr}{93} to \iso{Nb}{93}.

\subsection{Elements used for the classification}
The elements used for training and fitting the classifiers included every neutron-capture element available in the observational data set except Nb (which is problematic to fit, see discussion in Paper II). As in Paper II, `Peak 1' contains Rb, Sr, Y, Zr, Mo, and Ru; `Peak 2' contains La, Ce, Nd, Sm, and Eu; while we have no observations for Peak 3. 
The abundances of most of these elements in the Solar System are mainly produced by the $s$ process with a smaller contribution from the rapid neutron capture process ($r$ process). The most prominent exception is Eu, which is thought to have 94\% $r$-process origin in the Solar System \citep{s-r-contribution-Bisterzo}.

For the assumed systems of Ba stars, the $r$ process component of these elements originates from the initial composition, and the additional enrichment is solely attributed to the polluting AGB star. 
The carbon-enriched metal-poor (CEMP) stars are usually classified to be enhanced in $r$-process material if their [Eu/Fe] abundances are greater than 1 (e.g. \citealt{EuFe1-Beers, EuFe1-Jonsell, 2012lugaro}). For our stars, the maximum of the [Eu/Fe] values is 0.8, while their mean is 0.2, which is well below this criterion.
These [Eu/Fe] values are not high relative to the average of stars in the thin disk, which is around 0.2 dex (see \citealt{galcomposition}, Fig. 3).

We experimented with different element sets for training the classifiers by excluding some elements from this list. In particular, we studied two element sets in detail, introduced in Paper II as set A (all elements except Nb) and set F (all elements except Y, Nb, Mo, and La). Using different element sets slightly changed the identified AGB models for some stars. On a statistical scale (i.e. for the violin plots in Sect. \ref{sec:boxplot} and the correlations in Sect. \ref{sec:correl}) the patterns changed almost imperceptibly, only the spread decreased by using set F. 
However, the mass distribution (see Sect. \ref{subsec:distr-mass}) of our new RF classifier has an unexpected peak at 4 \msun~when we use the element set A, which disappears when we choose set F. As the Rb abundance is extremely sensitive to the AGB mass (see discussion in Sect. \ref{sec:boxplot}), this mass peak introduces a larger scatter in the residuals of Rb for set A than for set F. Otherwise, the quantitative behaviour of the quantities in the violin plots and correlations remain the same for all elements even in this case.
This means that the results of our algorithm are independent of the elements used for classification among those we studied. We therefore include the full set of elements in our algorithm, as this is the least arbitrary choice and contains the most information about the stars.

The features used for classification are as follows. Both the observations and the AGB models include the [Fe/H] and the abundances relative to iron (i.e. for element $A$, [$A$/Fe]), hereafter referred to as the `elemental abundances'. The relative abundances of the $s$-process peaks can be better described by the differences of elemental abundances (i.e. for elements $A$ and $B$, $[A$/Fe$] - [B$/Fe] = [$A/B$]), hereafter referred to as `elemental ratios'.
These features were used for both the training and the evaluation of the classifiers, as well as for the statistical analysis.
The derivation of second-order differences ($[A/B] - [C/D] $) did not significantly improve the classification performance, but made it difficult to interpret the physical meaning of the results. Therefore, only the [Fe/H], the elemental abundances, and the elemental ratios were used for the classification, leading to a total of 67 features.

\subsection{Classification techniques} 
For the two algorithms developed in Paper II, we created a pool of abundances by diluting all AGB models with $\delta$ values on the range of $\delta \in [0, 0.9]$ with increments of 0.002. This provided a large enough population (450 to 1000 entities for each AGB model, depending on how many different dilutions were rejected for having almost the same composition) to use machine learning. We then introduced the goodness of fit, which quantifies the difference between the fitting models and the observations based on a modified $\chi^2$ probe. Of the two classifiers described in Paper II, the first selected the models with the best goodness of fit, using a nearest neighbour classifier, here called the closest method (classifier CM). The second used a neural network (classifier NN), trained separately for the FRUITY and Monash models. Both classifiers predicted the best-fit AGB models with similar mass and metallicity for each star.

The set of neural networks (NN) included 5 independent networks trained with the same input parameters but with different randomisation. In Paper II, only the most probable AGB model from each network was selected, and all of these 5 were considered as a classified model, that is, as a possible polluter of the Ba star. In most cases this resulted in 1-2 classified models, as the results from the networks often overlapped.
In this work, we modified the NN method by allowing more than one polluter AGB model to be found by each network. Here, any AGB model that has a probability greater than 0.001 in at least 3 of the 5 networks is considered to be a classified model. We note that this probability limit is arbitrary, and it was chosen to result in 1-2 classified models on average, as in Paper II. This limit is a small number because the output layer of the network uses a softmax function (as opposed to the sigmoid function used in Paper II), which gives high probability to only a few models, but is the most popular choice for multi-label classification ML problems.

Here we also extended our ML techniques by adding a Random Forest (RF) classifier alongside the CM and NN. 
This method uses an ensemble of decision trees (800 in this work) and provides as high accuracy as the NN classifier when trained and tested on the diluted AGB models. For the RF classifier, we also used 5 independently trained Random Forests and considered those AGB models to fit the Ba star that gave a probability higher than 0.1 in at least 3 of the Random Forests. The probability limit is two orders of magnitude higher than for the NN because of the different metrics\footnote{The random forest method defines the probability by the number of decision trees that voted for that AGB model, normalised by the number of decision trees; while our neural network uses the output of the softmax function as a probability measure.} for calculating probabilities.
Further details of the classifier can be found in Appendix \ref{sec:appendix-rf}. 
We uploaded the codes to GitHub for the RF classifier as well as the necessary plotting routines for the figures in this paper, extending the online repository of Paper II.\footnote{\url{https://github.com/vblanka24/Ba_star_classification_PaperIII}}

Another advantage of using Random Forests is that we can obtain the `importance' of each feature used. This quantifies the influence of each feature on the result during the training of the classifier. We used the `Mean Decrease in Impurity' method, which is based on calculating the fraction of nodes that contain a criterion for each specific feature, weighted by the number of samples in each node \citep{RF-MDI}. In other words, the importance is high if, after training the Random Forest, that specific feature is included in many decisions (`question' asked about that feature).
We show the feature importances of the RF classifier for the Monash models in Fig. \ref{fig:importance}.1. In general, the metallicity is by far the most important, and elemental abundances are less important than elemental ratios. This is expected, because the elemental ratios, which describe the relative shape of the peaks, are indeed more descriptive for our problem than the absolute value of the elemental abundances (i.e. the height of the peaks). Usually the ratios of neighbouring elements of the same peak have similar importances. There are also large differences in the amplitude of importances for elemental ratios containing elements from different peaks. This shows that the classifier is sensitive enough to discriminate between elements from different peaks without any background information on them.

\section{Results}
\subsection{Systematic residuals of the observed and the modelled abundances} 
\label{sec:boxplot}
To test whether there are systematic deviations between the model predictions and the observations for specific elements, we used the residual of the features, defined as the difference between the observationally derived value of the Ba star and the diluted AGB model value (for the metallicity, the elemental abundances and elemental ratios). For each of the three classifiers, we took all classified models for all stars and calculated the residual for each feature. In the case of the RF classifier, we derived statistics from 533 data points, as typically $2-4$ models were classified for each of the 180 stars.

\begin{figure*}
  \begin{minipage}[t]{0.375\linewidth}
    \centering
    \includegraphics[width=\linewidth]{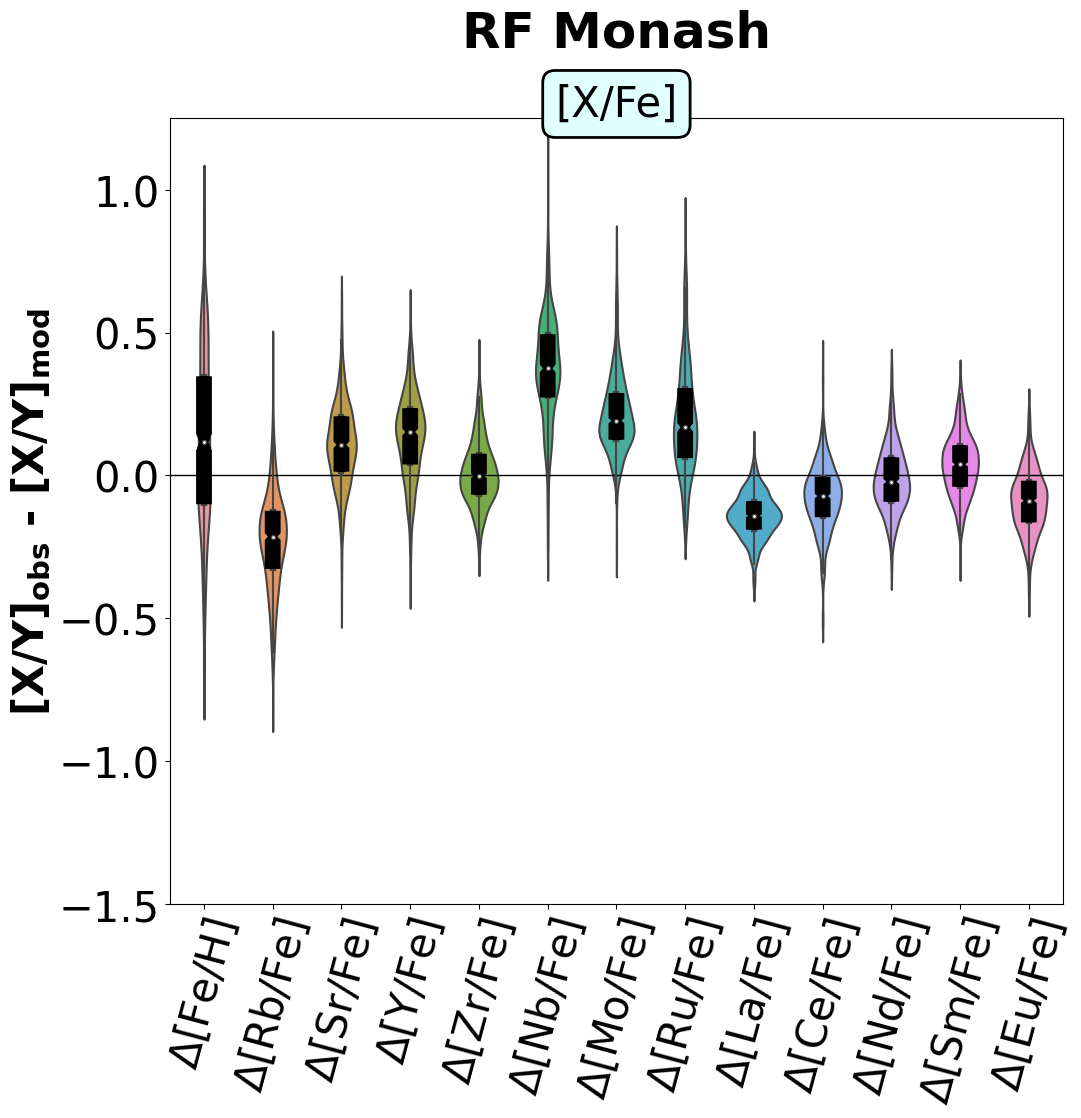}
   \caption{Violin plots of the residuals of the metallicity and the elemental abundances. The residuals are shown for the RF classifier of Monash models, ordered by increasing atomic number. 
   The black boxes above the violins show the interquartile region (IQR) of the residual distribution (i.e. the region where half of the data points lie around the median). Its notch and the corresponding white marker indicate the position of the median. 
   }%
   \label{fig:boxplot-abs}
  \end{minipage}
  \hfill
  \begin{minipage}[t]{0.61\linewidth}
      \centering
   \includegraphics[width=\linewidth]{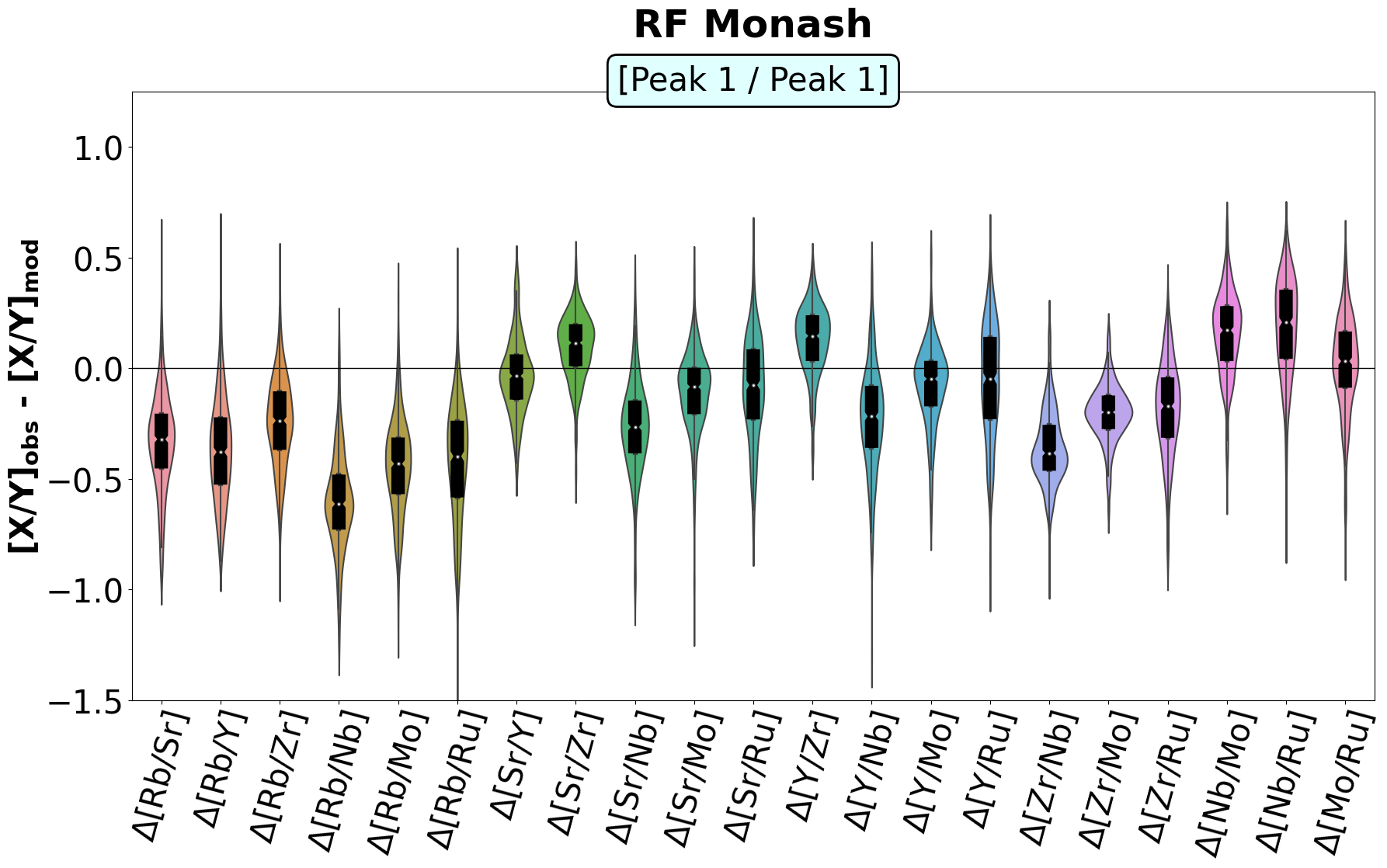}
   \caption{Same as Fig. \ref{fig:boxplot-abs}, but for elemental ratios of Peak 1 elements relative to each other. The ratios are ordered from the lightest to the heaviest element at the numerator.} 
   \label{fig:boxplot-p1p1}
  \end{minipage}
\end{figure*}

\begin{figure*}
\centering
 \includegraphics[width=\linewidth]{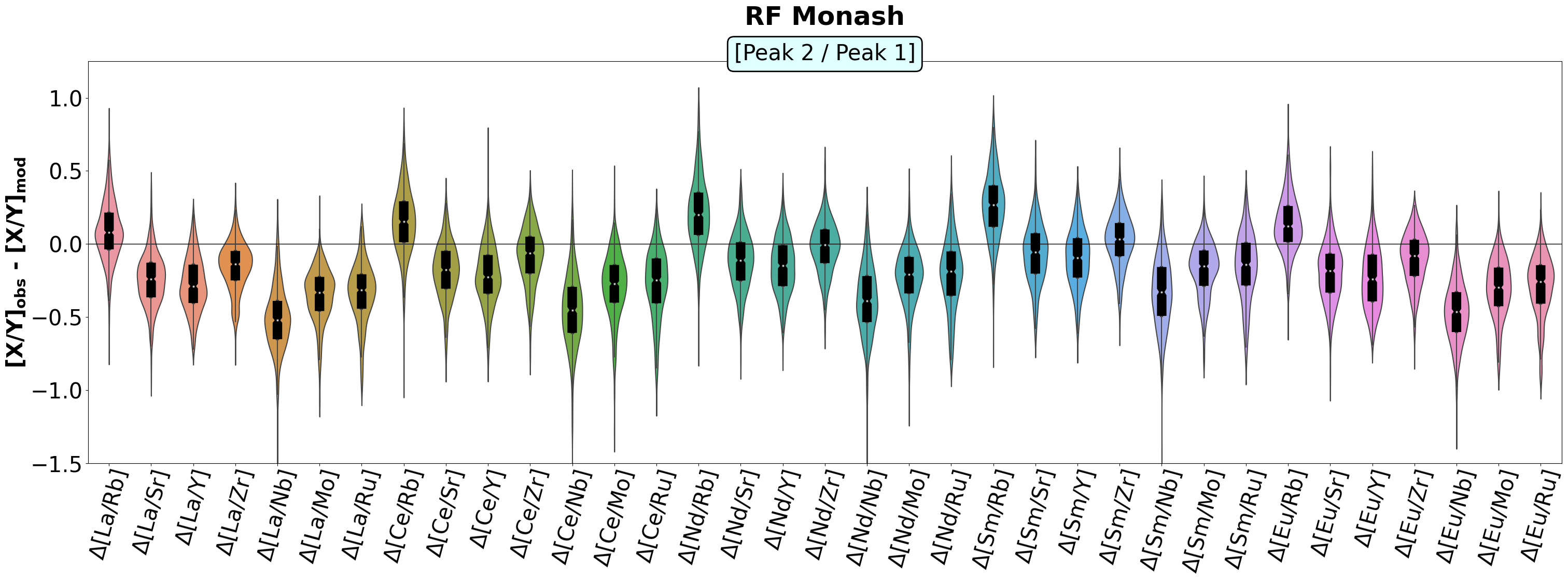}
\caption{Same as Fig. \ref{fig:boxplot-p1p1}, but for elemental ratios of Peak 2 relative to Peak 1 elements.}
\label{fig:boxplot-p2p1}
\end{figure*}

\begin{figure}
\centering
\includegraphics[width=0.65\linewidth]{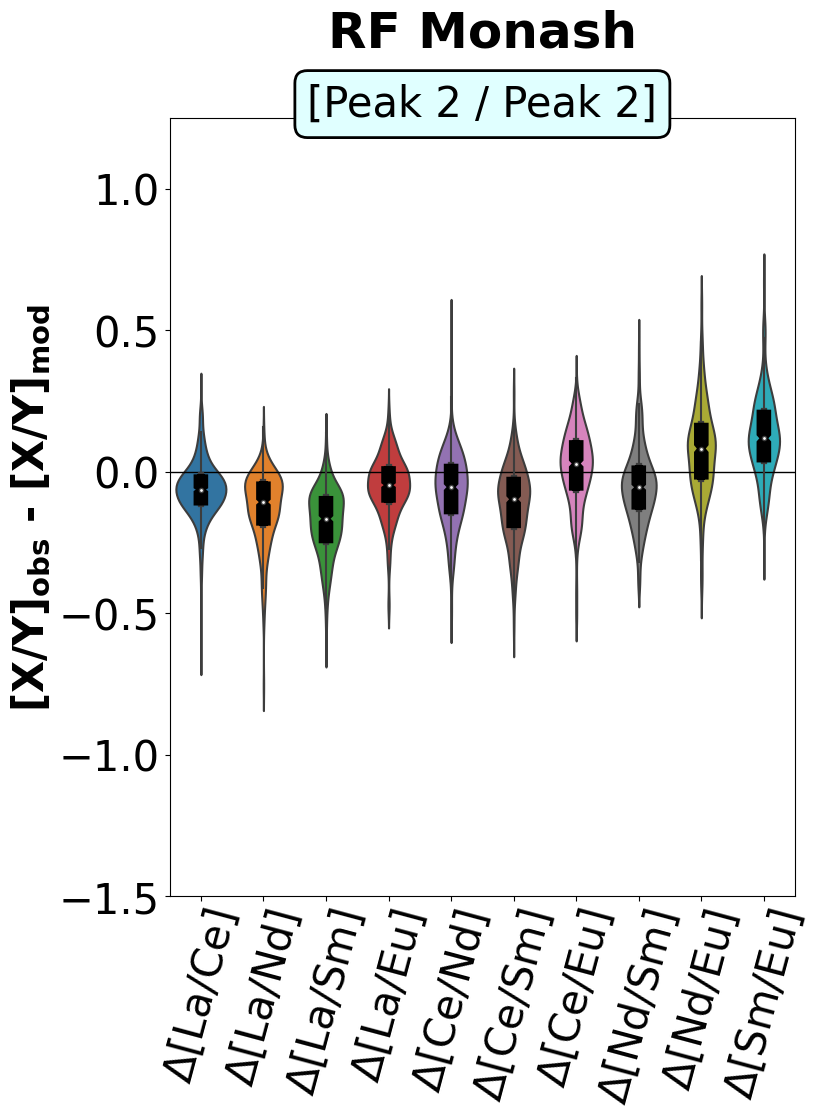}
\caption{Same as Fig. \ref{fig:boxplot-p1p1}, but for elemental ratios of Peak 2 elements relative to each other.}
\label{fig:boxplot-p2p2}
\end{figure}

Using all of these classified models, we created violin plots that demonstrate the distribution of the derived residuals for each feature. The `violins' are in fact the one-dimensional kernel density estimation (KDE) of the corresponding feature. Within the violin plots we can see boxplots showing the interquartile region (IQR) of the distribution (i.e. the region where half of the data points lie). The boxes have a notch and a marker at the median of the data set. We note that the plots are asymmetric. We use the violin plots to illustrate the distributions, and the boxplots for the quantitative analysis. 
We grouped the features based on the $s$-peak to which they belong (or are close to), since different physics may affect their shape (for further discussion, see Sect. 6.3 in Paper II). 
Fig. \ref{fig:boxplot-abs} shows the metallicity and elemental abundances, then Figs. \ref{fig:boxplot-p1p1}, \ref{fig:boxplot-p2p1} and \ref{fig:boxplot-p2p2} present the elemental ratios of different peaks relative to each other, for the RF classifier of Monash models.
We note that even though Nb was excluded when training the classifiers, we include it for statistical analysis. 

Looking at the 1D KDEs (the `violins') of the features instead of their corresponding boxplots gives a better insight into the shape of the distribution of the residuals.\footnote{The relative amplitude of the violins corresponding to the different features may be misleading. The distributions only can be used to illustrate whether there is a discontinuity in the data.}
This allows us to identify stars that are particularly enriched in some elements relative to the others. For example, if there was a clear discontinuity in the residuals, this would indicate a distinct group of stars that are peculiarly enriched in some elements. However, the distribution of the residuals for each elemental abundance and ratio is continuous (or only slightly bimodal), thus we cannot identify any group of peculiar outlier stars.

The position of the boxplots shows a similar pattern for the different classifiers (RF, CM and NN) and for both the Monash and FRUITY models (for their comparison see Fig. \ref{fig:box-big-compare}.3). The medians of the CM and RF classifiers for elemental ratios containing Nb, Mo and Ru are consistent for Monash models, while the residuals of the NN classifier may lie up to $\sim$0.2~dex above. For FRUITY models, the NN and RF elemental ratios are consistent, while the CM may lie below.
The RF classifier predicts the most of these heavy Peak 1 elements relative to the other classifiers. However, the overall pattern is qualitatively the same for all three classifiers. Therefore, in Figs. \ref{fig:boxplot-abs}, \ref{fig:boxplot-p1p1}, \ref{fig:boxplot-p2p1} and \ref{fig:boxplot-p2p2} we  only show the detailed analysis for one classifier, namely RF on the Monash AGB models.

\subsubsection{Metallicity and abundances (Fig. \ref{fig:boxplot-abs})}
\label{subsec:boxpl-absabund}
Figure \ref{fig:boxplot-abs} shows the violin- and boxplots created for the [Fe/H] and the elemental abundances. 
The quantitative statistical measures of the boxplots shown in Fig. \ref{fig:boxplot-abs} (median, mean, limits of the IQR) can be found in Table \ref{tab:boxpl-abs}.
We divided the features presented into sub-categories based on the median of their residual and the range of their IQR (their boxes). For the latter, we use the lower and upper values of the IQR, denoted as `Q1' and `Q3'. The categories for the features are as follows: 
\begin{itemize}
    \item Highly overestimated (median $<0$, Q3 $<0$): \\
        \indent\hspace{0.5cm} [Rb/Fe], [La/Fe];
    \item Slightly overestimated (median $<0$, Q3 $\approx 0$): \\
        \indent\hspace{0.5cm} [Eu/Fe], [Ce/Fe];
    \item Best reproduced (median $\approx 0$, Q1 $<0$, Q3 $>0$): \\
        \indent\hspace{0.5cm} [Nd/Fe], [Zr/Fe], [Sm/Fe];
    \item Slightly underestimated (median $>0$, Q1 $\approx 0$): \\
        \indent\hspace{0.5cm} [Fe/H], [Sr/Fe], [Y/Fe];
    \item Highly underestimated (median $>0$, Q1 $>0$): \\
        \indent\hspace{0.5cm} [Ru/Fe], [Mo/Fe], [Nb/Fe].
\end{itemize}
In other words, overestimated elements are more abundant in the (diluted) models than in the observed composition of Ba stars. Underestimated elements, on the other hand, are underproduced by the models relative to the observations.
The IQR of the highly over- and underestimated elements are so far from zero that, based on our method, these features cannot be reproduced by the diluted models to match the observed composition of Ba stars.

\begin{table}[htbp]
\caption{Statistical measures of the boxplots in Fig. \ref{fig:boxplot-abs}, ordered by the increasing median of the residual. }
\begin{center}
{\small
\begin{tabular}{rrrrr}
\textbf{} & \multicolumn{1}{c}{\textbf{Median}} & \multicolumn{1}{c}{\textbf{Mean}} & \multicolumn{1}{c}{\textbf{Q1}} & \multicolumn{1}{c}{\textbf{Q3}} \\ \hline
\textbf{$\boldsymbol{\Delta}$[Rb/Fe]} & $-0.21$ & $-0.23$ & $-0.32$ & $-0.13$ \\ 
\textbf{$\boldsymbol{\Delta}$[La/Fe]} & $-0.14$ & $-0.14$ & $-0.19$ & $-0.09$ \\ \hline
\textbf{$\boldsymbol{\Delta}$[Eu/Fe]} & $-0.09$ & $-0.09$ & $-0.16$ & $-0.02$  \\ 
\textbf{$\boldsymbol{\Delta}$[Ce/Fe]} & $-0.07$ & $-0.08$ & $-0.14$ & $-0.01$ \\ \hline
\textbf{$\boldsymbol{\Delta}$[Nd/Fe]} & $-0.02$ & $-0.01$ & $-0.09$ & $0.06$  \\ 
\textbf{$\boldsymbol{\Delta}$[Zr/Fe]} & $0.00$ & $0.01$ & $-0.06$ & $0.07$  \\ 
\textbf{$\boldsymbol{\Delta}$[Sm/Fe]} & $0.04$ & $0.04$ & $-0.04$ & $0.10$  \\ \hline
\textbf{$\boldsymbol{\Delta}$[Sr/Fe]} & $0.11$ & $0.11$ & $0.02$ & $0.20$  \\ 
\textbf{$\boldsymbol{\Delta}$[Fe/H]} & $0.12$ & $0.12$ & $-0.10$ & $0.34$  \\ 
\textbf{$\boldsymbol{\Delta}$[Y/Fe]} & $0.15$ & $0.14$ & $0.04$ & $0.23$  \\ \hline
\textbf{$\boldsymbol{\Delta}$[Ru/Fe]} & $0.17$ & $0.20$ & $0.06$ & $0.30$  \\ 
\textbf{$\boldsymbol{\Delta}$[Mo/Fe]} & $0.19$ & $0.21$ & $0.13$ & $0.28$  \\ 
\textbf{$\boldsymbol{\Delta}$[Nb/Fe]} & $0.37$ & $0.37$ & $0.28$ & $0.49$  \\ \hline
\end{tabular}
}
\end{center}
\tablefoot{The columns show the median and mean of the entire residual data set, as well as the lower and upper limits of the IQR (i.e. the boxes in Fig. \ref{fig:boxplot-abs}, Q1 and Q3). All quantities are indicated in dex.}

\label{tab:boxpl-abs}
\end{table}

Rubidium is overestimated, as previously reported in Papers I and II.
Since the abundance of Rb depends on the neutron source operating in the polluter AGB star, Rb can serve as a proxy to determine the AGB mass. Lower Rb values compared to neighbouring elements such as Zr suggest that the \iso{C}{13}($\alpha$,n)\iso{O}{16} reaction is the main neutron source, as opposed to the \iso{Ne}{22} reaction \citep{Rb-Abia}. 
The latter, operating in AGB stars above $\approx\,$4~\msun~(for Monash models at these metallicities), would give [Rb/Zr] ratios greater than 0, thus lower-mass AGB stars are strongly favoured here. 
Another source of this discrepancy may be due to uncertainties in nuclear physics. (See \citealt{roriz21a} for a detailed discussion of the Rb abundances in the sample stars used in this study and \citealt{Rb-vanRaai} for a detailed discussion of Rb nucleosynthesis.) 

Lanthanum is significantly overestimated compared to the observations, possibly due to a systematic observational error (see \citealt{cseh18}), although this has been revised and improved by \cite{roriz21a}. The deviation is much smaller than that of Rb, around half of that value, and the distribution spread is also much smaller.
Furthermore, La and Ce are both second peak elements with overestimated values, indicating a possible common source for this difference.
We note that the cross section of \iso{Ce}{140} has been recently measured to be 40\% higher, that would result in 20\% less Ce, which could account for our overestimation of Ce \citep{Ce-Amaducci}.

The elements that are slightly over-- or underestimated do not prove to be problematic to fit in general. Such differences may arise from various sources, of uncertainties including the AGB models, nuclear physics, observational errors, and our method of classifying the potential polluter AGB star (including the fact that our model grid does not cover all possible input parameters). 
However, elements that are highly underestimated (Nb, Mo, Ru) cannot be explained by these uncertainties alone, as their deviation is at least around 0.2 dex. This difference could be due to a missing process in the AGB models that introduces additional amounts of these elements (see discussion in Sect. \ref{sec:conclusions}).


Outlying data points may not accurately represent the real behaviour of the objects, but rather reflect the coarseness of our grid on AGB model input parameters, such as the metallicity. 
If the algorithm cannot find a model with a metallicity close enough to the true value, then the elemental abundances and ratios may have an offset because the correct model is not included in our set.
We analysed the boxplots excluding the most outlying data points in metallicity (|$\Delta$[Fe/H]| < 0.3 dex, $\approx 200$ points dropped). As a result, all the residuals had narrower boxes, but the median changed only slightly, by not more than 0.02 dex. 
Therefore, the pattern of the elemental ratios still demonstrate the shape described above, and we have decided not to remove any points from the analysis.

\subsubsection{Peak 1 / Peak 1 ratios (Fig. \ref{fig:boxplot-p1p1}) }
When taking ratios of the Peak 1 elements relative to each other, we notice a regular pattern in the residuals: with a fixed element in the numerator, as elements in the denominator increase in atomic mass, the residuals drop when Nb appears in the denominator. The residuals then slowly rise back to around 0 and fall again when Nb appears in the denominator. The lowest residual is seen for [Rb/Nb] with a value of $-$0.5~dex, but all residuals containing Nb in the denominator are at maximum $-$0.2~dex. 
This means that, compared to the observations, the AGB models statistically underproduce the elements from Nb, to Mo and Ru and the deficit is decreasing from Nb to Ru. The underestimation of the abundances for these elements is already visible in Fig. \ref{fig:boxplot-abs}. 


\subsubsection{Peak 2 / Peak 1 ratios (Fig. \ref{fig:boxplot-p2p1}) }
Another periodic pattern appears also when plotting the ratios of elements from the two different peaks. 
There is a sharp (between $-$0.5 and $-$0.4~dex) drop when Nb appears in the denominator, which decreases while moving to Ru. 
A more modest drop (less than half than in the case of Nb) can be seen also if Sr or Y are as denominator: Peak 2 elements compared to Rb or Zr may have residuals $0.1-0.2$ dex higher than those elemental ratios including Sr and Y. This indicates that Rb and Zr are produced in a greater amount than Sr and Y, compared to the Peak 2 elements, relative to the observations. 
Already in Fig. \ref{fig:boxplot-p1p1}, the overproduction of Rb and Zr relative to Sr and Y (or the underproduction of Sr and Y) is visible by the residual of their ratios.

\subsubsection{Peak 2 / Peak 2 ratios (Fig. \ref{fig:boxplot-p2p2})} 
The residuals of Peak 2 elements relative to each other exhibit a distribution with substantially less scatter than those of the Peak 1 elements, as already seen for the elemental abundances in Fig. \ref{fig:boxplot-abs}. This results in very narrow Peak 2 / Peak 2 elemental ratio distributions as well. While these narrow IQR ranges may not always intersect with 0, none of them have a median with an absolute value greater than 0.2~dex. 
The elemental ratios containing La in the numerator are slightly overestimated, as already discussed. Overall, the Peak 2 ratios confirm that the models satisfy the production of Peak 2 elements relative to each other.


\subsection{Correlations}
\label{sec:correl}
We investigated whether there are correlations between the different quantities involved in the study: the observed\footnote{In this section, by observed quantities we mean the elemental abundances derived from observations (by modelling their spectra) as well as their differences, the elemental ratios.} [Fe/H] of the Ba stars; the observed elemental abundances and ratios; the residuals between the classified models and the Ba star abundances; and the $\delta$ parameter. When the observed quantities are plotted on the $y$ axis, 180 data points are visible, one for each star. If residuals are used instead, the number of points is equal to that of the classified models (533, using the RF Monash models in the figures). 

To better visualise the distribution of the points, we also show their two-dimensional kernel density estimation (2D KDE). 
This way, the colouring of the clouds demonstrates the value of the probability density function on a fine grid (a blue-green colour map is used for plots with 180 points, while yellow-pink is used for plots with 533 points). For each subplot, the KDE has been normalised by its maximum value, to allow for direct comparison of the shape of the KDE between subplots containing different elements, without considering their magnitude.

To quantify the degree of correlation, we calculated the Spearman's rank correlation coefficient $r_S$. 
The absolute value of this coefficient is high (see below for exact values) if the points follow a monotonic function (not necessarily linear, as in the case of Pearson's correlation coefficient). If its value is positive, it represents a positive correlation, while negative coefficients imply an anti-correlation between the quantities.

We have also fitted a line to the data as another visual measure of the strength of the relationship and for visualisation purposes. The limits of the two axes are different, so the $a=1$ line does not appear to lie on the 45$^\circ$ diagonal, the values of $a$ are given in the labels. 

For the linear fit, we use Ordinary Least Squares (OLS) in cases where [Fe/H] or $\delta$ is on the $x$ axis, and Orthogonal Distance Regression (ODR) in the other cases. OLS minimises the distance of the points from the line only in the $y$ direction, while ODR minimises the distance on both the $x$ and $y$ axes. 
In this work we investigate the dependence of the amount of $s$-process elements on the parameters of the AGB star, such as the metallicity. This means that we treat the $s$-elemental abundances as independent variables to fit, with errors due to both potential unknown processes and the observations, while [Fe/H] is treated as a quantity with only the observational error. It is therefore justified to treat [Fe/H] as fixed when fitting the line. 
However, when elemental abundances or ratios are on the $x$ axis, we should treat them as independent variables as the other quantity on the $y$ axis, and the fitting with ODR is justified.
In addition, in cases where the slope was high ($a \gtrsim 2$) and there is not the [Fe/H] on the $x$-axis, the OLS fit failed because a) it did not pass through the region with the highest KDE, and b) when the axes were swapped, the slopes of those lines were not the reciprocals of each other. 

It is difficult to state whether the figures show correlation or not. The two quantities plotted show a positive correlation if both $a$ and $r_S$ are positive, and an anticorrelation if both are negative. 
We follow the commonly used nomenclature for the strength of the correlation: 
For $|r_S| \lesssim 0.15$ there is no significant correlation; for $0.15 \lesssim |r_S| \lesssim 0.3$ the correlation is weak; for $0.3 \lesssim |r_S| \lesssim 0.6$ it is moderate; while for $0.6 \lesssim |r_S|$ it is strong. 
We also take into account the magnitude of $a$: if $|a| \lesssim 0.1$, there is no significant relationship; $a \approx 1$ implies a $y=x$ one-to-one correspondence; if $|a| \gtrsim 10$, this again implies no significant relationship (by swapping the axes we would get a slope of $|a| \lesssim 0.1$).
The definitions introduced above are arbitrary, but are intended to give a qualitative description of the results. In general, the correlations are not strong in our data (except for the elemental abundances as a function of each other), but it is important to keep in mind that there are many sources of uncertainty, as discussed in each subsection.

In this section we demonstrate the different types of plots, show examples, and discuss the most notable subplots containing specific $s$-process elements. The full set of plots can be found online at a Zenodo repository\footnote{See DOI \url{https://doi.org/10.5281/zenodo.11299088}}).
From now on, when referring to any elemental abundance or ratio of an $s$-process peak, we write \textit{P1} (for Peak 1 elements) or \textit{P2} (for Peak 2 elements), for example $\Delta\,$[\textit{P2/P1}] for residuals of Peak 2 / Peak 1 elements.

\begin{figure*}[h!]
\centering
 \includegraphics[width=\linewidth]{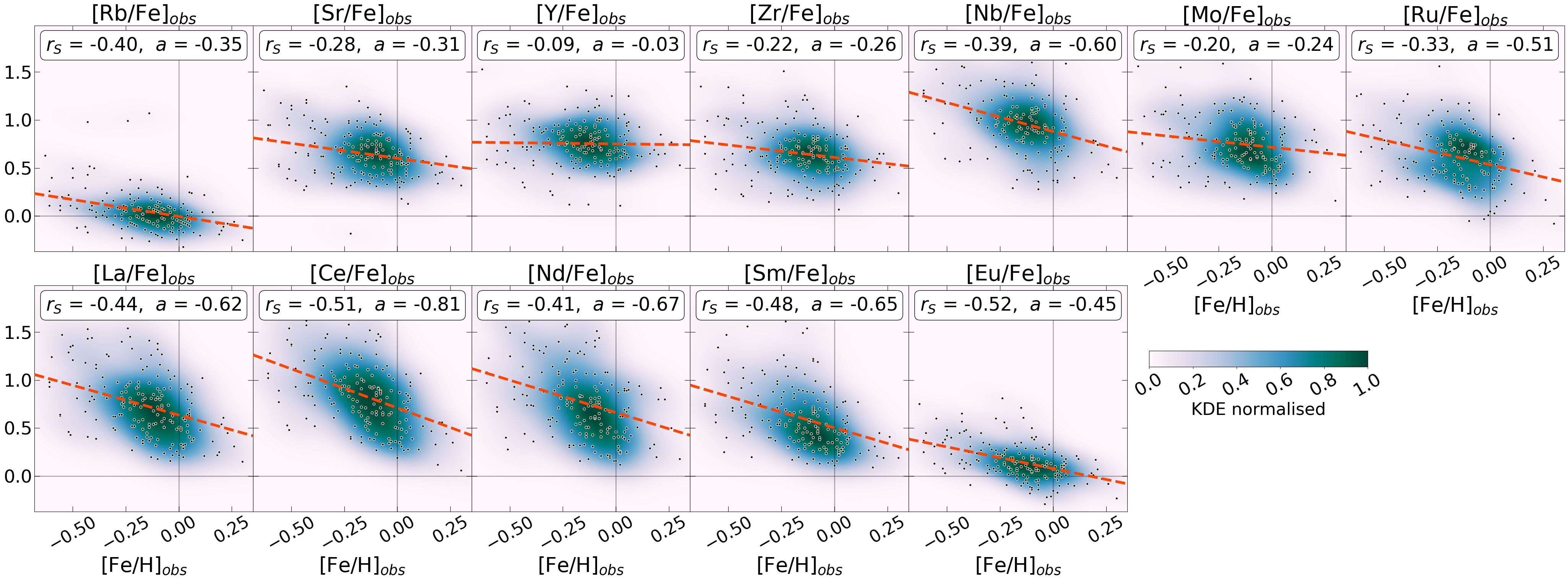}
\caption{Observed elemental abundances [$A$/Fe] (each denoted in the title of the subplot) as a function of the observed [Fe/H]. Here 180 points are shown, one for each star, the colouring corresponds to their 2D KDE.
The Spearman's correlation coefficient $r_S$ and the slope $a$ of the line are indicated. The linear fit was done with OLS.}
\label{fig:correl_obs_feh-Fe}
\end{figure*}

\subsubsection{Observed values as a function of [Fe/H] (Fig. \ref{fig:correl_obs_feh-Fe})}

In this subsection we show the observed elemental abundances and ratios as a function of the [Fe/H] of the Ba star. This approach tests whether the production of $s$-elements depends on the metallicity of the star.
As already \cite{s-vs-feh-busso} showed, the production of the first to the second peak $s$-process elements increases with metallicity. This can be explained by the fact that if there are more Fe seeds in the star, the neutron exposure decreases and consequently heavier $s$-elements are less produced. In contrast, in low-metallicity stars, most of the $s$-process products accumulate in the Pb peak, and therefore we can see a deficit in the first and second peak elements. The plots shown in this subsection generally prove this behaviour of AGB nucleosynthesis.
We note that here we use only observed values, so the results do not depend on our AGB and classification models.

The elemental abundances as a function of [Fe/H] can be seen in Fig. \ref{fig:correl_obs_feh-Fe}. There is a clear trend that [$A$/Fe] decreases with increasing metallicity for both peaks. A slight bimodality can also be seen in the distribution of Mo and Ru. 

The elemental ratios as a function of the [Fe/H] can be seen at the online repository.
The [\textit{P1/P1}] ratios show an almost flat dependence on metallicity (the absolute value of $a$ is low), and with insignificant to weak $r_S$. This indicates that the relative production of Peak 1 elements does not depend on metallicity.
The few exceptions that show a weak to moderate positive relationship are the [Nb/\textit{P1}] and [Y/\textit{P1}] ratios. This is because Y is the least dependent on metallicity among the Peak 1 elements, while Nb is the most affected, as already seen in Fig. \ref{fig:correl_obs_feh-Fe}. Therefore their elemental ratios with other elements depend on the metallicity.

The [\textit{P2/P1}] ratios show a considerable (typically $|a| \gtrsim 0.3$) and moderate to strong anticorrelation with metallicity, implying that the production is shifted towards Peak 1 elements in more metal-rich environments, as expected. A slight bimodality is also visible in the case of [Ce/Rb] and [Nd/Rb].

The [\textit{P2/P2}] ratios show a modest anticorrelation with metallicity. This implies that the heavier elements in Peak 2 are slightly accumulated over their lighter counterparts. This is the most pronounced in the case of Eu. A slight positive relationship is visible for the [La/\textit{P2}] ratios.

\begin{figure*}[h!]
\centering
 \includegraphics[width=\linewidth]{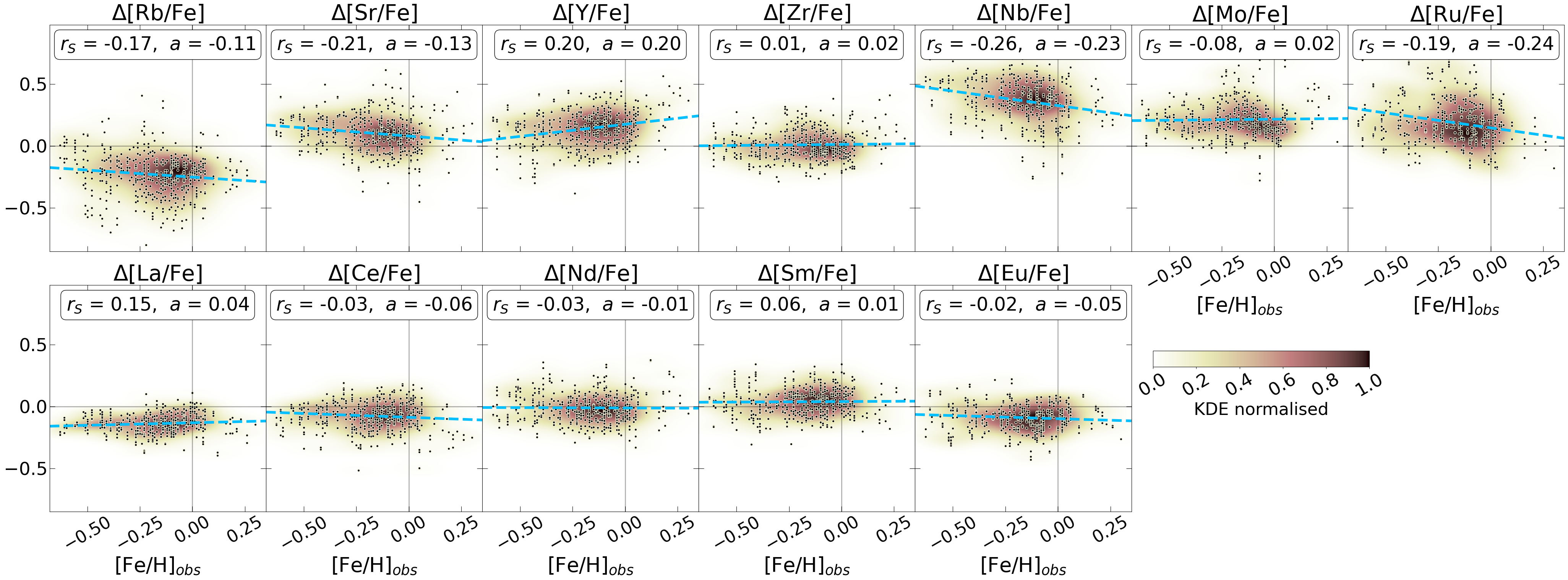}
\caption{Residuals $\Delta \,$[$A$/Fe] as a function of the observed [Fe/H]. Here 533 points are shown, one for each classified model. 
The other properties of the plot are the same as in Fig. \ref{fig:correl_obs_feh-Fe}.}
\label{fig:correl_res_feh-Fe}
\end{figure*}

\subsubsection{Residuals as a function of [Fe/H] (Fig. \ref{fig:correl_res_feh-Fe})} 
\label{subsec:res-feh-iproc}

In this subsection we analyse the correlations of the residuals between the observations and the classified models. The residuals of the Ba star abundances and the classified AGB models as a function of the Ba star [Fe/H] can be seen in Fig. \ref{fig:correl_res_feh-Fe}, while the elemental ratio residuals vs [Fe/H] are shown in
the online repository.

Correlations between the residuals and the metallicity may indicate that the potential extra production source is metallicity-dependent, which is usually reported for the $i$ process (see discussion in Sect. \ref{sec:conclusions}). Therefore, this analysis may be a test if our discrepancies are caused by the $i$ process, or other metallicity-dependent process. 
In general, we find that indeed the residuals of Nb and Ru are the most correlated with the metallicity. 
However, these residuals show a weak correlation, and thus we cannot report a clear metallicity dependence. It should be kept in mind that the residuals include the uncertainties related not only to observations, but also to the AGB models and our model selection (partly constrained by our finite number of AGB models). Therefore, the presence of a larger scatter in the points is inevitable and the $r_S$ coefficients are lower than in the case of using observations only. 

The $\Delta \,$[$A$/Fe] residuals of the elemental abundances show insignificant to weak anticorrelation with the [Fe/H] (Fig. \ref{fig:correl_res_feh-Fe}). 
The extent of negative relationship with the metallicity is the strongest for Nb, followed by Ru, Sr, Rb, Mo; while Zr and Y have zero to positive relationship. This order is further evidenced by the elemental ratios.


Of the $\Delta \,$[\textit{P1/P1}] residuals vs [Fe/H], Nb has a negative relationship with all 6 other Peak 1 elements (considering $\Delta$[Nb/\textit{A}]); Ru is only positively related to Nb; Sr and Rb are positively related to Nb and Ru, while having 0 correlation with each other; Mo is only negatively related to Zr and Y; Zr has only a negative relationship with Y; and Y is positively related to each \textit{P1} element. 

Regarding the $\Delta\,$[\textit{P2/P1}] residuals vs [Fe/H], every \textit{P1} element is negatively related to \textit{P2} elements as a function of metallicity, except for Zr and Y. The absolute value of the correlation and slope decreases in the same order as for $\Delta\,$[\textit{P1/P1}]. It is also worth noting that Ru shows a slight bimodality with respect to almost all \textit{P1} elements and Fe.
The $\Delta \,$[\textit{P2/P2}] residuals have almost no relationship with each other as a function of metallicity. 

If a residual $\Delta [A/B]$ vs [Fe/H] has a negative correlation (with the fitted line being above 0 for low metallicity), this means that at lower compared to higher metallicity, the residuals have a higher value, and thus the models underproduce the element $A$ relative to $B$. This is the case for most of the elements, meaning that they may be produced by a process that is slightly more efficient at lower metallicites and is dependent on metallicity.
Overall, we can see that although the relationships are very moderate, Nb and Ru consistently have the largest underproduction in the models as a function of metallicity. 
However, Mo consistently shows an opposite dependence on metallicity, which complicates the picture.

\begin{figure*}[h!]
\centering
 \includegraphics[width=\linewidth]{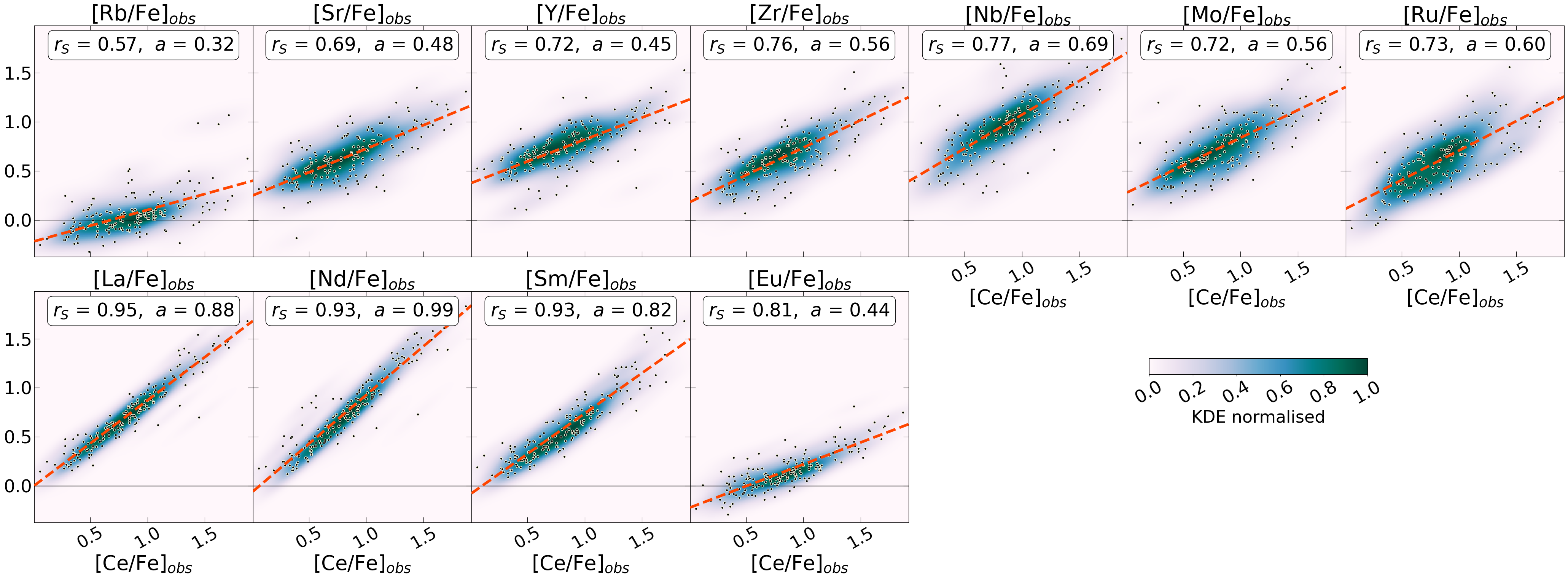}
\caption{Observed elemental abundances [$A$/Fe] as a function of the observed [Ce/Fe]. The colouring corresponds to the 2D KDE of the 180 points shown. The linear fit was performed with ODR.
Note that if we refer to [$A$/Fe] vs [$B$/Fe], but plot [$B$/Fe] vs [$A$/Fe], the sign of the relationship (correlation or anti-correlation) remains the same. In fact, the swap of the two axes results in a reflection around a diagonal line with a slope of 1. The resulting slope is therefore $1/a$, while $r_S$ is the same in both cases.
}
\label{fig:correl_obs_obs_ce}
\end{figure*}

\subsubsection{Observed values as a function of observed values (Fig. \ref{fig:correl_obs_obs_ce})}

The observed elemental abundances as a function of each other may test how the production of the $s$-process elements are related to each other. Deviations from a general trend could mean that some of the elements are partly produced in different processes than the others. This is an important test, as using observations only, we can directly compare the composition of different elements in the stars, without using the models.
We see that Nb, Mo and Ru have a different behaviour (lower correlations and lowest slopes) relative to other Peak 1 elements; while the trends of the Rb and Eu abundances also differ from that of the other elements.
The corresponding plots can be found in the online repository. As an example, the elemental abundances as a function of [Ce/Fe] are shown in Fig. \ref{fig:correl_obs_obs_ce}. 
In this and the following subsections, instead of using the elemental ratios (which would complicate the interpretation of the results), we examine the relative behaviour of two elements by plotting their elemental abundances as a function of each other.

In general, the Peak 1 elements have lower correlations and larger scatters than the Peak 2 elements. We have already pointed out based on the violin plots in Sect. \ref{sec:boxplot} that the elements in Peak 1 are much less consistently reproduced by our models than those in Peak 2 (which have the residual around 0, with less scatter), and this may be the result of a missing nucleosynthesis process affecting the first peak in the models (see discussion in Sect. \ref{sec:conclusions}). 
Here we see that even without using the models, Peak 1 elements do not follow the expected behaviour of $s$-process production, like Peak 2 elements.

The plots of [\textit{P1}/Fe] vs [\textit{P1}/Fe] show a moderate to strong correlation.
The most correlated elements are Nb relative to Zr and Mo. This may be because the Nb is the decay product of \iso{Zr}{93}, and proves that the abundance of Mo is connected to Nb, supposedly due to the same origin.
The second strongest correlations are seen for Zr relative to Sr and Y, that are elements generally well fitted (see Fig. \ref{fig:boxplot-abs}).
[Nb/Fe] vs [\textit{P1}/Fe] abundances have the highest slopes, which means that Nb changes the most with increasing the abundances of other Peak 1 elements. The Ru and Mo abundances also have higher slopes than other Peak 1 elements.
Rb and Ru have the lowest correlation values with all the other Peak 1 elements and each other.
Also, the increase in [Rb/Fe] with [\textit{P1}/Fe] is considerably less steep than for the other Peak 1 elements ($a \approx 0.3-0.4$ as opposed to $a \gtrsim 0.6$), because the range of [Rb/Fe] in the observations is much smaller than for the other elements. The reason for this behaviour supposedly lies in that i) the Rb abundance is almost exclusively sensitive to the AGB mass, and our sample AGB masses are restricted to a small range, ii) even at similar masses, the Rb abundance is much less dependent on metallicity than other Peak 1 elements (see Fig. \ref{fig:correl_obs_feh-Fe}). Relative to the Peak 2 elements, [Rb/Fe] also has a lower slope than the other [\textit{P1}/Fe] abundances (since [\textit{P2}/Fe] vs [Rb/Fe] has a higher slope than relative to any other Peak 1 element).

Almost all of the [\textit{P2}/Fe] vs [\textit{P1}/Fe] abundances show a moderate to strong correlation with slopes $a > 1$, which means that with increasing the abundance of Peak 1 elements, Peak 2 elements are produced even in a greater amount.
The elements with eccentric behaviour are again Rb and Eu, which demonstrate a lower correlation. Rb has higher slopes than the other Peak 1 elements relative to the Peak 2 elements, since the range of [Rb/Fe] abundances is narrow. In the case of Eu, the slopes are considerably lower than in the case of other Peak 2 elements, given a small range of abundances, as in the case of Rb.

The [\textit{P2}/Fe] vs [\textit{P2}/Fe] plots show a high correlation, with much less scatter compared to the other two plots. 
Here, the only element with outlier behaviour is Eu, with a bit lower correlations and lower [Eu/Fe] vs [\textit{P2}/Fe] slopes than other Peak 2 elements, for the reasons discussed above. However, the primordial component of Eu, originating mostly of the $r$ process is not negligible, and therefore this outlier behaviour may not be due to the $s$-process component.

\begin{figure}[h!]
\centering
\includegraphics[width=\linewidth]{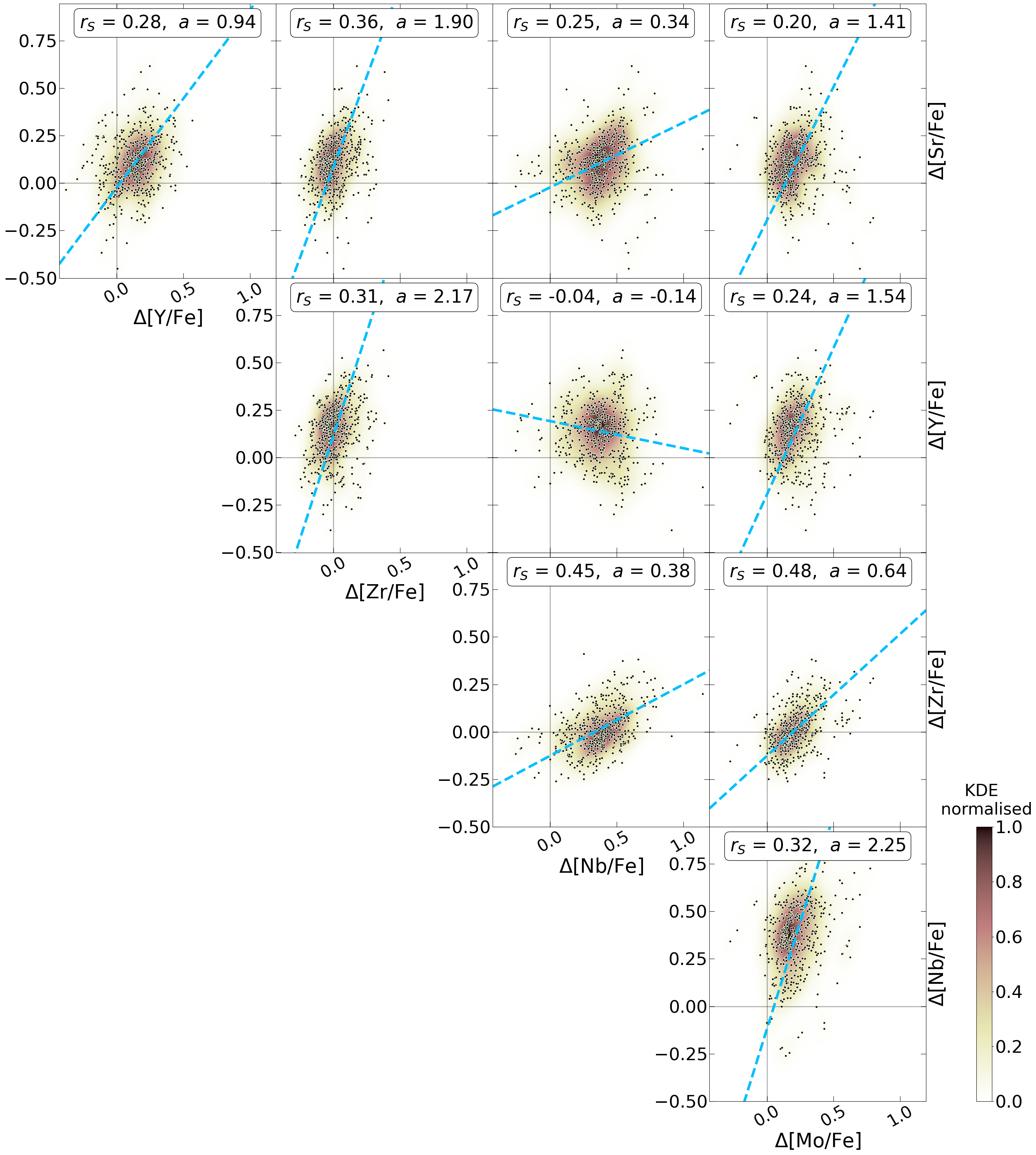}
\caption{Subset of Peak 1 residual plots ($\Delta \,$[\textit{P1}/Fe] vs $\Delta \,$[\textit{P1}/Fe] that contain the elements Sr, Y, Zr, Nb, and Mo. The colouring corresponds to the 2D KDE of the 533 points shown. The linear fit was performed with ODR.
}
\label{fig:correl_res_res_5elem}
\end{figure}

\subsubsection{Residuals as a function of residuals (Fig. \ref{fig:correl_res_res_5elem})}
\label{subsec:correl-res-res}
Plotting residuals as a function of other residuals shows whether different elements are poorly fitted for the same combination of stars and classified models.
If the residuals of some elements are positively correlated with each other, this could mean that our inability to fit them is caused by the same physical mechanism, or there are similar systematics affecting the residuals of both elements. In contrast, if there is an anticorrelation between them, this indicates that as the fit of one element improves, the models fail to reproduce the other. 
We find that the residual of Nb is the most correlated with all the other residuals, meaning that Nb is the most sensitive element to the poorness of fit in any other element. This is similar in the case of Mo. We also find that the residuals of Sr, Y, and Zr are correlated to the residuals of Nb and Mo, which may imply that a potential process causing the mismatch of these two elements (see discussion in Sect. \ref{sec:conclusions}) may affect even lighter elements.
The residuals of the elemental ratios as a function of the other residuals can be seen in the online repository. Fig. \ref{fig:correl_res_res_5elem} shows the subset containing Sr, Y, Zr, Nb, and Mo of the Peak 1 vs Peak 1 residuals.

The residuals of Peak 1 elements ($\Delta \,$[\textit{P1}/Fe] vs $\Delta \,$[\textit{P1}/Fe]) usually have a positive or no correlation with each other. 
The residual of Rb is uncorrelated with those of the other Peak 1 elements, except for Nb, with which there is a clear positive correlation. 
The residuals of Sr, Y and Zr are clearly correlated with each other as well as with Nb and Mo. The correlation of Zr with Nb is expected since the only stable isotope of Nb is formed as a decay product of \iso{Zr}{93}.
The relationship of Sr and Y with Nb and Mo is not expected and suggests that Sr and Y are also part of the effect that results in the underproduction of Nb, Mo and Ru (discussed in Sect. \ref{sec:conclusions}).
The residual of Nb is positively correlated with all elements in Peak 1, which means that it is the most sensitive to the misfit of other elements: if any other fit is poor, Nb is the most prone to have a poor fit as well. 
Mo behaves similarly, except that there is no correlation with Rb.
Instead, Ru correlates only with Nb and Mo, and has a slight anticorrelation with Y.

The residuals of the Peak 2 and Peak 1 elements ($\Delta \,$[\textit{P2}/Fe] vs $\Delta \,$[\textit{P1}/Fe]) show a clear anticorrelation in almost all cases. This means that when we have a better fit for one peak, the fit to the other peak gets worse. This may be due to our classifier method and the Nb problem: if there is a lack of Nb in the models, the algorithm may favour those models that produce more Peak 1 elements, but if the amplitude of Peak 1 is increased, Peak 2 will decrease.

Peak 2 element residuals ($\Delta \,$[\textit{P2}/Fe] vs $\Delta \,$[\textit{P2}/Fe]) show a clear positive relation with each other. This is not surprising, since we can consistently reproduce the shape of the second $s$-process peak: in Fig. \ref{fig:boxplot-p2p2} almost all residuals of the $\Delta\,$[\textit{P2/P2}] ratios are 0. Thus, if one of the Peak 2 elements has a poor fit, the other Peak 2 elements will also have higher residuals. (The only exception is Nd relative to Eu, where the correlation is 0).

\begin{figure*}
\centering
 \includegraphics[width=\linewidth]{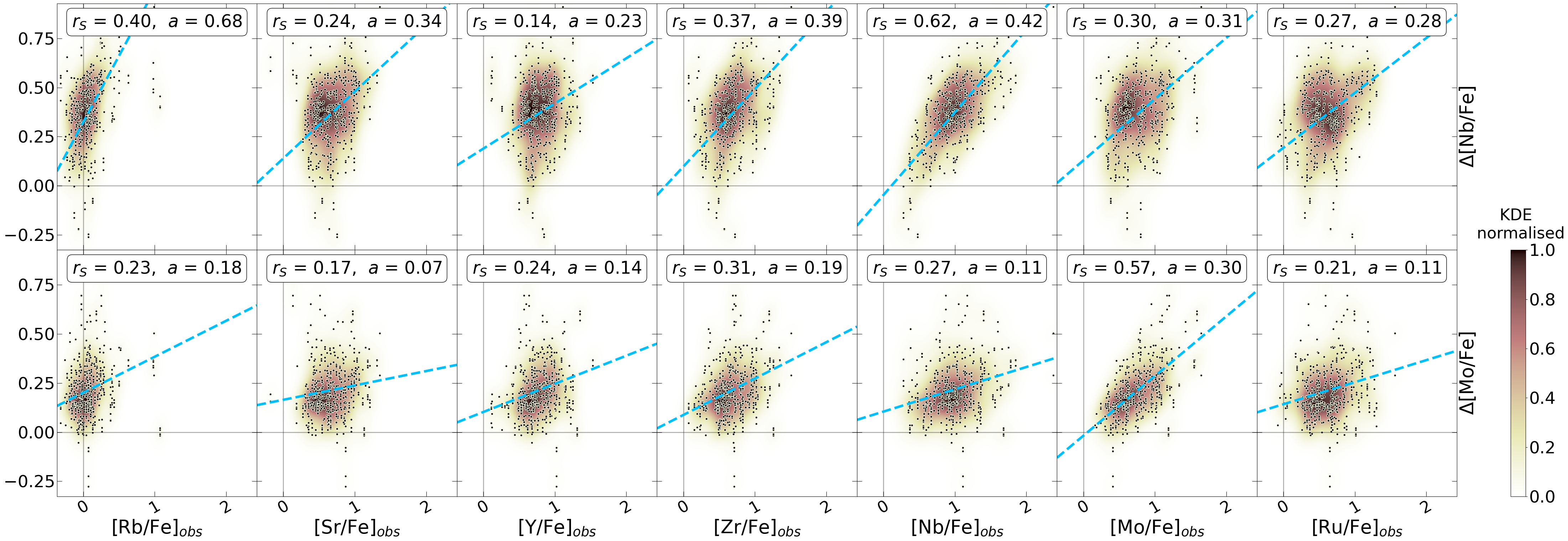}
\caption{Residuals of Nb and Mo as a function of [\textit{P1}/Fe] abundances.
The colouring corresponds to the 2D KDE of the 533 points shown. The linear fit was performed with ODR.}
\label{fig:correl_res_obs_NbMo}
\end{figure*}

\subsubsection{Residuals as a function of observed values (Fig. \ref{fig:correl_res_obs_NbMo})}
\label{sec:correl-res-obs}

The efficiency of the $s$ process is generally in relation with the abundances of the $s$ elements.
Therefore, the analysis of the residuals against the observed abundances can reveal whether our offset increases with the production of certain elements. 
The Peak 1 elements show the highest correlation with their own abundances; the residuals of Nb and Mo are also slightly correlated to all other Peak 1 abundances. The residuals of these two elements also show a correlation with the abundances of Peak 2 elements, implying that the poorness of fit worsens with all of the $s$-elemental abundances.
We find again evidence that the two peaks cannot be reproduced simultaneously.
The corresponding plots can be seen in the online repository. As an example, the residuals of Nb and Mo as a function of the [\textit{P1}/Fe] abundances are shown in Fig. \ref{fig:correl_res_obs_NbMo}.


The residuals $\Delta [$\textit{P1}/Fe]  remarkably show the highest positive correlation with their own [\textit{P1}/Fe] abundances.
Rb, Sr and Zr show no correlation with other \textit{P1} elements.
Y is anticorrelated with all \textit{P1} elements and has a lower residual value at higher Peak 1 production, but it has a positive correlation with itself, making the fit worse at higher Y abundances.
Nb and Mo (see Fig. \ref{fig:correl_res_obs_NbMo}) show a weak correlation with all \textit{P1} elements, and the fit gets worse at higher Peak 1 abundances. These elements are also the most correlated with their respective abundances.
Ru has a non-significant correlation with Rb, Zr, Nb and Mo, and the strongest relationship is with itself.

The residuals $\Delta [$\textit{P1}/Fe] vs [\textit{P2}/Fe] generally show no correlation, except for Eu. 
However, Nb, Mo and Ru show low to moderate correlations with Peak 2 elements, their offset increasing with higher Peak 2 abundances. This also supports our deduction in the previous section that when our algorithm tries to fit a high second peak, the Nb abundances cannot be explained, as already presented for $\Delta\,$[\textit{P2}/Fe] vs $\Delta\,$[\textit{P1}/Fe].

Almost all $\Delta [$\textit{P2}/Fe] vs [\textit{P1}/Fe] residuals show an anti-correlation, with Peak 2 elements having larger offset at higher Peak 1 abundances. This could be due to the same effect as seen in the previous plot (i.e. the first and second peaks cannot be matched simultaneously).
Nd shows a slightly opposite trend to no correlation, as does Eu for some Peak 1 elements.

$\Delta [$\textit{P2}/Fe] vs [\textit{P2}/Fe] generally shows no correlation to a slight anti-correlation, with larger offsets at larger Peak 2 abundances. Nd, however, has a slight positive correlation, but with zero residuals at intermediate values of Peak 2 elements.

\begin{figure*}
\centering
 \includegraphics[width=\linewidth]{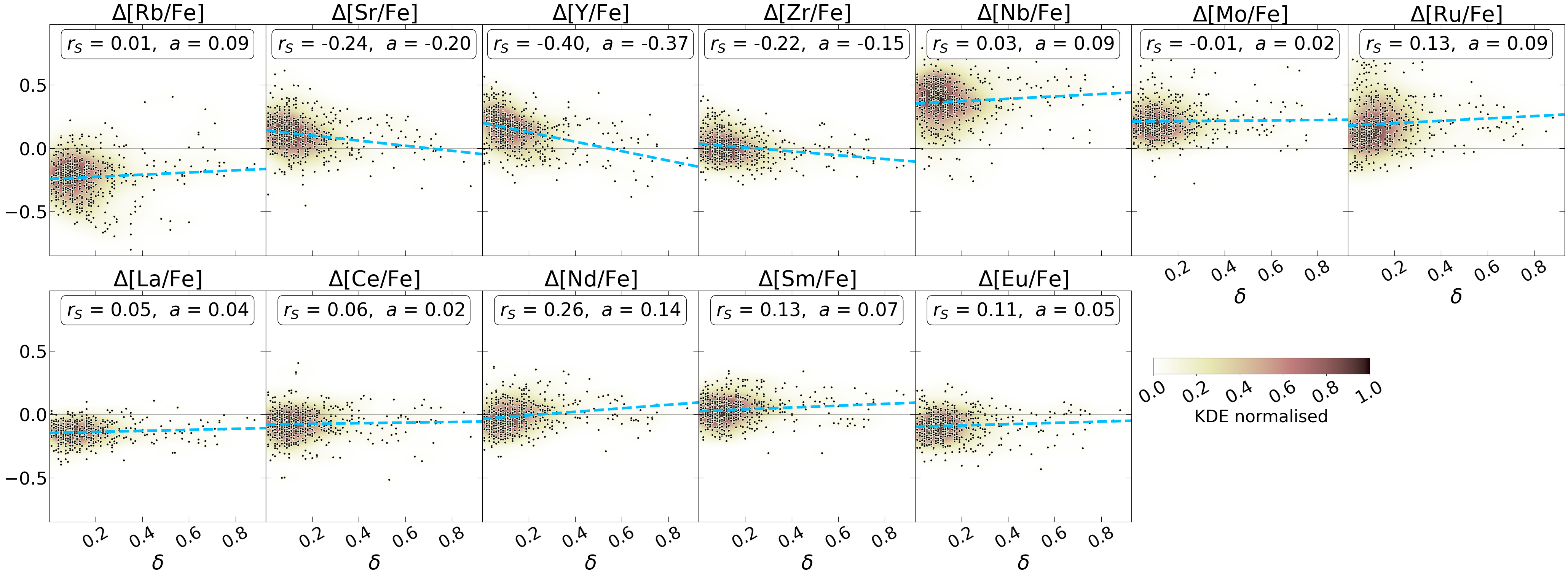}
\caption{Residuals $\Delta\,$[$A$/Fe] as a function of the $\delta$ parameter.
The colouring corresponds to the 2D KDE of the 533 points shown. The linear fit was performed with OLS.}
\label{fig:correl_res_dil}
\end{figure*}

\subsubsection{Dependence on $\delta$ (Fig. \ref{fig:correl_res_dil})}
\label{subsec:correl-dil}

Residuals as a function of $\delta$ show whether our inability to fit the observations depends on the $\delta$ parameter. In Fig. \ref{fig:correl_res_dil} we can see that there is no significant correlation for most elements. 
However, Sr, Y and Zr have a weak to moderate anti-correlation with $\delta$ and have slightly lower residuals at higher $\delta$ values.
If there were high residuals at extreme $\delta$ values, this could mean that our algorithm can only fit the abundances well with low or high $\delta$ values, which is not the case. 
On the other hand, high $\delta$ models fit the general trend well and do not have extremely high residuals. This suggests that binary systems with such high accreted mass could explain the discrepancies and may therefore occur in nature.

The elemental abundances show a clear and strong positive relationship with $\delta$ (see the online repository). This is not surprising: to achieve a higher observed $s$-elemental abundance, more AGB material is needed in the Ba star envelope. 
The observed [Fe/H] shows no dependence on $\delta$ ($r_S = -0.09$, $a=-0.15$).

\subsection{Distribution of model parameters}
\label{sec:distributions}
\subsubsection{Distribution of AGB model masses}
\label{subsec:distr-mass}
\begin{figure}[h!]
\centering
\includegraphics[width=.9\linewidth]{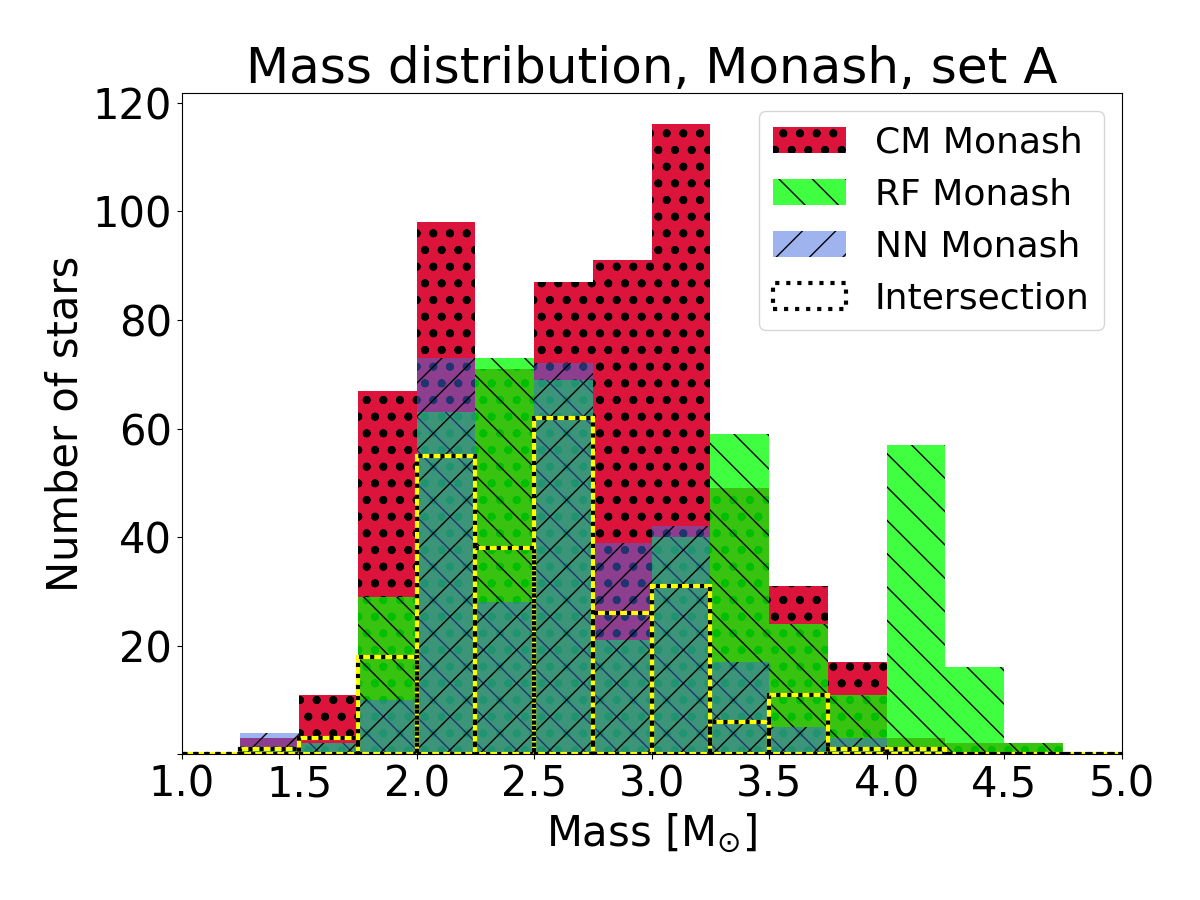}
\caption{Mass distribution histogram of the polluting AGB stars based on Monash models. Here we used element set A (all elements except Nb). Each star is counted in bins in the range of the classified minimum and maximum masses. Different colours indicate the models of different classifiers. The bin width is 0.25 \msun. The dashed line labelled 'Intersection' shows the histogram for stars classified in the same bin by at least two of the classifiers.}
\label{fig:distribution-mass-mon-setA}
\end{figure}

\begin{figure}[h!]
\centering
\includegraphics[width=.9\linewidth]{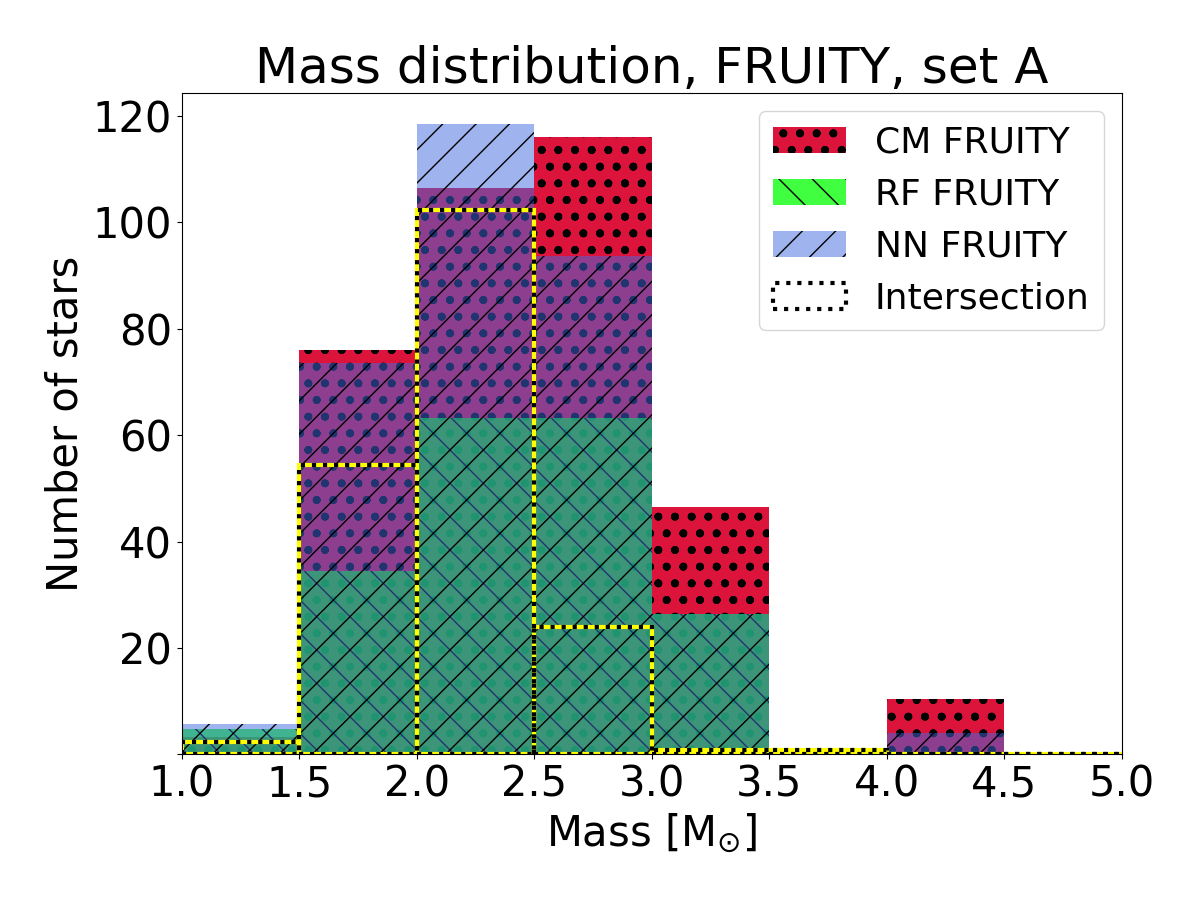}
\caption{Same as Fig. \ref{fig:distribution-mass-mon-setA}, but for classifiers on FRUITY models. The bin width is 0.5 \msun~because this is the resolution of the FRUITY model masses. The bin at 3.5 \msun~is empty because FRUITY models have no models within that mass range. }
\label{fig:distribution-mass-fruity}
\end{figure}

The distribution of the classified AGB model masses provides a valuable insight into the typical mass ranges of potential polluters in the Ba star systems.
Figs. \ref{fig:distribution-mass-mon-setA} and \ref{fig:distribution-mass-mon-setF} show the distribution for Monash models, using element sets A and F respectively, while Fig. \ref{fig:distribution-mass-fruity} shows the distribution for FRUITY models with element set A. All three classifiers are displayed in each plot. 

In these histograms, each star is assigned to a range of bins based on their minimum and maximum masses of classified AGB models.  The star is counted in the bin of its minimum and maximum classified mass, and every bin in between. As a result, stars may appear in more than one bin if their masses fall within the range of multiple bins. 
We are therefore only interested in the shape of the histograms from the different classifiers. The height of the histograms is not important because it scales a) with the average number of models identified for each star by that classifier, b) with the range of masses identified, since if the mass identification is uncertain, the stars can be counted many times, and therefore cause a higher histogram than for other classifiers. 

We defined an intersection of the bins combining the results of the three classifiers. A star is counted in the intersection bin if at least two classifiers agree that the polluter AGB is in that regime. The intersection of the classifiers follows a similar shape to the individual classification distributions but takes into account the results of all three classifiers.

The distribution using Monash models with element set A in Fig. \ref{fig:distribution-mass-mon-setA} shows an unrealistic peak at 4 \msun~ for the RF classifier only. The peak disappears when we use element set F (see Fig. \ref{fig:distribution-mass-mon-setF}). This bias may be due to the training of the classifier, although we tried to avoid overfitting. This is the only major difference between using element set A and F, all the violin plots and correlations change only slightly quantitatively, but not qualitatively.

The different classifiers give differently shaped distributions. The two ML classifiers (RF and NN) show a distribution that peaks between 2 and 2.5~\msun, decreasing more slowly towards the higher masses, while the CM classifier gives a flatter maximum between 2 and 3.5~\msun. Although the exact shape is different for the classifiers, they all agree that the majority of polluter AGBs are of low mass, typically not exceeding 4~\msun. This is also supported by the analysis of the correlations between the abundances. As discussed in Sect. \ref{sec:correl}, the Rb abundances are typically lower than those of the other Peak 1 elements, indicating a low-mass AGB origin due to the nature of the neutron source.
This is consistent with the WD masses of Ba star sample derived from orbital parameters by \cite{Escorza-newmasses}. Based on their analysis, there should be no $s$-process producing AGB stars with masses lower than 1 \msun, while some systems require masses higher than 3 \msun~(the exact masses depend on the initial-final mass relation of the AGB stars and WDs). These authors also find a high-mass tail, which may coincide with our bump using RF set A, however, our bump still appears to be unrealistically high.

In Paper II we presented the mass distribution of the classified AGB models for the intersection of the CM and NN classifiers. The distribution derived here is slightly different because we have plotted a histogram for all three classifiers individually, and the classification criterion for the NN has been slightly changed. Qualitatively, the two distributions agree in having a small number of AGB masses below 1.75 \msun~and above 4 \msun, while their maxima are close to 2 \msun.

The mass distribution based on the FRUITY models is shown in Fig. \ref{fig:distribution-mass-fruity}. In agreement with the distribution based on the Monash models, it peaks between 2 and 2.5 \msun~and essentially does not exceed 3.5 \msun. 

\subsubsection{Distribution of the $\delta$ parameter}
\begin{figure}
\centering
\includegraphics[width=.9\linewidth]{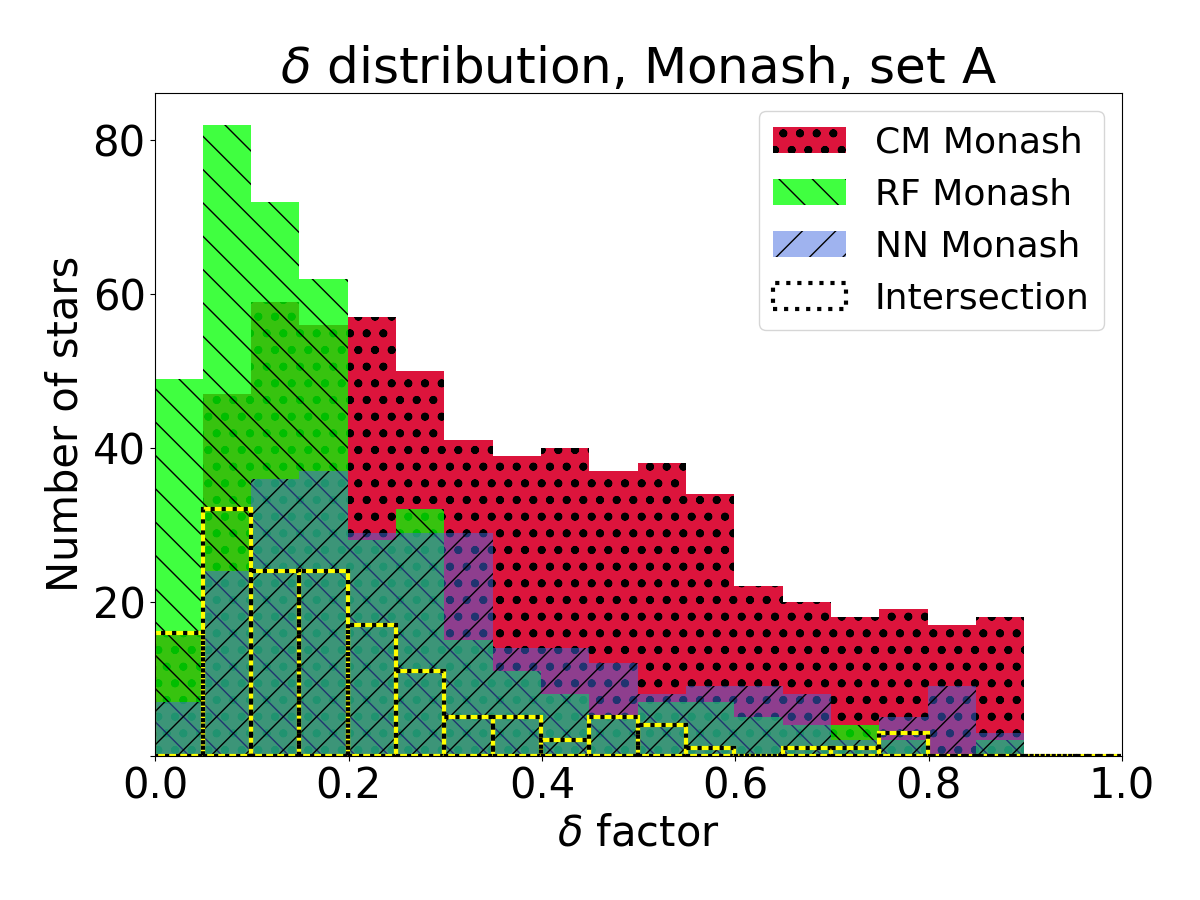}
\caption{Same as Fig. \ref{fig:distribution-mass-mon-setA}, but for the $\delta$ parameter using Monash models with element set A. The bin size is 0.05.}
\label{fig:distribution-dil-mon-setA}
\end{figure}

The distribution of the $\delta$ parameter is useful for studying the efficiency of the past accretion event in the binary systems of Ba stars. (For detailed works on the binary evolution of Ba stars, see for instance \citealt{binary-Jorissen}, \citealt{binary-Pols}, \citealt{binary-Izzard}, \citealt{binary-Stancliffe}, and \citealt{binary-Escorza}.)
Higher $\delta$ values corresponds to Ba stars composed of relatively more AGB material.  The $\delta$ distribution for Monash models using element set A can be seen in Fig. \ref{fig:distribution-dil-mon-setA}, while for FRUITY models in Fig. \ref{fig:distribution-dil-fru-setA}. The height of the bins and their intersection is determined in the same way as for the mass distribution.

All the three classifiers agree in the case of the Monash models (Fig. \ref{fig:distribution-dil-mon-setA}) that high $\delta$ systems are much rarer (but not impossible) than low $\delta$ systems, and the distribution gradually decreases with $\delta$. The classifiers RF and NN have their maxima between $\delta = (0.05 - 0.2)$, with a steeper decrease than the CM, which has a plateau between $\delta = (0.3-0.6)$.
When using the element set F, the shape of the distribution is qualitatively the same, but the plateau in the distribution of the CM disappears and the distribution of the RF decreases less steeply towards high $\delta$ values.

For FRUITY models (Fig. \ref{fig:distribution-dil-fru-setA}), the distributions have a similar shape but shifted towards higher $\delta$ by about 0.1. This is presumably due to the fact that FRUITY models eject fewer $s$-process elements compared to Monash models with the same parameters (see Fig. 4. in Paper II). A higher $\delta$ is therefore required to match the observed composition of the Ba star in general. The distribution of the CM classifier is shifted even further, peaking in the region $\delta = (0.4-0.6)$.

The distribution is well constrained only in the middle region, since near the extreme $\delta$ values effects due to observational biases and our method can change its shape. The low-$\delta$ end of the distribution ($\delta \lesssim 0.2$) may be higher in reality because Ba stars with lower enrichment in $s$-process elements are not identified as Ba stars in spectroscopic surveys. Thus, we are likely to miss many stars in this regime due to selection effects.
The high-$\delta$ end of the distribution is artificially truncated at $\delta = 0.9$, as we assumed that above this value the solution is not realistic and our algorithm rejects these models (see Paper I and II).

We can expect the $\delta$ distribution to be somewhat correlated with the accreted mass distribution, but it is important to note that they do not follow each other exactly. The accreted mass is the product of $\delta$ and the initial mass of the Ba star envelope, since we assume homogeneous mixing in the envelopes of Ba giants due to convection. 
The Ba star envelope mass is unknown for the systems studied and is usually difficult to estimate, since it requires the initial mass of the Ba star and the effect of accretion in a stellar structure analysis. 
In the case of systems with low $\delta$, we can assume that accretion has not altered the evolution of the star much. 
However, Ba stars in systems with high $\delta$ changed their evolutionary path considerably due to mass growth, which is an interesting application of future work.

\section{The \iso{Zr}{93} -- \iso{Nb}{93} thermometer of the \textit{s} process}
\label{sec:NbZr}
\subsection{The method}
The [Zr/Fe] and [Nb/Fe] abundances of chemically peculiar stars extrinsically enriched with $s$-process material (contaminated by a companion star, such as Ba stars) can help to estimate the $s$-process temperature, as shown by \cite{ZrNb-Neysk}. Nb has only one stable isotope, \iso{Nb}{93}, which is the decay product of \iso{Zr}{93}, with a half-life of $\sim$1.5 Myr. For extrinsic stars, we can assume that all the \iso{Zr}{93} has decayed to \iso{Nb}{93} because it is most likely that the mass accretion stopped a longer time ago than the decay time.

Assuming steady-state $s$ process, the [Zr/Fe] abundance can be expressed as a function of the [Nb/Fe] abundance by a linear equation (with slope 1 by definition given the use of logarithms), as follows (using the nomenclature of \citeauthor{ZrNb-Neysk}):
\begin{align}
    \left[\frac{\mathrm{Zr}}{\mathrm{Fe}}\right]&=\left[\frac{\mathrm{Nb}}{\mathrm{Fe}}\right]+\log \left(\omega^*\right)-\log \frac{N_{\odot}(\mathrm{Zr})}{N_{\odot}(\mathrm{Nb})}, \label{eq:NbZr}\\
    \omega^*&=\left\langle\sigma_{93}\right\rangle\left(\frac{1}{\left\langle\sigma_{90}\right\rangle}+\frac{1}{\left\langle\sigma_{91}\right\rangle}+\frac{1}{\left\langle\sigma_{92}\right\rangle}+\frac{1}{\left\langle\sigma_{94}\right\rangle}\right), \label{eq:omegastar}
\end{align}
where $\sigma_i$ is the cross section of the Zr isotope \iso{Zr}{i} and $N_\odot (A)$ is the number of nuclei $A$ in the Sun. The solar fraction $\log \frac{N_{\odot}(\mathrm{Zr})}{N_{\odot}(\mathrm{Nb})}$ is equal to 1.18, using \cite{solar-g98} values, which are the solar values used by our source of observations. In the more recent solar abundances by \citealt{solar-asp09}, this ratio differs in the second decimal only, which would change \omast by about 10\% in the later analysis.

The term \omast depends on the temperature of the $s$ process through the temperature dependence of the cross sections of the Zr isotopes. Therefore, the experimental measurement of the cross sections can be used to constrain the relation between \omast and the temperature.
We note that these equations only hold if the abundances are in nuclear statistical equilibrium ($\langle \sigma_i \rangle N_i$ is roughly constant) and if the neutron density is not high enough for the branching point at the unstable \iso{Sr}{89}, \iso{Sr}{90} and \iso{Zr}{95} to reduce or produce significantly \iso{Zr}{90} or \iso{Zr}{96}, respectively (see discussion in Sect. \ref{subsec:ZrNb_comparison}).

According to the above, the [Zr/Fe] vs [Nb/Fe] abundances of Ba stars should lie approximately on a line with slope 1 and intercept $(\log\,\omega^* - 1.18)$. This intercept depends on the temperature and can therefore constrain it under the assumption of steady state and the use of available laboratory measurements of the Zr isotope cross sections. In other words, each parallel line represent different $s$-process temperature and can be referred to as an `isothermal'. 

\subsection{The \omast -- $T$ relation}
\begin{figure}
\centering
\includegraphics[width=\linewidth]{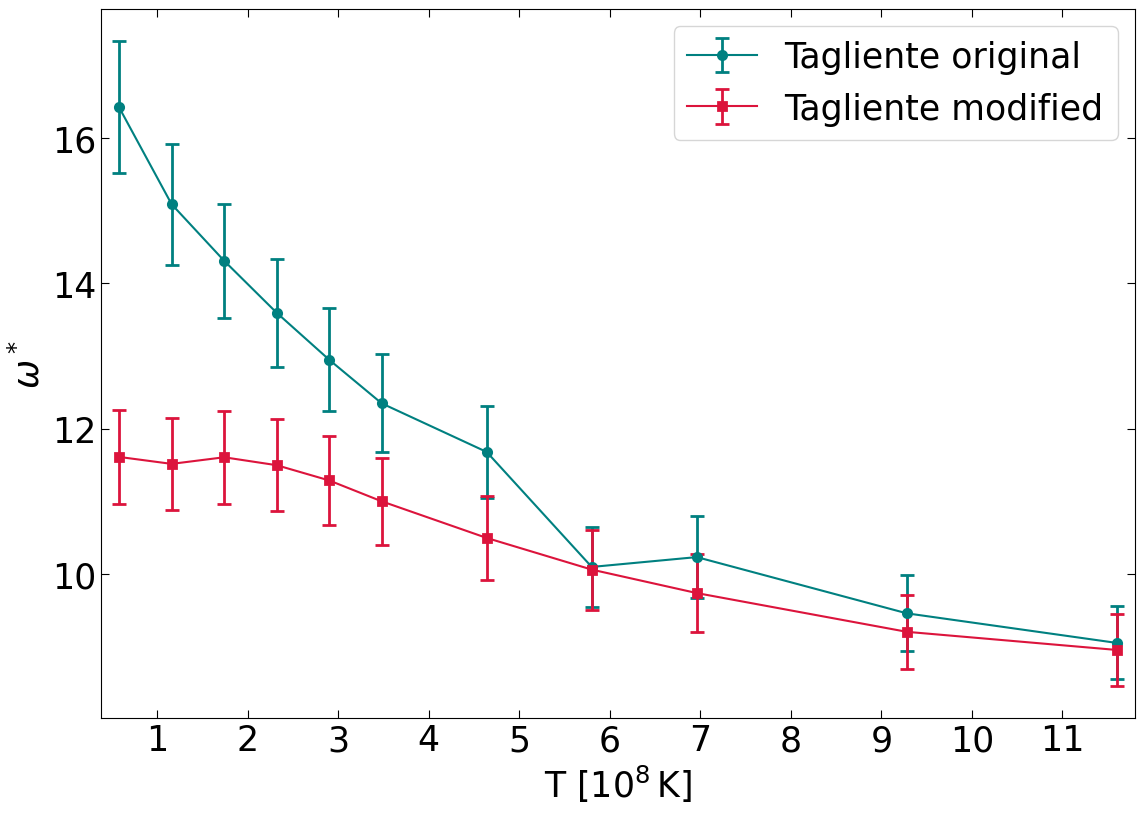}
\caption{Values of $\omega^*$ calculated from the Zr isotopic cross sections (see Eq. \ref{eq:omegastar}) as a function of the temperature.  The `Tagliente original' points are the values published in the Tagliente et al. papers listed in the main text, while the `Tagliente modified' points are derived by scaling the theoretical kernels by 0.65. The error bars are derived propagating the relative error of 5\% on all cross sections.}
\label{fig:omega-T}
\end{figure}

To calculate the \omast -- $T$ relation, we have used the most recent cross sections measured at the CERN n\_TOF experiment. The data come from the works of Tagliente et al: \citeyear{crossec-Zr90} for \iso{Zr}{90}; \citeyear{crossec-Zr91} for \iso{Zr}{91};
\citeyear{crossec-Zr92} and \citeyear{crossec-Zr92-2} for \iso{Zr}{92};
\citeyear{crossec-Zr93} for \iso{Zr}{93};
\citeyear{crossec-Zr94} for \iso{Zr}{94}, as summarised in \cite{Lugaro14}.
However, it should be noted that the value of $\sigma_{93}$ at low temperatures may be up to 30\% lower than published. The reason is that the n\_TOF measurements for \iso{Zr}{93} were limited to energies below 8 keV and had to be complemented with theoretical data. The final cross section can be be therefore calculated by scaling or not the theoretical data by the same difference of 35\% found between the theoretical and the measured capture kernels below 8 keV. In Fig.~\ref{fig:omega-T} we plot the \omast as function of temperature (Eq. \ref{eq:omegastar}) for the two possible cases, including their error bars.\footnote{Assuming independent and uncorrelated variables, the absolute error of \omast (after calculating the partial derivatives and rearraging):
\begin{equation*}
    \Delta \omega^* = \sqrt{\sum_{i=90}^{94} \left(\frac{\partial \omega^*}{\partial \sigma_i}\right)^2 \Delta \sigma_i} =
    \sigma_{93} \cdot \delta \sigma \cdot \sqrt{\left(\sum_{i\neq93} \frac{1}{\sigma_i} \right)^2
    + \sum_{i\neq93} \frac{1}{\sigma_i^2}},
\end{equation*}
where $\delta \sigma = 0.05$ is the typical relative error of the cross sections.}

In AGB stars, the \iso{Ne}{22} neutron source becomes dominant over the \iso{C}{13} neutron source at about 250 MK, corresponding to $\omega^* \approx 13.0$ or 11.5 (see the two set of points in Fig. \ref{fig:omega-T}).  AGB models can reach temperatures of at most around $350-400$ MK ($\omega^* \approx 11.5$ or 10.5) at the $s$ process site, but barely above 300 MK in the low-mass AGB stars that we identify as possible polluters of the Ba stars (see Sect. \ref{subsec:distr-mass} and e.g. \citealt{maxtemp_Busso01}, \citealt{maxtemp_Lugaro03}, \citealt{maxtemp_Kappeler}).

\subsection{Deriving AGB abundances from the Ba star abundances}
\label{subsec:NbZr-AGB}
We show the [Zr/Fe] vs [Nb/Fe] abundances for our sample of Ba stars in Fig. \ref{fig:NbZr-twofig}, panel a). The different lines with slope 1 correspond to different isothermals, as discussed in Sect. \ref{sec:NbZr-temp}. However, because of the accretion and mixing of the AGB material into the Ba star envelope, a direct comparison of the Ba star abundances and the theoretical $s$-process isothermal lines is in principle inaccurate, since the points and the theoretical lines do not represent the same quantity. 
Previous works (\citealt{ZrNb-Neysk}, \citealt{ZrNb-Kar} and \citealt{roriz21b}, see in detail in Sect. \ref{subsec:ZrNb_comparison}) used the \iso{Zr}{93} -- \iso{Nb}{93} thermometer based on the Ba star abundances, because they did not estimate the amount of dilution. Thanks to our classification method, we can remove the effect of the dilution of the AGB star material in the Ba star to calculate the original abundance in the AGB star and check if the conversion from Ba star to AGB abundances may affect conclusions on the $s$-process temperature.

We have an estimate for the value of the $\delta$ parameter for each of the models that match each Ba star, based on the goodness of fit of the classified models.
Therefore, we can calculate the original abundance of the AGB star, given the definition of $\delta$ (the fraction of the material in the form of AGB material) and $(1-\delta)$ (the fraction from the initial composition of the Ba star, see Eq. (1) in Paper II, rearranged):
\begin{equation}
    [A\mathrm{/Fe]_{AGB}} = \log \left( \frac{1}{\delta} \Bigl\{
    10^{[A\mathrm{/Fe]_{obs}}} - 
    (1-\delta) \cdot 10^{[A\mathrm{/Fe]_{ini}}}
    \Bigl\} \right),
    \label{eq:AGB_fromBa}
\end{equation}
where $[A\mathrm{/Fe]_{AGB}}$ is the original abundance of the element $A$ in the AGB star ($A$ in this case is Nb or Zr), $[A\mathrm{/Fe]_{obs}}$ is the abundance of the Ba star now observed, while $[A\mathrm{/Fe]_{ini}}$ is the initial abundance of the Ba star before the accretion event. 
In other words, the $[A/\mathrm{Fe]_{AGB}}$ values are the Ba star abundances, reversed into the AGB stars abundances using the $\delta$ values derived from the fitting of the AGB models. The result is shown in Fig.~\ref{fig:NbZr-twofig}, panel b) as coloured points, while the grey crosses represent the original abundances of the Ba stars, that are in panel a). As there are multiple AGB models, therefore, multiple $\delta$ values that fit the same Ba star, more data points are present as AGB star abundances than original Ba star abundances. 
This shift is largest at low $\delta$, when the Ba star is heavily enriched in $s$-process material even though only its small fraction is originating from the AGB material.

The shift from Ba to AGB star compositions is close but not strictly along on the slope-1 (isothermal) lines. It would be exactly along the isothermal lines if the observed [Zr/Fe] and [Nb/Fe] abundances were equal (and consequently the [Zr/Nb] ratios were equal to 0), however, this is not the case. Nevertheless, the deviation from the isothermal lines is not too large: since the abundances are of similar magnitude (and [Zr/Nb] is close to 0 in the Ba star abundances, typically -0.2), the [Zr/Nb] ratio is almost the same in the AGB star material and in the Ba star after accretion. 
In conclusion, the Ba star points do not deviate much from the isothermal lines when converted to AGB star abundances and the temperature range drawn from the Ba star abundances are qualitatively the same as those derived from the AGB abundances. 

A complication with Eq. (\ref{eq:AGB_fromBa}) is that we do not know the initial composition of the Ba star, $[A\mathrm{/Fe]_{ini}}$. We can reasonably assume that the relative abundance of the heavy elements is similar to solar ($[A\mathrm{/Fe}] \approx 0$), especially at such high metallicities as for Ba stars.
\cite{galcomposition} find that most galactic stars in the metallicity range of our sample have [Zr/Fe] between -0.3 and 0.3 dex, but no data are available for [Nb/Fe]. Fortunately, this dispersion does not lead to large differences in $[A/\mathrm{Fe]_{AGB}}$, even assuming extreme initial values. As an example, with a relatively low observed $[A/\mathrm{Fe]} = 0.5$ and $\delta = 0.01$, the difference in $[A/\mathrm{Fe]_{AGB}}$ is 0.26 dex, assuming $[A\mathrm{/Fe]_{ini}} = 0$ or 0.3 dex. For higher $\delta$ or higher observed abundances the difference between $[A/\mathrm{Fe]_{AGB}}$ is dramatically reduced and is typically less than 0.1 dex. Furthermore, since Nb is a decay product of Zr, we can assume that the initial composition of the two elements may differ from 0 in a similar way, again producing a shift around the isothermal line. This means that the uncertainties due to the unknown initial abundances cannot explain offsets much larger than 0.1 dex.



\subsection{The estimated \textit{s}-process temperatures}
\label{sec:NbZr-temp}

\begin{figure}[h!]
\centering
\includegraphics[width=\linewidth]{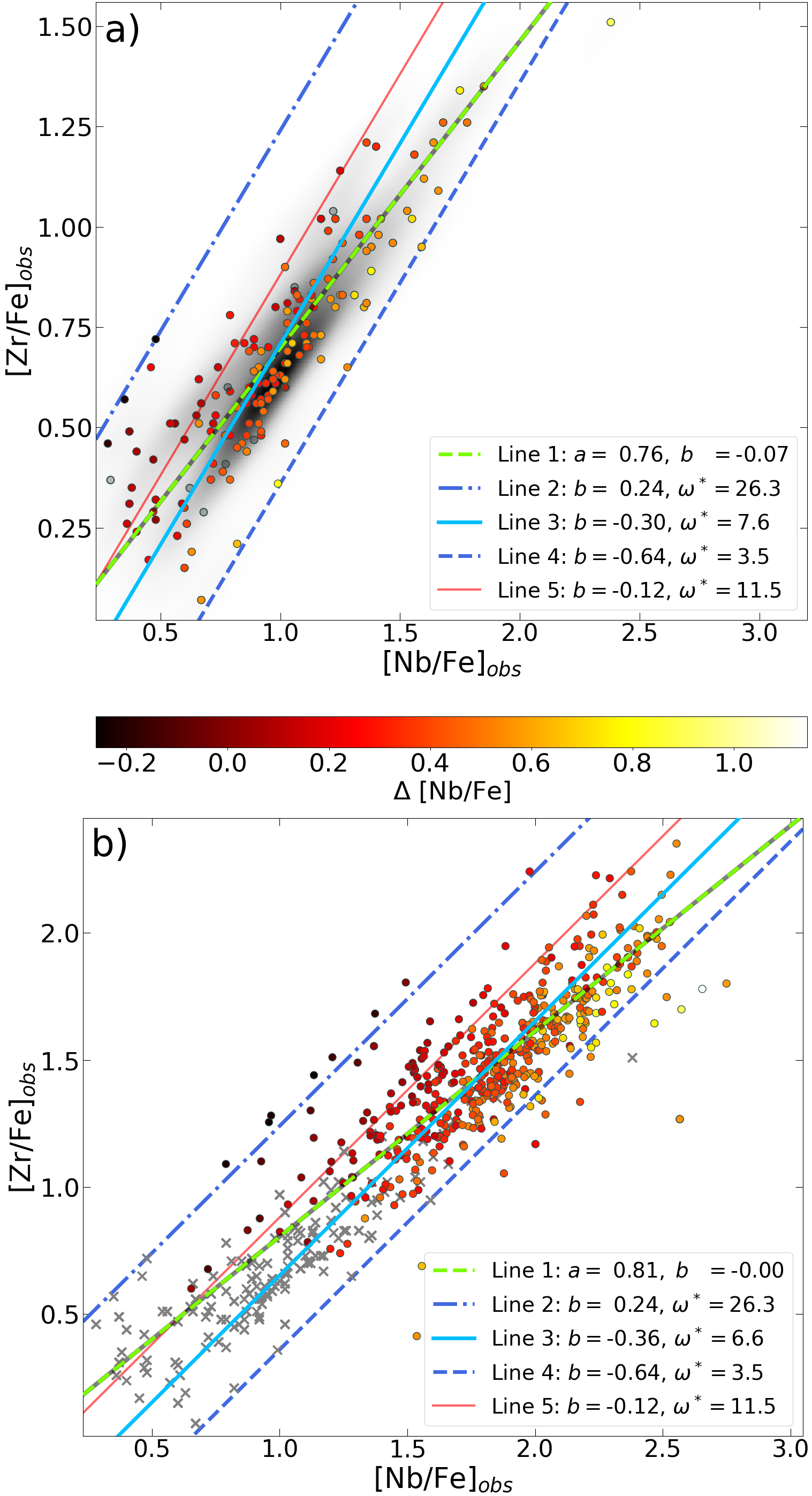}
\caption{Abundances [Zr/Fe] as a function of [Nb/Fe] and the corresponding linear fits. \textbf{a)} [Zr/Fe] abundances as a function of [Nb/Fe] for our sample of Ba stars. (The figure includes 162 points, because for 18 stars Nb abundances are not available.) The points are colour-coded to show the residual $\Delta\,$[Nb/Fe] for each star, averaged over all its classified models (the 9 stars with grey, semi-transparent points have no residual, as no models were classified for these stars). Line 1 is a fit to the data, with the slope $a$ as a free parameter; Line 3 is an isothermal line (with slope fixed to 1) fitted to the data; Lines 2 and 4 are isothermal lines passing through the stars with the (second) lowest and highest $b$ intercept, respectively; Line 5 is the isothermal corresponding to \omast=10.5, the lowest value achievable for AGB temperatures (at most 450 MK) based on Fig.~\ref{fig:omega-T}. The greyscale colouring in the background represents the 2D KDE of the points. \textbf{b)} Same as panel a), but including the 533 calculated AGB star abundances (there is more than one $\delta$ for each Ba star) from Eq. (\ref{eq:AGB_fromBa}). The original Ba star abundances are marked with grey crosses, while the AGB abundances are coloured on the same scale as in panel a). The lines have the same meaning as in panel a), but fitted to the AGB star abundances.
}
\label{fig:NbZr-twofig}
\end{figure}

\begin{figure}
    \centering
    \includegraphics[width=\linewidth]{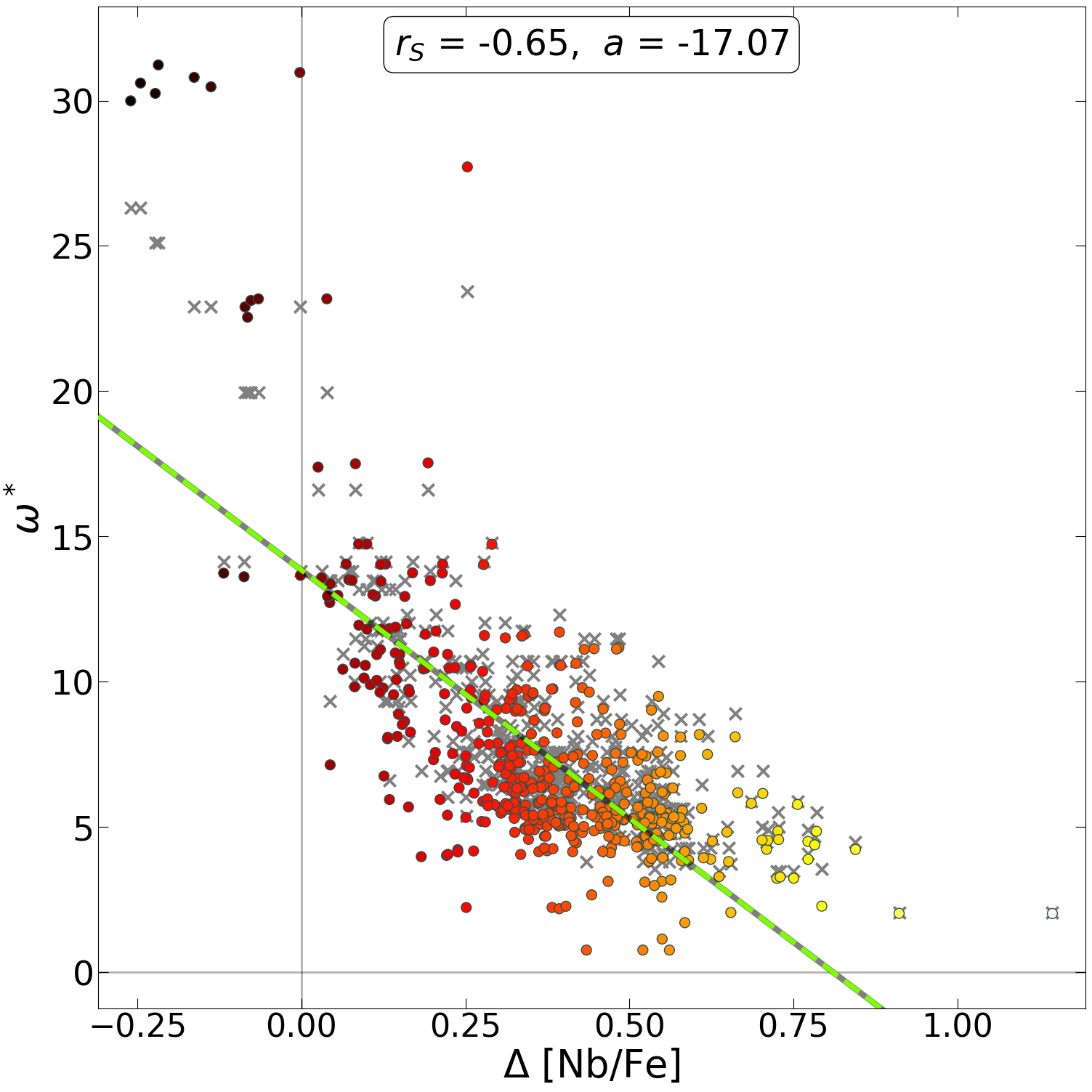}
    \caption{Values of $\omega^*$ for each Ba star as a function of the residuals $\Delta\,$[Nb/Fe].
    The grey crosses represent the points calculated from the Ba star abundances. The coloured points correspond to the data recalculated to the AGB values, colour-coded to the value of the residual, with the same scale is in Fig. \ref{fig:NbZr-twofig}. The fitted line is to the AGB values, the $r_S$ coefficient and the slope $a$ of the fitted line to this relation are shown above. The same fit for the Ba star abundances gives $r_S = -0.74$ and $a = -15.38$.}
    \label{fig:omres}
\end{figure}

In Fig.~\ref{fig:NbZr-twofig} we fitted an isothermal line (i.e. with a slope fixed at 1) to the stellar abundances (Line 3; fitted with ODR, as for other plots representing observations vs observations). 
The distribution of Ba stars (panel a) and AGB stars (panel b) and the derived \omast values qualitatively match each other and we present them together in the following. 

From the value of the intercept $b$ of Line 3 we find $\omega^*$ to be 7.6 and 6.6 (for the Ba stars and the AGB stars, respectively). These values are not only unrealistic for AGB conditions (where temperatures are always below 400-450 MK, see Fig.~\ref{fig:omega-T}), but also the corresponding lines clearly do not represent the observed distribution. The deviation of many stars from this fitted isothermal line can exceed 0.35 dex, which is well outside the observational uncertainties.

A resolution to this contradiction is to use several isothermal lines of different temperatures, covering the observed range.
The Ba (AGB) stars leading to the highest and second lowest (the lowest being an outlier) $\omega^*$ are 26.3 (31.2) and 3.5 (0.8) respectively, with a high number of the Ba (AGB) stars having \omast lower than the fit of 7.6 (6.6).
However, if we compare these \omast values with the temperature dependence in Fig. \ref{fig:omega-T}, we see that only a small fraction of the observed stars (the few just above the solid red line, Line 5) can be explained by reasonable AGB temperatures between 90 and 450 MK, corresponding to \omast values between roughly 10 and 17. 
All stars located below the red solid Line 5 (that is the vast majority) cannot be explained within the framework of assuming the steady-state $s$ process, as they fall outside the realistic AGB temperature range. 
It is worth noting that the use of the AGB star abundances, which take into account the dilution of the material, cannot account for the low \omast values, it actually aggravates the discrepancy with the \omast -- $T$ relation, since it leads to even lower \omast values.

Another explanation could be that the process that produces Zr leads to a relationship other than a line with a slope of 1, meaning that the assumption of steady-state is not valid. We also fitted a line to the Ba stars with a slope not fixed at 1 (a higher order curve seems inappropriate for our data and would lead to overfitting): this gave a slope of $a = 0.76$ $(0.81)$ for the Ba (AGB) stars and a better qualitative representation of the data. From this line, there is still a large natural scatter of the stars, but about half of that present in the case of the isothermal fit.


We consider the residual of Nb to be the most sensitive indicator of the inability to explain the observations. We seek to see if the temperature estimate worsens with this residual. In Fig. \ref{fig:NbZr-twofig} it is clear that the residual $\Delta\,$[Nb/Fe] grows at points with lower \omast. 
In order to examine this relationship explicitly, we have plotted \omast as a function of the residual $\Delta\,$[Nb/Fe] (Fig. \ref{fig:omres}). 
The relationship resembles a scatter around a line, but it deviates to much higher \omast values at low residuals.
There is a clear anticorrelation between the \omast and the residual of Nb ($r_S = -0.74$ and $-0.65$, for the Ba and AGB stars, respectively). 
This anticorrelation suggests that our inability to fit the observed [Nb/Fe] abundances becomes progressively worse for the \omast values corresponding to the temperatures outside the typical AGB $s$-process temperatures. Thus, the residual of Nb proves to be a good indicator of the stars that cannot be explained by the steady-state $s$ process.

\subsection{Comparison to literature and discussion}
\label{subsec:ZrNb_comparison}
The same plot as in Fig. \ref{fig:NbZr-twofig} a) was created by \cite{ZrNb-Neysk} for 9 extrinsic stars. These authors fitted an isothermal line to the two stars in their sample most enriched in $s$-process elements. Based on these two stars, they found the \omast value  between 16 and 17, about twice as large as for our fit of 7.6. The other 7 stars presented in that work would have \omast values even larger than 17. Based on their \omast -- $T$ relation, this fit gave temperatures below (100--250) MK, consistent with the \iso{C}{13} neutron source.
The stars studied in that work would be in the regime of 
our stars with the highest \omast values in Fig.~\ref{fig:NbZr-twofig}. This discrepancy between results could either mean that their sample was underrepresented and selected by chance from this regime, or that there are systematic deviations between the two sources of observations. A clear difference is that our sample reaches significantly higher Nb enrichment, up to (1.3 -- 1.5) dex, than the sample used by \citet{ZrNb-Neysk}, where the two most-enriched stars reach 0.8 dex.

The same plot was made by \citeauthor{ZrNb-Kar} (\citeyear{ZrNb-Kar}, see their Fig.~14), for 20 Ba stars and \citeauthor{roriz21b} (\citeyear{roriz21b}, see their Fig. 6), for the same 180 stars that we use here. These authors both showed the slope-1 band corresponding to a typical AGB temperature range of $(1-3)\cdot 10^8\,$K on their plots.
\citeauthor{ZrNb-Kar} found that most of their stars lie in or close to this band, while roughly a third of their sample is below it, and would require a higher temperature. To account for non-equilibrium processes and the complex behaviour inside AGB stars, they also calculated STAREVOL models instead of the simple steady-state equation we also use (see their Fig. 15). However, these models could not reproduce stars below the diagonal band of $(1-3)\cdot 10^8\,$K, but led to points above it, resulting in an even larger gap between observations and theoretical predictions. Their models with 2--3 \msun~resulted in lines with \omast$=15.8$, while the models with 4--5 \msun~were even above and could not reach sufficiently high abundances. 

As these authors have already discussed, the breakdown of the local equilibrium may have three reasons. 
One possibility is that the marginal neutron flux in the thermal pulse is strong enough to reset the Zr and Nb abundances (produced by the neutron flux due to the \iso{C}{13} over long timescales of the order of 10,000 yr). If the timescale of this process is faster than the neutron-capture timescale for the Zr isotopes, equilibrium cannot be established. 
The neutron-capture timescale of the Zr isotopes, for a typical neutron density of the order of $10^9$ cm$^{-3}$, is between 2 (for \iso{Zr}{93}) and 8 years (for \iso{Zr}{90}), so this may be a possibility, but no AGB model shows its realisation.  
The other two explanations are related to the activation of two branching points. If the branching point at \iso{Zr}{95} was considerably activated, then \iso{Zr}{96} would be significantly produced and would have to be added in Eq.~\ref{eq:NbZr}. This would change the calculation so that the same \omast would be obtained at slightly lower temperatures, and so more stars would be included in the realistic temperature regime (Line 5 in Fig. \ref{fig:NbZr-twofig} would shift downwards, to lower $b$ intercepts). 
Instead, \iso{Zr}{90} would be bypassed by activating the branching points at \iso{Sr}{89,90}. This would again change the \omast -- T relationship, but in the opposite direction and by a greater amount (given that $\sigma_{90} \simeq 2 \cdot \sigma_{96}$). 
Furthermore, solutions relying on branching points are unlikely because our sample stars show no enrichment of Rb relative to Sr and Zr \citep{roriz21a}, indicating no significant activation of branching points in the AGB companions, and hence low neutron densities in the thermal pulse due to the \iso{Ne}{22} neutron source and low initial mass, as discussed in detail in Sect.~\ref{subsec:boxpl-absabund} as well as Papers I and II.


In summary, we determined the \omast values corresponding to our stars, and point out that the vast majority of stars clearly cannot be explained by a steady-state $s$ process. Following \citeauthor{roriz21b} (\citeyear{roriz21b}, see their Sect.~5), we propose and discuss the potential reasons for the discrepancy between the distribution of our stars and their $s$-process temperature (in order of increasing likeliness):
\begin{itemize}

    \item Different sources of solar abundances can result in variations of \omast by about 10\% only, which is insufficient to explain our offset.

    \item Systematically wrong cross sections can have a large impact on the derived temperatures. However, to reach the desired regime, the cross sections should have $\approx$ 100\% error in the case of \iso{Zr}{93} or even larger in the case of other isotopes, which is unrealistic.
    
    \item The initial composition of the Ba stars may affect their observed abundances.  However, the initial abundance of $s$-process elements in Ba stars is so low compared to that of AGB stars that the final mixture has almost no memory of the initial abundance. Even with extremely low $\delta$ and extremely high initial abundances (within the limits of \citealt{galcomposition}), this effect can cause an offset in the abundances of no more than 0.3 dex, and typically of the order of 0.1 dex. This is not sufficient to explain the deviation, especially since the effect is thought to be similar for both Zr and Nb, shifting the data points along the isothermal lines.
    
    \item We have studied in detail the effect of the dilution of different AGB models in the Ba star envelopes (see Fig. \ref{fig:NbZr-twofig}, panel b). Although there is a larger scatter in the derived AGB abundances than in the original Ba abundances, the \omast values are not higher, but usually even lower, than when using the original Ba star abundances. This trend is in the opposite direction to the deviation that needs to be explained.
    
    \item It may happen that all our points are shifted due to systematic errors in the spectroscopic analysis, such as not including non-LTE (NLTE) corrections, as further detailed in Sect. \ref{sec:conclusions}.
    However, as pointed out by \cite{roriz21b}, there are clear correlations between the [Nb, Mo, Ru/Sr] abundances (see their Fig. 11), which makes it difficult to invoke the same problems for all three elements acting in a similar way.
    
    \item The simplistic analytical treatment of the $s$ process from Eqs. (\ref{eq:NbZr}) and (\ref{eq:omegastar}) should be substituted by predictions from AGB models. This possibility was investigated in detail by \cite{ZrNb-Kar} using STAREVOL models and did not improve the fit. 
    Furthermore, we confirm that our models systematically underproduce Nb relative to Zr (see Sect.~\ref{fig:boxplot-p1p1}). Therefore, this possible explanation is not to be present in AGB models based on currently computed structure and evolution.
    
\end{itemize}

We thus suggest that the most possible explanation for this discrepancy is that the assumptions of Eq. (\ref{eq:NbZr}) fail and therefore \iso{Zr}{93} is not solely produced by steady-state $s$ process (see discussion in Sect. \ref{sec:conclusions}).



\section{Discussion and conclusions}
\label{sec:conclusions}
We studied the statistical properties of a population of 180 Ba stars. 
To find the possible polluter AGB star, we used the classifiers introduced in Paper II, as well as a newly developed classifier using the Random Forest ML algorithm (RF).
This RF classifier performed similarly well during training to the neural network presented in Paper II. It identified slightly different AGB models for each star, but statistically gave similar results to the other two classifiers from Paper II, even without excluding any specific element. This confirms that our method is well constructed, since another independent technique gave similar results to those published previously. 
We also showed the feature importance of the RF classifier, from which we can see that the algorithm identifies the features of different peaks with different weights, without any input about the underlying nature of the data. This suggests that the algorithm was trained to identify physically meaningful patterns.

Based on the violin plots shown in Sect. \ref{sec:boxplot}, we see that the heavier $s$-process elements in and just after the first peak (Nb, Mo, and Ru) are significantly underproduced in the AGB nucleosynthesis models relative to the Ba star abundances. The additional source of these elements may be an additional physical mechanism not yet considered in the models. This could be the trace of the intermediate neutron-capture process ($i$ process), a late thermal pulse, or an unidentified nucleosynthesis or mixing effect. 
This process should not produce an excess of Rb, since that element is already significantly overproduced in our models. As the Peak 2 elements can generally be fitted with much more precision than Peak 1 elements, the second peak should not be affected either.

Using the [Zr/Fe] -- [Nb/Fe] abundances as a thermometer in Sect. \ref{sec:NbZr}, we find that most of our Ba star population cannot be reproduced with reasonable $s$-process temperatures. 
The discrepancy is so significant that it is unlikely to be resolved by steady-state $s$ processes alone. 
As above, this suggests that the currently predicted AGB structure and evolution misses some, perhaps transient features, which could lead to neutron fluxes with different properties from those found in the current models. This could lead to an enhanced production of Nb (as well as Mo and Ru, but not Rb), possibly as a local effect starting from the abundances of first $s$-process peak elements. As the abundance of Nb is dominated by the decay of \iso{Zr}{93}, this discrepancy should affect the \iso{Zr}{93} production as well.
This is also supported by the violin plots in Sect. \ref{sec:boxplot} and the correlations in Sect. \ref{sec:correl} (therein specifically Figs. \ref{fig:correl_res_feh-Fe}, \ref{fig:correl_res_res_5elem} and \ref{fig:correl_res_obs_NbMo}).





Based on the correlations presented in Sect. \ref{subsec:res-feh-iproc}, the residuals do not show a strong dependence on [Fe/H]. However, Nb and Ru relative to all other elements are moderately correlated with the metallicity. This may indicate that the unknown process producing these generally problematic elements is dependent on the metallicity. 
These results provide indications that the $i$ process, if that is the missing nucleosynthesis component in our stars, is  slightly dependent on metallicity, and is also efficient at  solar metallicites. 
Several authors have reported that the $i$ process should be metallicity dependent, and much more efficient at lower metallicity. 
For example, \cite{iproc-Cristallo} and \cite{iproc-met-Choplin} presented AGB nucleosynthesis models that, at low metallicites, undergo a proton ingestion event (PIE) and produce neutron densities representative of the $i$ process.
\cite{CEMP-Hampel} investigated carbon-enriched metalpoor stars (CEMP stars, specifically CEMP-$r/s$ that show overabundances in both $s$- and $r$-process elements) in the halo with [Fe/H] $ < -1$. The abundance pattern of these stars could only be reproduced with the $i$ process, at neutron densities $n \approx 10^{14} \,\mathrm{cm^{-3}}$. 
\cite{CEMP-Abate} also found indications of a missing process or missing physical ingredients from the $s$-process models for CEMP stars in the halo.
\cite{postAGB-lugaro} arrived to the same conclusion by the examination of low-metallicity post-AGB stars.
Our work provides an indication that the $i$ process may be efficient at solar metallicites as well, based on a large population of 180 Ba stars. \cite{iproc-highmet-Kar} also found 15 high-metallicity stars (in the same metallicity region as our sample) for which a pure $s$-process pattern is insufficient and an $i$-process component is required. This was also the case for $\sim$25\% of our sample in Paper II.  The challenge will be to find a mechanism for the $i$ process at high metallicity.


In Sect. \ref{subsec:correl-res-res}, we showed that the residuals of Nb, Mo and Ru as a function of each other increase together. The residuals of Sr, Y and Zr also rise with those of Nb, Mo and Ru. This may indicate that the possible missing process already affects lower atomic numbers than Nb. 
In general, our inability to fit the first peak means that the fit of the second peak becomes worse. This discrepancy could be solved by finding the source of the missing Nb, Mo and Ru.

The residuals as a function of the observed abundances (Sect. \ref{sec:correl-res-obs}) tend to be most correlated with their own elemental abundances. Nb, Mo and Ru show a positive correlation with almost all the abundances, their offset increasing with higher abundances. This means that the discrepancy of these elements grows with increasing $s$-process production.

We also examined the AGB parameter distributions for the classified models in Sect. \ref{sec:distributions}. The mass distribution of the polluter AGB stars shows a peak at low masses, and the maximum barely exceeds 4 \msun. 
This confirms that Ba stars are mainly polluted by low-mass AGB stars.
The distribution of $\delta$ shows a preference for low accreted mass,\footnote{For the $\delta$ -- period diagram of the stars that have measured period, see Paper I, Fig. 17.} 
which is realistic. However, there are a few cases where only high $\delta$ values can explain the observed elemental distribution. The systems with high $\delta$ parameter are not a source of high residuals, as investigated in Sect. \ref{subsec:correl-dil}.
Hence some Ba stars may indeed have accreted a high fraction of their envelope material from the AGB star, in an extreme binary interaction that may only be explained by Roche-lobe overflow or wind Roche-lobe overflow (see e.g. \citealt{wind-Abate1} and \citealt{wind-Abate2}).
We point out the need for proper envelope masses of Ba stars to better understand the mass accretion history. In addition, the accreted material could significantly change the structure of the Ba star, which is an interesting stellar evolution problem to be investigated.

Our conclusions rely on the derived abundances, that have multiple sources of uncertainties.
\cite{roriz21b} provided the uncertainties of their abundances (see their Sect. 3.3), which is typically 0.25 dex for all elements. These errors however should not introduce systematic deviations, as these are derived from the random errors of the abundance estimation, and in principle should mainly increase the scatter in our data points.
We did not consider any NLTE corrections in this work. 
Specifically, for Rb, \cite{NLTE-Rb-Korotin} calculated new solar values that differ in about 0.2 dex from the solar abundances used in this paper, while \cite{NLTE-Rb-Takeda} found this value to be smaller than 0.1 dex. Such a correction would result in higher Rb values, and thus make the difference between the models and the observations smaller, but is not sufficient to explain our deviations.
For Y and Eu, the recent calculations of \cite{NLTE-Y-Sun} show that the solar Y and Eu abundance changes by around 0.1 dex compared to the solar abundances used in this paper. This is again a weak effect, and the difference is less than the typical errors on the individual abundance measurements. 
For Sr, \cite{NLTE-Sr-1} calculated NLTE corrections, for our stars this is about 0.2 dex (see their Fig. 5). However, this would move the observational data further from model abundances, making the residuals even larger.
All other elements used have no available NLTE corrections, but it would be essential to revisit their abundances using NLTE in the future. These corrections may modify our results, but it seems unlikely that it would explain the discrepancy for the elements with the highest residuals. Also, current studies indicate that at this high metallicities ([Fe/H] $\gtrsim$ -0.6), due to effects such as the decreasing overionisation, elements tend to have small NLTE corrections (see e.g. \citealt{NLTE-FeHdependence-Lind}, \citealt{NLTE-Fehdependence-Guiglion}, \citealt{Mishenina-NbRu} and \citealt{NLTE-Fehdependence-Thevenin}).

In order to rule out possible systematic errors in the observations and reveal the real limits of current $s$-process nucleosynthesis models, it would be very beneficial to have abundances derived from high-resolution spectra of even more Ba stars, and confirmed by different analyses. Our algorithm can handle any number of Ba stars and we could easily refine the statistics using more observations.
Furthermore, the correlations and the slope of the lines in Sect. \ref{sec:correl} may be better constrained by extending the range of metallicites for the observed Ba stars with CH and CEMP stars.
For example, most recently, \citet{roriz24} reported W abundances for 94 Ba stars. Although the uncertainties are large, there are a few stars with clearly higher [W/hs] (where $\mathrm{hs = La+Ce+Nd}$) than predicted by the $s$-process models. The star with the highest [W/hs] $\simeq$ 0.4 is BD $-$09$^\circ$4337, which also shows the highest [Nb/Fe], [Zr/Fe], and $\Delta\,$[Nb/Fe] residual (top panel of Fig.~\ref{fig:NbZr-twofig}). For the star with the second highest [W/hs] (also $\simeq$ 0.4, HD 114678) the Nb abundance is not available. Clearly more observational data are needed to confirm potentially interesting stars and including the abundance of Pb, which is crucial for understanding the whole $s$-process nucleosynthesis including the third magic peak, and its relationship with the lighter elements.

\begin{acknowledgements}
This project has been supported by the Lend\"ulet Program LP2023-10 and by the LP2021-9 Lend\"ulet grant of the Hungarian Academy of Sciences. We also thank the ChETEC-INFRA (G.A. no. 101008324) project for support.
M.J. gratefully acknowledges funding of MATISSE: \textit{Measuring Ages Through Isochrones, Seismology, and Stellar Evolution}, awarded through the European 
Commission's Widening Fellowship.  
This project has received funding from the European Union's Horizon 2020 research and innovation programme. M.L. was also supported by the NKFIH excellence grant TKP2021-NKTA-64. 
A.K. was supported by the Australian Research Council Centre of Excellence for All Sky Astrophysics in 3 Dimensions (ASTRO 3D), through project number
CE170100013.
We thank Marco Pignatari, Ádám Kadlecsik and László Molnár for the helpful ideas and discussions.\\
Software used: Scikit-learn \citep{scikit-learn}, Scipy \citep{Scipy}, Numpy \citep{Numpy}, Matplotlib \citep{Matplotlib}.
\end{acknowledgements}

\bibliographystyle{aa}
\bibliography{bibliography}

\begin{thebibliography}{79}
\expandafter\ifx\csname natexlab\endcsname\relax\def\natexlab#1{#1}\fi

\bibitem[{{Abate} {et~al.}(2015{\natexlab{a}}){Abate}, {Pols}, {Izzard}, \&
  {Karakas}}]{wind-Abate2}
{Abate}, C., {Pols}, O.~R., {Izzard}, R.~G., \& {Karakas}, A.~I.
  2015{\natexlab{a}}, \aap, 581, A22

\bibitem[{{Abate} {et~al.}(2015{\natexlab{b}}){Abate}, {Pols}, {Karakas}, \&
  {Izzard}}]{CEMP-Abate}
{Abate}, C., {Pols}, O.~R., {Karakas}, A.~I., \& {Izzard}, R.~G.
  2015{\natexlab{b}}, \aap, 576, A118

\bibitem[{{Abate} {et~al.}(2015{\natexlab{c}}){Abate}, {Pols}, {Karakas}, \&
  {Izzard}}]{wind-Abate1}
{Abate}, C., {Pols}, O.~R., {Karakas}, A.~I., \& {Izzard}, R.~G.
  2015{\natexlab{c}}, \aap, 576, A118

\bibitem[{{Abia} {et~al.}(2001){Abia}, {Busso}, {Gallino}, {Dom{\'\i}nguez},
  {Straniero}, \& {Isern}}]{Rb-Abia}
{Abia}, C., {Busso}, M., {Gallino}, R., {et~al.} 2001, \apj, 559, 1117

\bibitem[{{Amaducci} {et~al.}(2024){Amaducci}, {Colonna}, {Cosentino},
  {Cristallo}, {Finocchiaro}, {Krti{\v{c}}ka}, {Massimi}, {Mastromarco},
  {Mazzone}, {Maugeri}, {Mengoni}, {Roederer}, {Straniero}, {Valenta},
  {Vescovi}, {Aberle}, {Alcayne}, {Andrzejewski}, {Audouin}, {Babiano-Suarez},
  {Bacak}, {Barbagallo}, {Bennett}, {Berthoumieux}, {Billowes}, {Bosnar},
  {Brown}, {Busso}, {Caama{\~n}o}, {Caballero-Ontanaya}, {Calvi{\~n}o},
  {Calviani}, {Cano-Ott}, {Casanovas}, {Cerutti}, {Chiaveri}, {Cort{\'e}s},
  {Cort{\'e}s-Giraldo}, {Damone}, {Davies}, {Diakaki}, {Dietz},
  {Domingo-Pardo}, {Dressler}, {Ducasse}, {Dupont}, {Dur{\'a}n}, {Eleme},
  {Fern{\'a}ndez-Dom{\'\i}nguez}, {Ferrari}, {Furman}, {G{\"o}bel}, {Garg},
  {Gawlik-Ramiega}, {Gilardoni}, {Gon{\c{c}}alves}, {Gonz{\'a}lez-Romero},
  {Guerrero}, {Gunsing}, {Harada}, {Heinitz}, {Heyse}, {Jenkins}, {Junghans},
  {K{\"a}ppeler}, {Kadi}, {Kimura}, {Knapov{\'a}}, {Kokkoris}, {Kopatch},
  {Kurtulgil}, {Ladarescu}, {Lederer-Woods}, {Leeb}, {Lerendegui-Marco},
  {Lonsdale}, {Macina}, {Manna}, {Mart{\'\i}nez}, {Masi}, {Mastinu}, {Mendoza},
  {Michalopoulou}, {Milazzo}, {Mingrone}, {Moreno-Soto}, {Musumarra}, {Negret},
  {Nolte}, {Og{\'a}llar}, {Oprea}, {Patronis}, {Pavlik}, {Perkowski},
  {Petrone}, {Piersanti}, {Pirovano}, {Porras}, {Praena}, {Quesada},
  {Ramos-Doval}, {Rauscher}, {Reifarth}, {Rochman}, {Rubbia},
  {Sabat{\'e}-Gilarte}, {Saxena}, {Schillebeeckx}, {Schumann}, {Sekhar},
  {Smith}, {Sosnin}, {Sprung}, {Stamatopoulos}, {Tagliente}, {Tain},
  {Tarife{\~n}o-Saldivia}, {Tassan-Got}, {Thomas}, {Torres-S{\'a}nchez},
  {Tsinganis}, {Ulrich}, {Urlass}, {Vannini}, {Variale}, {Vaz}, {Ventura},
  {Vlachoudis}, {Vlastou}, {Wallner}, {Woods}, {Wright}, {{\v{Z}}ugec}, \& {n
  TOF Collaboration}}]{Ce-Amaducci}
{Amaducci}, S., {Colonna}, N., {Cosentino}, L., {et~al.} 2024, \prl, 132,
  122701

\bibitem[{{Asplund} {et~al.}(2009){Asplund}, {Grevesse}, {Sauval}, \&
  {Scott}}]{solar-asp09}
{Asplund}, M., {Grevesse}, N., {Sauval}, A.~J., \& {Scott}, P. 2009, \araa, 47,
  481

\bibitem[{{Beers} \& {Christlieb}(2005)}]{EuFe1-Beers}
{Beers}, T.~C. \& {Christlieb}, N. 2005, \araa, 43, 531

\bibitem[{{Bergemann} {et~al.}(2012){Bergemann}, {Hansen}, {Bautista}, \&
  {Ruchti}}]{NLTE-Sr-1}
{Bergemann}, M., {Hansen}, C.~J., {Bautista}, M., \& {Ruchti}, G. 2012, \aap,
  546, A90

\bibitem[{{Bidelman} \& {Keenan}(1951)}]{Bafirst}
{Bidelman}, W.~P. \& {Keenan}, P.~C. 1951, \apj, 114, 473

\bibitem[{{Bisterzo} {et~al.}(2014){Bisterzo}, {Travaglio}, {Gallino},
  {Wiescher}, \& {K{\"a}ppeler}}]{s-r-contribution-Bisterzo}
{Bisterzo}, S., {Travaglio}, C., {Gallino}, R., {Wiescher}, M., \&
  {K{\"a}ppeler}, F. 2014, \apj, 787, 10

\bibitem[{{Breiman}(2001)}]{RF-MDI}
{Breiman}, L. 2001, Machine Learning, 45, 5

\bibitem[{{Burbidge} {et~al.}(1957){Burbidge}, {Burbidge}, {Fowler}, \&
  {Hoyle}}]{bbfh}
{Burbidge}, E.~M., {Burbidge}, G.~R., {Fowler}, W.~A., \& {Hoyle}, F. 1957,
  Reviews of Modern Physics, 29, 547

\bibitem[{{Busso} {et~al.}(2001{\natexlab{a}}){Busso}, {Gallino}, {Lambert},
  {Travaglio}, \& {Smith}}]{s-vs-feh-busso}
{Busso}, M., {Gallino}, R., {Lambert}, D.~L., {Travaglio}, C., \& {Smith},
  V.~V. 2001{\natexlab{a}}, \apj, 557, 802

\bibitem[{{Busso} {et~al.}(2001{\natexlab{b}}){Busso}, {Gallino}, {Lambert},
  {Travaglio}, \& {Smith}}]{maxtemp_Busso01}
{Busso}, M., {Gallino}, R., {Lambert}, D.~L., {Travaglio}, C., \& {Smith},
  V.~V. 2001{\natexlab{b}}, \apj, 557, 802

\bibitem[{{Busso} {et~al.}(1999){Busso}, {Gallino}, \& {Wasserburg}}]{busso99}
{Busso}, M., {Gallino}, R., \& {Wasserburg}, G.~J. 1999, \araa, 37, 239

\bibitem[{{Choplin} {et~al.}(2022){Choplin}, {Siess}, \&
  {Goriely}}]{iproc-met-Choplin}
{Choplin}, A., {Siess}, L., \& {Goriely}, S. 2022, \aap, 667, A155

\bibitem[{{Cristallo} {et~al.}(2011){Cristallo}, {Piersanti}, {Straniero},
  {Gallino}, {Dom{\'\i}nguez}, {Abia}, {Di Rico}, {Quintini}, \&
  {Bisterzo}}]{2011cristallo}
{Cristallo}, S., {Piersanti}, L., {Straniero}, O., {et~al.} 2011, \apjs, 197,
  17

\bibitem[{{Cristallo} {et~al.}(2009{\natexlab{a}}){Cristallo}, {Piersanti},
  {Straniero}, {Gallino}, {Dom{\'\i}nguez}, \&
  {K{\"a}ppeler}}]{iproc-Cristallo}
{Cristallo}, S., {Piersanti}, L., {Straniero}, O., {et~al.} 2009{\natexlab{a}},
  \pasa, 26, 139

\bibitem[{{Cristallo} {et~al.}(2009{\natexlab{b}}){Cristallo}, {Straniero},
  {Gallino}, {Piersanti}, {Dom{\'{\i}}nguez}, \& {Lederer}}]{cristallo09}
{Cristallo}, S., {Straniero}, O., {Gallino}, R., {et~al.} 2009{\natexlab{b}},
  \apj, 696, 797

\bibitem[{{Cristallo} {et~al.}(2009{\natexlab{c}}){Cristallo}, {Straniero},
  {Gallino}, {Piersanti}, {Dom{\'\i}nguez}, \& {Lederer}}]{2009cristallo}
{Cristallo}, S., {Straniero}, O., {Gallino}, R., {et~al.} 2009{\natexlab{c}},
  \apj, 696, 797

\bibitem[{{Cristallo} {et~al.}(2015){Cristallo}, {Straniero}, {Piersanti}, \&
  {Gobrecht}}]{2015cristallo}
{Cristallo}, S., {Straniero}, O., {Piersanti}, L., \& {Gobrecht}, D. 2015,
  \apjs, 219, 40

\bibitem[{{Cseh} {et~al.}(2018){Cseh}, {Lugaro}, {D'Orazi}, {de Castro},
  {Pereira}, {Karakas}, {Moln{\'a}r}, {Plachy}, {Szab{\'o}}, {Pignatari}, \&
  {Cristallo}}]{cseh18}
{Cseh}, B., {Lugaro}, M., {D'Orazi}, V., {et~al.} 2018, \aap, 620, A146

\bibitem[{{Cseh} {et~al.}(2022){Cseh}, {Vil{\'a}gos}, {Roriz}, {Pereira},
  {D'Orazi}, {Karakas}, {So{\'o}s}, {Drake}, {Junqueira}, \&
  {Lugaro}}]{paper1-Ce}
{Cseh}, B., {Vil{\'a}gos}, B., {Roriz}, M.~P., {et~al.} 2022, \aap, 660, A128

\bibitem[{{de Castro} {et~al.}(2016){de Castro}, {Pereira}, {Roig}, {Jilinski},
  {Drake}, {Chavero}, \& {Sales Silva}}]{deC}
{de Castro}, D.~B., {Pereira}, C.~B., {Roig}, F., {et~al.} 2016, \mnras, 459,
  4299

\bibitem[{{den Hartogh} {et~al.}(2023){den Hartogh}, {Yag{\"u}e L{\'o}pez},
  {Cseh}, {Pignatari}, {Vil{\'a}gos}, {Roriz}, {Pereira}, {Drake}, {Junqueira},
  \& {Lugaro}}]{paper2-ML}
{den Hartogh}, J.~W., {Yag{\"u}e L{\'o}pez}, A., {Cseh}, B., {et~al.} 2023,
  \aap, 672, A143

\bibitem[{{Escorza} \& {De Rosa}(2023)}]{Escorza-newmasses}
{Escorza}, A. \& {De Rosa}, R.~J. 2023, \aap, 671, A97

\bibitem[{{Escorza} {et~al.}(2020){Escorza}, {Siess}, {Van Winckel}, \&
  {Jorissen}}]{binary-Escorza}
{Escorza}, A., {Siess}, L., {Van Winckel}, H., \& {Jorissen}, A. 2020, \aap,
  639, A24

\bibitem[{{Fishlock} {et~al.}(2014){Fishlock}, {Karakas}, {Lugaro}, \&
  {Yong}}]{2014fishlock}
{Fishlock}, C.~K., {Karakas}, A.~I., {Lugaro}, M., \& {Yong}, D. 2014, \apj,
  797, 44

\bibitem[{{Forsberg} {et~al.}(2019){Forsberg}, {J{\"o}nsson}, {Ryde}, \&
  {Matteucci}}]{galcomposition}
{Forsberg}, R., {J{\"o}nsson}, H., {Ryde}, N., \& {Matteucci}, F. 2019, \aap,
  631, A113

\bibitem[{{Grevesse} \& {Sauval}(1998)}]{solar-g98}
{Grevesse}, N. \& {Sauval}, A.~J. 1998, \ssr, 85, 161

\bibitem[{{Guiglion} {et~al.}(2024){Guiglion}, {Bergemann}, {Storm}, {Lian},
  {Cescutti}, \& {Serenelli}}]{NLTE-Fehdependence-Guiglion}
{Guiglion}, G., {Bergemann}, M., {Storm}, N., {et~al.} 2024, \aap, 683, A73

\bibitem[{{Hampel} {et~al.}(2016){Hampel}, {Stancliffe}, {Lugaro}, \&
  {Meyer}}]{CEMP-Hampel}
{Hampel}, M., {Stancliffe}, R.~J., {Lugaro}, M., \& {Meyer}, B.~S. 2016, \apj,
  831, 171

\bibitem[{{Han} {et~al.}(1995){Han}, {Eggleton}, {Podsiadlowski}, \&
  {Tout}}]{accretion-Han}
{Han}, Z., {Eggleton}, P.~P., {Podsiadlowski}, P., \& {Tout}, C.~A. 1995,
  \mnras, 277, 1443

\bibitem[{Harris {et~al.}(2020)Harris, Millman, van~der Walt, Gommers,
  Virtanen, Cournapeau, Wieser, Taylor, Berg, Smith, Kern, Picus, Hoyer, van
  Kerkwijk, Brett, Haldane, del R{\'{i}}o, Wiebe, Peterson,
  G{\'{e}}rard-Marchant, Sheppard, Reddy, Weckesser, Abbasi, Gohlke, \&
  Oliphant}]{Numpy}
Harris, C.~R., Millman, K.~J., van~der Walt, S.~J., {et~al.} 2020, Nature, 585,
  357

\bibitem[{Hunter(2007)}]{Matplotlib}
Hunter, J.~D. 2007, Computing in Science \& Engineering, 9, 90

\bibitem[{{Izzard} {et~al.}(2010){Izzard}, {Dermine}, \&
  {Church}}]{binary-Izzard}
{Izzard}, R.~G., {Dermine}, T., \& {Church}, R.~P. 2010, \aap, 523, A10

\bibitem[{{Jonsell} {et~al.}(2006){Jonsell}, {Barklem}, {Gustafsson},
  {Christlieb}, {Hill}, {Beers}, \& {Holmberg}}]{EuFe1-Jonsell}
{Jonsell}, K., {Barklem}, P.~S., {Gustafsson}, B., {et~al.} 2006, \aap, 451,
  651

\bibitem[{{Jorissen} {et~al.}(2019){Jorissen}, {Boffin}, {Karinkuzhi}, {Van
  Eck}, {Escorza}, {Shetye}, \& {Van Winckel}}]{jorissen19}
{Jorissen}, A., {Boffin}, H.~M.~J., {Karinkuzhi}, D., {et~al.} 2019, A\&A, 626,
  A127

\bibitem[{{Jorissen} {et~al.}(1998{\natexlab{a}}){Jorissen}, {Van Eck},
  {Mayor}, \& {Udry}}]{accretion-Jori1}
{Jorissen}, A., {Van Eck}, S., {Mayor}, M., \& {Udry}, S. 1998{\natexlab{a}},
  \aap, 332, 877

\bibitem[{{Jorissen} {et~al.}(1998{\natexlab{b}}){Jorissen}, {Van Eck},
  {Mayor}, \& {Udry}}]{binary-Jorissen}
{Jorissen}, A., {Van Eck}, S., {Mayor}, M., \& {Udry}, S. 1998{\natexlab{b}},
  \aap, 332, 877

\bibitem[{{K{\"a}ppeler} {et~al.}(2011{\natexlab{a}}){K{\"a}ppeler}, {Gallino},
  {Bisterzo}, \& {Aoki}}]{kaeppeler11}
{K{\"a}ppeler}, F., {Gallino}, R., {Bisterzo}, S., \& {Aoki}, W.
  2011{\natexlab{a}}, Reviews of Modern Physics, 83, 157

\bibitem[{{K{\"a}ppeler} {et~al.}(2011{\natexlab{b}}){K{\"a}ppeler}, {Gallino},
  {Bisterzo}, \& {Aoki}}]{maxtemp_Kappeler}
{K{\"a}ppeler}, F., {Gallino}, R., {Bisterzo}, S., \& {Aoki}, W.
  2011{\natexlab{b}}, Reviews of Modern Physics, 83, 157

\bibitem[{{Karakas} \& {Lattanzio}(2014)}]{Karakas-Lattanzio-review}
{Karakas}, A.~I. \& {Lattanzio}, J.~C. 2014, \pasa, 31, e030

\bibitem[{{Karakas} \& {Lugaro}(2016)}]{2016karakas}
{Karakas}, A.~I. \& {Lugaro}, M. 2016, \apj, 825, 26

\bibitem[{{Karakas} {et~al.}(2018){Karakas}, {Lugaro}, {Carlos}, {Cseh},
  {Kamath}, \& {Garc{\'\i}a-Hern{\'a}ndez}}]{2018karakas}
{Karakas}, A.~I., {Lugaro}, M., {Carlos}, M., {et~al.} 2018, \mnras, 477, 421

\bibitem[{{Karakas} {et~al.}(2006){Karakas}, {Lugaro}, {Wiescher},
  {G{\"o}rres}, \& {Ugalde}}]{Karakas-Lugaro-22Ne}
{Karakas}, A.~I., {Lugaro}, M.~A., {Wiescher}, M., {G{\"o}rres}, J., \&
  {Ugalde}, C. 2006, \apj, 643, 471

\bibitem[{{Karinkuzhi} {et~al.}(2023){Karinkuzhi}, {Van Eck}, {Goriely},
  {Siess}, {Jorissen}, {Choplin}, {Escorza}, {Shetye}, \& {Van
  Winckel}}]{iproc-highmet-Kar}
{Karinkuzhi}, D., {Van Eck}, S., {Goriely}, S., {et~al.} 2023, \aap, 677, A47

\bibitem[{{Karinkuzhi} {et~al.}(2018){Karinkuzhi}, {Van Eck}, {Jorissen},
  {Goriely}, {Siess}, {Merle}, {Escorza}, {Van der Swaelmen}, {Boffin},
  {Masseron}, {Shetye}, \& {Plez}}]{ZrNb-Kar}
{Karinkuzhi}, D., {Van Eck}, S., {Jorissen}, A., {et~al.} 2018, \aap, 618, A32

\bibitem[{{Korotin}(2020)}]{NLTE-Rb-Korotin}
{Korotin}, S.~A. 2020, Astronomy Letters, 46, 541

\bibitem[{{Lind} \& {Amarsi}(2024)}]{NLTE-FeHdependence-Lind}
{Lind}, K. \& {Amarsi}, A.~M. 2024, arXiv e-prints, to appear in Annual Reviews
  of Astronomy and Astrophysics, arXiv:2401.00697

\bibitem[{{Lugaro} {et~al.}(2015){Lugaro}, {Campbell}, {Van Winckel}, {De
  Smedt}, {Karakas}, \& {K{\"a}ppeler}}]{postAGB-lugaro}
{Lugaro}, M., {Campbell}, S.~W., {Van Winckel}, H., {et~al.} 2015, \aap, 583,
  A77

\bibitem[{{Lugaro} {et~al.}(2003{\natexlab{a}}){Lugaro}, {Herwig}, {Lattanzio},
  {Gallino}, \& {Straniero}}]{lugaro03}
{Lugaro}, M., {Herwig}, F., {Lattanzio}, J.~C., {Gallino}, R., \& {Straniero},
  O. 2003{\natexlab{a}}, \apj, 586, 1305

\bibitem[{{Lugaro} {et~al.}(2003{\natexlab{b}}){Lugaro}, {Herwig}, {Lattanzio},
  {Gallino}, \& {Straniero}}]{maxtemp_Lugaro03}
{Lugaro}, M., {Herwig}, F., {Lattanzio}, J.~C., {Gallino}, R., \& {Straniero},
  O. 2003{\natexlab{b}}, \apj, 586, 1305

\bibitem[{{Lugaro} {et~al.}(2012){Lugaro}, {Karakas}, {Stancliffe}, \&
  {Rijs}}]{2012lugaro}
{Lugaro}, M., {Karakas}, A.~I., {Stancliffe}, R.~J., \& {Rijs}, C. 2012, \apj,
  747, 2

\bibitem[{{Lugaro} {et~al.}(2023){Lugaro}, {Pignatari}, {Reifarth}, \&
  {Wiescher}}]{Lugaro-review}
{Lugaro}, M., {Pignatari}, M., {Reifarth}, R., \& {Wiescher}, M. 2023, Annual
  Review of Nuclear and Particle Science, 73, 315

\bibitem[{{Lugaro} {et~al.}(2014){Lugaro}, {Tagliente}, {Karakas}, {Milazzo},
  {K{\"a}ppeler}, {Davis}, \& {Savina}}]{Lugaro14}
{Lugaro}, M., {Tagliente}, G., {Karakas}, A.~I., {et~al.} 2014, \apj, 780, 95

\bibitem[{{McClure}(1983)}]{RV2-McC}
{McClure}, R.~D. 1983, \apj, 268, 264

\bibitem[{{McClure} {et~al.}(1980){McClure}, {Fletcher}, \&
  {Nemec}}]{RV1-McCFN}
{McClure}, R.~D., {Fletcher}, J.~M., \& {Nemec}, J.~M. 1980, \apjl, 238, L35

\bibitem[{{Mishenina} {et~al.}(2019){Mishenina}, {Pignatari}, {Gorbaneva},
  {Travaglio}, {C{\^o}t{\'e}}, {Thielemann}, \& {Soubiran}}]{Mishenina-NbRu}
{Mishenina}, T., {Pignatari}, M., {Gorbaneva}, T., {et~al.} 2019, \mnras, 489,
  1697

\bibitem[{{Neyskens} {et~al.}(2015){Neyskens}, {van Eck}, {Jorissen},
  {Goriely}, {Siess}, \& {Plez}}]{ZrNb-Neysk}
{Neyskens}, P., {van Eck}, S., {Jorissen}, A., {et~al.} 2015, \nat, 517, 174

\bibitem[{Pedregosa {et~al.}(2011)Pedregosa, Varoquaux, Gramfort, Michel,
  Thirion, Grisel, Blondel, Prettenhofer, Weiss, Dubourg, Vanderplas, Passos,
  Cournapeau, Brucher, Perrot, \& Duchesnay}]{scikit-learn}
Pedregosa, F., Varoquaux, G., Gramfort, A., {et~al.} 2011, Journal of Machine
  Learning Research, 12, 2825

\bibitem[{{Pereira} {et~al.}(2011){Pereira}, {Sales Silva}, {Chavero}, {Roig},
  \& {Jilinski}}]{Pereira11}
{Pereira}, C.~B., {Sales Silva}, J.~V., {Chavero}, C., {Roig}, F., \&
  {Jilinski}, E. 2011, \aap, 533, A51

\bibitem[{{Piersanti} {et~al.}(2013){Piersanti}, {Cristallo}, \&
  {Straniero}}]{2013piersanti}
{Piersanti}, L., {Cristallo}, S., \& {Straniero}, O. 2013, \apj, 774, 98

\bibitem[{{Pols} {et~al.}(2003){Pols}, {Karakas}, {Lattanzio}, \&
  {Tout}}]{binary-Pols}
{Pols}, O.~R., {Karakas}, A.~I., {Lattanzio}, J.~C., \& {Tout}, C.~A. 2003, in
  Astronomical Society of the Pacific Conference Series, Vol. 303, Symbiotic
  Stars Probing Stellar Evolution, ed. R.~L.~M. {Corradi}, J.~{Mikolajewska},
  \& T.~J. {Mahoney}, 290

\bibitem[{{Roriz} {et~al.}(2024){Roriz}, {Lugaro}, {Junqueira}, {Sneden},
  {Drake}, \& {Pereira}}]{roriz24}
{Roriz}, M.~P., {Lugaro}, M., {Junqueira}, S., {et~al.} 2024, \mnras, 528, 4354

\bibitem[{{Roriz} {et~al.}(2021{\natexlab{a}}){Roriz}, {Lugaro}, {Pereira},
  {Drake}, {Junqueira}, \& {Sneden}}]{roriz21a}
{Roriz}, M.~P., {Lugaro}, M., {Pereira}, C.~B., {et~al.} 2021{\natexlab{a}},
  \mnras, 501, 5834

\bibitem[{{Roriz} {et~al.}(2021{\natexlab{b}}){Roriz}, {Lugaro}, {Pereira},
  {Sneden}, {Junqueira}, {Karakas}, \& {Drake}}]{roriz21b}
{Roriz}, M.~P., {Lugaro}, M., {Pereira}, C.~B., {et~al.} 2021{\natexlab{b}},
  \mnras, 507, 1956

\bibitem[{{Stancliffe} {et~al.}(2007){Stancliffe}, {Glebbeek}, {Izzard}, \&
  {Pols}}]{binary-Stancliffe}
{Stancliffe}, R.~J., {Glebbeek}, E., {Izzard}, R.~G., \& {Pols}, O.~R. 2007,
  \aap, 464, L57

\bibitem[{{Storm} {et~al.}(2024){Storm}, {Barklem}, {Yakovleva}, {Belyaev},
  {Palmeri}, {Quinet}, {Lodders}, {Bergemann}, \& {Hoppe}}]{NLTE-Y-Sun}
{Storm}, N., {Barklem}, P.~S., {Yakovleva}, S.~A., {et~al.} 2024, \aap, 683,
  A200

\bibitem[{{Tagliente} {et~al.}(2008{\natexlab{a}}){Tagliente}, {Fujii},
  {Milazzo}, {Moreau}, {Aerts}, {Abbondanno}, {{\'A}lvarez}, {Alvarez-Velarde},
  {Andriamonje}, {Andrzejewski}, {Assimakopoulos}, {Audouin}, {Badurek},
  {Baumann}, {Be{\v{c}}v{\'a}{\v{r}}}, {Berthoumieux}, {Bisterzo},
  {Calvi{\~n}o}, {Calviani}, {Cano-Ott}, {Capote}, {Carrapi{\c{c}}o},
  {Cennini}, {Chepel}, {Chiaveri}, {Colonna}, {Cortes}, {Couture}, {Cox},
  {Dahlfors}, {David}, {Dillman}, {Domingo-Pardo}, {Dridi}, {Duran},
  {Eleftheriadis}, {Embid-Segura}, {Ferrant}, {Ferrari}, {Ferreira-Marques},
  {Furman}, {Gallino}, {Goncalves}, {Gonzalez-Romero}, {Gramegna}, {Guerrero},
  {Gunsing}, {Haas}, {Haight}, {Heil}, {Herrera-Martinez}, {Igashira},
  {Jericha}, {K{\"a}ppeler}, {Kadi}, {Karadimos}, {Karamanis}, {Kerveno},
  {Koehler}, {Kossionides}, {Krti{\v{c}}ka}, {Lamboudis}, {Leeb}, {Lindote},
  {Lopes}, {Lozano}, {Lukic}, {Marganiec}, {Marrone}, {Mart{\'\i}nez},
  {Massimi}, {Mastinu}, {Mengoni}, {Mosconi}, {Neves}, {Oberhummer}, {O'Brien},
  {Pancin}, {Papachristodoulou}, {Papadopoulos}, {Paradela}, {Patronis},
  {Pavlik}, {Pavlopoulos}, {Perrot}, {Pigni}, {Plag}, {Plompen}, {Plukis},
  {Poch}, {Praena}, {Pretel}, {Quesada}, {Rauscher}, {Reifarth}, {Rubbia},
  {Rudolf}, {Rullhusen}, {Salgado}, {Santos}, {Sarchiapone}, {Savvidis},
  {Stephan}, {Tain}, {Tassan-Got}, {Tavora}, {Terlizzi}, {Vannini}, {Vaz},
  {Ventura}, {Villamarin}, {Vincente}, {Vlachoudis}, {Vlastou}, {Voss},
  {Walter}, {Wendler}, {Wiescher}, \& {Wisshak}}]{crossec-Zr90}
{Tagliente}, G., {Fujii}, K., {Milazzo}, P.~M., {et~al.} 2008{\natexlab{a}},
  \prc, 77, 035802

\bibitem[{{Tagliente} {et~al.}(2022){Tagliente}, {Kopecky}, {Heyse},
  {Krti{\v{c}}ka}, {Massimi}, {Mengoni}, {Milazzo}, {Plompen}, {Schillebeeckx},
  {Valenta}, {Wynants}, {Altstadt}, {Andrzejewski}, {Audouin}, {B{\'e}cares},
  {Barbagallo}, {Be{\v{c}}v{\'a}{\v{r}}}, {Belloni}, {Berthoumieux},
  {Billowes}, {Boccone}, {Bosnar}, {Brugger}, {Calvi{\~n}o}, {Calviani},
  {Cano-Ott}, {Carrapi{\c{c}}o}, {Cerutti}, {Chiaveri}, {Chin}, {Colonna},
  {Cort{\'e}s}, {Cort{\'e}s-Giraldo}, {Cristallo}, {Diakaki}, {Domingo-Pardo},
  {Dressler}, {Dur{\'a}n}, {Eleftheriadis}, {Ferrari}, {Fraval}, {Furman},
  {G{\"o}bel}, {G{\'o}mez-Hornillos}, {Ganesan}, {Garc{\'\i}a}, {Giubrone},
  {Gon{\c{c}}alves}, {Gonz{\'a}lez-Romero}, {Goverdovski}, {Griesmayer},
  {Guerrero}, {Gunsing}, {Heftrich}, {Hern{\'a}ndez-Prieto}, {Jericha},
  {K{\"a}ppeler}, {Kadi}, {Karadimos}, {Katabuchi}, {Ketlerov}, {Khryachkov},
  {Kivel}, {Kokkoris}, {Kroll}, {Lampoudis}, {Langer}, {Leal-Cidoncha},
  {Lederer}, {Leeb}, {Leong}, {Losito}, {Lugaro}, {Mallick}, {Manousos},
  {Marganiec}, {Mart{\'\i}nez}, {Mastinu}, {Mastromarco}, {Mendoza},
  {Mingrone}, {Mirea}, {Paradela}, {Pavlik}, {Perkowski}, {Praena}, {Quesada},
  {Rauscher}, {Reifarth}, {Riego-Perez}, {Robles}, {Rubbia}, {Ryan},
  {Sabat{\'e}-Gilarte}, {Sarmento}, {Saxena}, {Schmidt}, {Schumann},
  {Sedyshev}, {Tain}, {Tarife{\~n}o-Saldivia}, {Tarr{\'\i}o}, {Tassan-Got},
  {Tsinganis}, {Vannini}, {Variale}, {Vaz}, {Ventura}, {Vermeulen}, {Vescovi},
  {Vlachoudis}, {Vlastou}, {Wallner}, {Ware}, {Weigand}, {Weiss}, {Wright},
  {{\v{Z}}ugec}, \& {n TOF Collaboration}}]{crossec-Zr92-2}
{Tagliente}, G., {Kopecky}, S., {Heyse}, J., {et~al.} 2022, \prc, 105, 025805

\bibitem[{{Tagliente} {et~al.}(2010){Tagliente}, {Milazzo}, {Fujii},
  {Abbondanno}, {Aerts}, {{\'A}lvarez}, {Alvarez-Velarde}, {Andriamonje},
  {Andrzejewski}, {Audouin}, {Badurek}, {Baumann}, {Be{\v{c}}v{\'a}{\v{r}}},
  {Belloni}, {Berthoumieux}, {Bisterzo}, {Calvi{\~n}o}, {Calviani}, {Cano-Ott},
  {Capote}, {Carrapi{\c{c}}o}, {Cennini}, {Chepel}, {Chiaveri}, {Colonna},
  {Cortes}, {Couture}, {Cox}, {Dahlfors}, {David}, {Dillmann}, {Domingo-Pardo},
  {Dridi}, {Duran}, {Eleftheriadis}, {Embid-Segura}, {Ferrari},
  {Ferreira-Marques}, {Furman}, {Gallino}, {Goncalves}, {Gonzalez-Romero},
  {Gramegna}, {Guerrero}, {Gunsing}, {Haas}, {Haight}, {Heil},
  {Herrera-Martinez}, {Igashira}, {Jericha}, {K{\"a}ppeler}, {Kadi},
  {Karadimos}, {Karamanis}, {Kerveno}, {Kossionides}, {Krti{\v{c}}ka},
  {Lamboudis}, {Leeb}, {Lindote}, {Lopes}, {Lozano}, {Lukic}, {Marganiec},
  {Marrone}, {Mart{\'\i}nez}, {Massimi}, {Mastinu}, {Mengoni}, {Moreau},
  {Mosconi}, {Neves}, {Oberhummer}, {O'Brien}, {Pancin}, {Papachristodoulou},
  {Papadopoulos}, {Paradela}, {Patronis}, {Pavlik}, {Pavlopoulos}, {Perrot},
  {Pigni}, {Plag}, {Plompen}, {Plukis}, {Poch}, {Praena}, {Pretel}, {Quesada},
  {Rauscher}, {Reifarth}, {Rosetti}, {Rubbia}, {Rudolf}, {Rullhusen},
  {Salgado}, {Santos}, {Sarchiapone}, {Savvidis}, {Stephan}, {Tain},
  {Tassan-Got}, {Tavora}, {Terlizzi}, {Vannini}, {Vaz}, {Ventura},
  {Villamarin}, {Vincente}, {Vlachoudis}, {Vlastou}, {Voss}, {Walter},
  {Wendler}, {Wiescher}, \& {Wisshak}}]{crossec-Zr92}
{Tagliente}, G., {Milazzo}, P.~M., {Fujii}, K., {et~al.} 2010, \prc, 81, 055801

\bibitem[{{Tagliente} {et~al.}(2011){Tagliente}, {Milazzo}, {Fujii},
  {Abbondanno}, {Aerts}, {{\'A}lvarez}, {Alvarez-Velarde}, {Andriamonje},
  {Andrzejewski}, {Audouin}, {Badurek}, {Baumann}, {Be{\v{c}}v{\'a}{\v{r}}},
  {Belloni}, {Berthoumieux}, {Bisterzo}, {Calvi{\~n}o}, {Calviani}, {Cano-Ott},
  {Capote}, {Carrapi{\c{c}}o}, {Cennini}, {Chepel}, {Chiaveri}, {Colonna},
  {Cortes}, {Couture}, {Cox}, {Dahlfors}, {David}, {Dillmann}, {Domingo-Pardo},
  {Dridi}, {Duran}, {Eleftheriadis}, {Embid-Segura}, {Ferrari},
  {Ferreira-Marques}, {Furman}, {Gallino}, {Goncalves}, {Gonzalez-Romero},
  {Gramegna}, {Guerrero}, {Gunsing}, {Haas}, {Haight}, {Heil},
  {Herrera-Martinez}, {Jericha}, {K{\"a}ppeler}, {Kadi}, {Karadimos},
  {Karamanis}, {Kerveno}, {Kossionides}, {Krti{\v{c}}ka}, {Lamboudis}, {Leeb},
  {Lindote}, {Lopes}, {Lozano}, {Lukic}, {Marganiec}, {Marrone},
  {Mart{\'\i}nez}, {Massimi}, {Mastinu}, {Mengoni}, {Moreau}, {Mosconi},
  {Neves}, {Oberhummer}, {O'Brien}, {Pancin}, {Papachristodoulou},
  {Papadopoulos}, {Paradela}, {Patronis}, {Pavlik}, {Pavlopoulos}, {Perrot},
  {Pigni}, {Plag}, {Plompen}, {Plukis}, {Poch}, {Praena}, {Pretel}, {Quesada},
  {Rauscher}, {Reifarth}, {Rosetti}, {Rubbia}, {Rudolf}, {Rullhusen},
  {Salgado}, {Santos}, {Sarchiapone}, {Savvidis}, {Stephan}, {Tain},
  {Tassan-Got}, {Tavora}, {Terlizzi}, {Vannini}, {Vaz}, {Ventura},
  {Villamarin}, {Vincente}, {Vlachoudis}, {Vlastou}, {Voss}, {Walter},
  {Wiescher}, \& {Wisshak}}]{crossec-Zr94}
{Tagliente}, G., {Milazzo}, P.~M., {Fujii}, K., {et~al.} 2011, \prc, 84, 015801

\bibitem[{{Tagliente} {et~al.}(2013){Tagliente}, {Milazzo}, {Fujii},
  {Abbondanno}, {Aerts}, {{\'A}lvarez}, {Alvarez-Velarde}, {Andriamonje},
  {Andrzejewski}, {Audouin}, {Badurek}, {Baumann}, {Be{\v{c}}v{\'a}{\v{r}}},
  {Belloni}, {Berthoumieux}, {Calvi{\~n}o}, {Calviani}, {Cano-Ott}, {Capote},
  {Carrapi{\c{c}}o}, {Cennini}, {Chepel}, {Chiaveri}, {Colonna}, {Cortes},
  {Couture}, {Dahlfors}, {David}, {Dillmann}, {Domingo-Pardo}, {Dridi},
  {Duran}, {Eleftheriadis}, {Embid-Segura}, {Ferrari}, {Ferreira-Marques},
  {Furman}, {Goncalves}, {Gonzalez-Romero}, {Gramegna}, {Guerrero}, {Gunsing},
  {Haas}, {Haight}, {Heil}, {Herrera-Martinez}, {Jericha}, {K{\"a}ppeler},
  {Kadi}, {Karadimos}, {Karamanis}, {Kerveno}, {Kossionides}, {Krti{\v{c}}ka},
  {Lamboudis}, {Leeb}, {Lindote}, {Lopes}, {Lukic}, {Marganiec}, {Marrone},
  {Mart{\'\i}nez}, {Massimi}, {Mastinu}, {Mengoni}, {Moreau}, {Mosconi},
  {Neves}, {Oberhummer}, {O'Brien}, {Papachristodoulou}, {Papadopoulos},
  {Paradela}, {Patronis}, {Pavlik}, {Pavlopoulos}, {Perrot}, {Pigni}, {Plag},
  {Plompen}, {Plukis}, {Poch}, {Praena}, {Pretel}, {Quesada}, {Reifarth},
  {Rosetti}, {Rubbia}, {Rudolf}, {Rullhusen}, {Salgado}, {Santos},
  {Sarchiapone}, {Savvidis}, {Stephan}, {Tain}, {Tassan-Got}, {Tavora},
  {Terlizzi}, {Vannini}, {Vaz}, {Ventura}, {Villamarin}, {Vincente},
  {Vlachoudis}, {Vlastou}, {Voss}, {Walter}, {Wiescher}, \&
  {Wisshak}}]{crossec-Zr93}
{Tagliente}, G., {Milazzo}, P.~M., {Fujii}, K., {et~al.} 2013, \prc, 87, 014622

\bibitem[{{Tagliente} {et~al.}(2008{\natexlab{b}}){Tagliente}, {Milazzo},
  {Fujii}, {Aerts}, {Abbondanno}, {{\'A}lvarez}, {Alvarez-Velarde},
  {Andriamonje}, {Andrzejewski}, {Assimakopoulos}, {Audouin}, {Badurek},
  {Baumann}, {Be{\v{c}}v{\'a}{\v{r}}}, {Belloni}, {Berthoumieux},
  {Calvi{\~n}o}, {Calviani}, {Cano-Ott}, {Capote}, {Carrapi{\c{c}}o},
  {Cennini}, {Chepel}, {Colonna}, {Cortes}, {Couture}, {Cox}, {Dahlfors},
  {David}, {Dillmann}, {Domingo-Pardo}, {Dridi}, {Duran}, {Eleftheriadis},
  {Embid-Segura}, {Ferrant}, {Ferrari}, {Ferreira-Marques}, {Furman},
  {Goncalves}, {Gonzalez-Romero}, {Gramegna}, {Guerrero}, {Gunsing}, {Haas},
  {Haight}, {Heil}, {Herrera-Martinez}, {Igashira}, {Jericha}, {K{\"a}ppeler},
  {Kadi}, {Karadimos}, {Karamanis}, {Kerveno}, {Koehler}, {Kossionides},
  {Krti{\v{c}}ka}, {Lamboudis}, {Leeb}, {Lindote}, {Lopes}, {Lozano}, {Lukic},
  {Marganiec}, {Marrone}, {Mart{\'\i}nez}, {Massimi}, {Mastinu}, {Mengoni},
  {Moreau}, {Mosconi}, {Neves}, {Oberhummer}, {O'Brien}, {Pancin},
  {Papachristodoulou}, {Papadopoulos}, {Paradela}, {Patronis}, {Pavlik},
  {Pavlopoulos}, {Perrot}, {Pigni}, {Plag}, {Plompen}, {Plukis}, {Poch},
  {Praena-Rodriguez}, {Pretel}, {Quesada}, {Rauscher}, {Reifarth}, {Rubbia},
  {Rudolf}, {Rullhusen}, {Salgado}, {Santos}, {Sarchiapone}, {Savvidis},
  {Stephan}, {Tain}, {Tassan-Got}, {Tavora}, {Terlizzi}, {Vannini}, {Vaz},
  {Ventura}, {Villamarin}, {Vincente}, {Vlachoudis}, {Vlastou}, {Voss},
  {Walter}, {Wiescher}, \& {Wisshak}}]{crossec-Zr91}
{Tagliente}, G., {Milazzo}, P.~M., {Fujii}, K., {et~al.} 2008{\natexlab{b}},
  \prc, 78, 045804

\bibitem[{{Takeda}(2021)}]{NLTE-Rb-Takeda}
{Takeda}, Y. 2021, Astronomische Nachrichten, 342, 515

\bibitem[{{Th{\'e}venin} \& {Idiart}(1999)}]{NLTE-Fehdependence-Thevenin}
{Th{\'e}venin}, F. \& {Idiart}, T.~P. 1999, \apj, 521, 753

\bibitem[{{van Raai} {et~al.}(2012){van Raai}, {Lugaro}, {Karakas},
  {Garc{\'\i}a-Hern{\'a}ndez}, \& {Yong}}]{Rb-vanRaai}
{van Raai}, M.~A., {Lugaro}, M., {Karakas}, A.~I., {Garc{\'\i}a-Hern{\'a}ndez},
  D.~A., \& {Yong}, D. 2012, \aap, 540, A44

\bibitem[{Virtanen {et~al.}(2020)Virtanen, Gommers, Oliphant, Haberland, Reddy,
  Cournapeau, Burovski, Peterson, Weckesser, Bright, {van der Walt}, Brett,
  Wilson, Millman, Mayorov, Nelson, Jones, Kern, Larson, Carey, Polat, Feng,
  Moore, {VanderPlas}, Laxalde, Perktold, Cimrman, Henriksen, Quintero, Harris,
  Archibald, Ribeiro, Pedregosa, {van Mulbregt}, \& {SciPy 1.0
  Contributors}}]{Scipy}
Virtanen, P., Gommers, R., Oliphant, T.~E., {et~al.} 2020, Nature Methods, 17,
  261

\end{thebibliography}


\begin{appendix}
\section{Details of the Random Forest (RF) classifier}
\label{sec:appendix-rf}

To build the Random Forest classifier, we used the Scikit-Learn library of Python3. 
After reading in the data for all stars, we calculated the elemental ratios by subtracting the abundances from each other. We then normalised each feature (elemental abundances and ratios) by subtracting the mean value of the feature from each data point and dividing by its standard deviation.

The classifier is trained on sets of AGB model outputs, each diluted by a different amount. In order to mimic the diluted material in the Ba star envelope, we mixed AGB material and solar-like material with different values of $\delta$, in 0.002 increments. This leads to a set of hundreds of versions (between 100 and 550, typically around 450) of each AGB model. The exact number of variations of an AGB model is different, as versions that have very similar composition to their counterparts with neighbouring $\delta$ are deleted from the set. Each AGB model is considered as a class, and all the diluted versions create the population behind each class. The training set takes out 80\% of the data set, the other 20\% is used for testing the accuracy. 

Building a too complex classifier can lead to overfitting the training data; however, a too primitive model can not be trained with high performance. For finding the optimal complexity in between these two cases, we performed the optimisation of the hyperparameters using 5-fold cross validation. 
    The final hyperparameters of the classifier are the following:
\begin{itemize}
    \item The number of trees in the ensemble. (\textit{n\_estimators = 800})
    \item The maximal number of levels for each decision tree. (\textit{max\_depth = 15})
    \item The minimal number of sample for which the tree still can make a branching. (\textit{min\_samples\_split = 2})
    \item The minimal number of sample for which the tree can end in a leaf. (\textit{min\_samples\_leaf = 2})
    \item The maximal number of features used in each branching, where `auto' means no limit. (\textit{max\_features = "auto"})
    \item The data set for building the trees is sampled with replacement. (\textit{bootstrap = True})
    \item The size of the data set used for training each tree equals to the size of the whole sample. (\textit{max\_samples = None})
\end{itemize}

All the other hyperparametrs are set to the default value. 
We built 5 different Random Forest classifiers with different random inputs. At the end, those AGB models that are predicted with at least 0.1 probability by 3 of the 5 classifiers are selected to provide an eligible fit. This limit is arbitrary, but leads to around 2--3 fitted models on average for each star. 
The feature importances averaged for the five random forests can be seen in Fig. \ref{fig:importance}.1.

\vspace{.1cm}
\begin{minipage}{\textwidth}
    \includegraphics[width=0.9\linewidth]{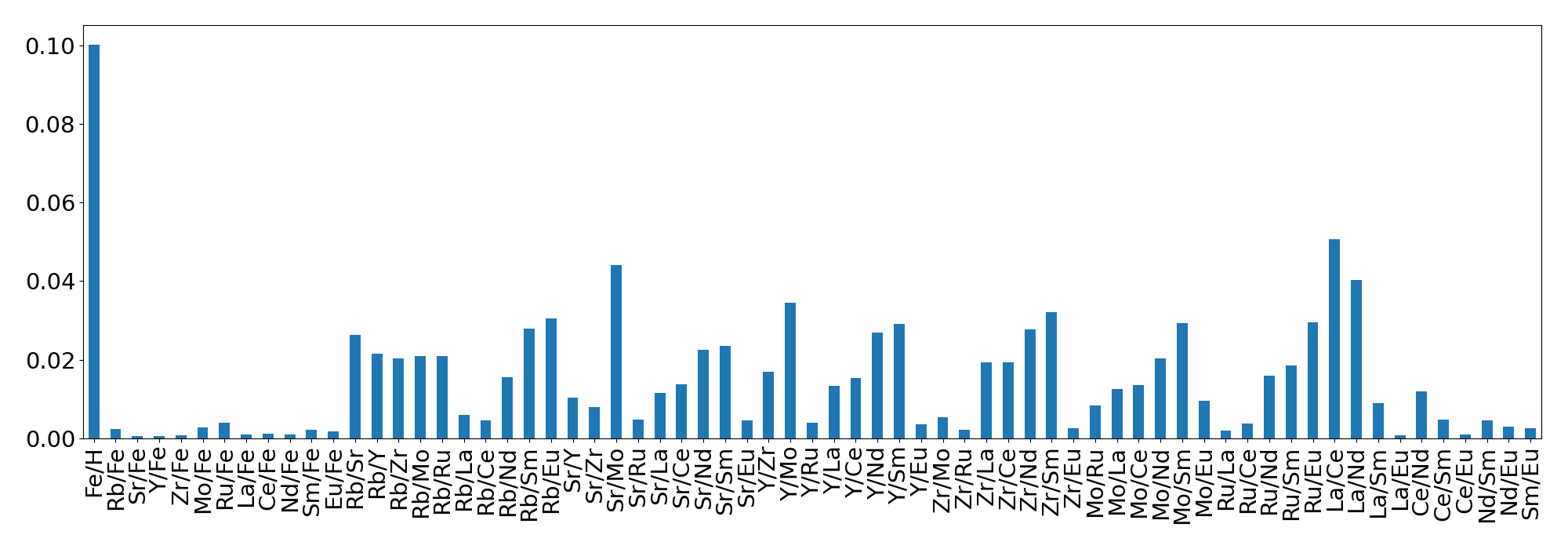}
    \vspace{.1cm} \\
    \noindent \textbf{Fig. \ref{fig:importance}.1.} Feature importances averaged for the five Random Forest classifiers on Monash models. The metallicity is recognised to be by far the most important feature, and the elemental ratios are usually more important than the abundances relative to Fe.
    \label{fig:importance}
    
\end{minipage}

\section{Comparison of results for the different classifiers and sets of AGB models}
\begin{figure}[h!]
\centering
\includegraphics[width=.9\linewidth]{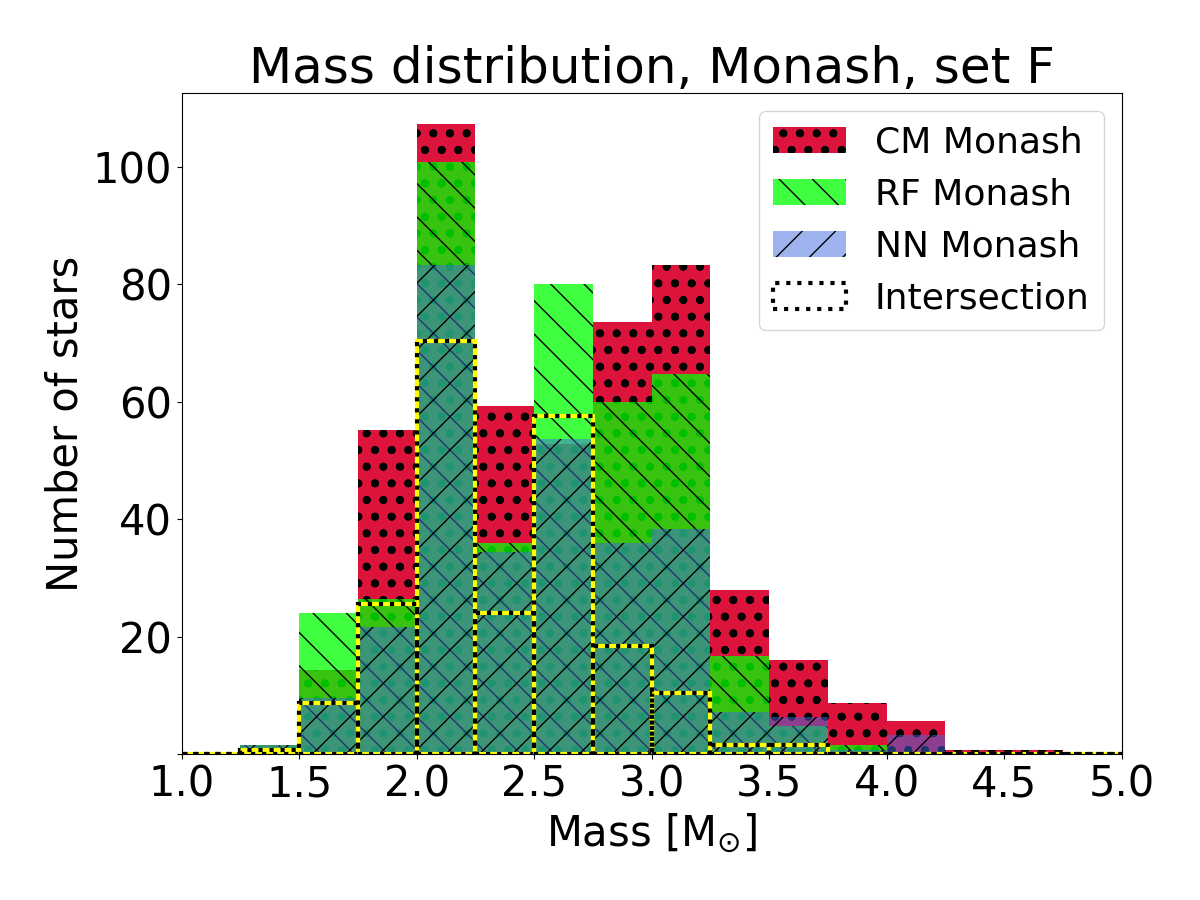}
\vspace{-.3cm}
\caption{Mass distribution histogram of the polluter AGB stars based on Monash models as in Fig. \ref{fig:distribution-mass-mon-setA}, but with element set F (all elements except for Y, Nb, Mo, and La) used for training the classifiers.}
\label{fig:distribution-mass-mon-setF}
\end{figure}

The same mass distribution as in Fig. \ref{fig:distribution-mass-mon-setA} can be seen in Fig. \ref{fig:distribution-mass-mon-setF}, but using the elemet set F for training, where the peak at 4 \msun~disappears from the distribution of the RF.

\begin{figure}[h!]
\centering
\includegraphics[width=.9\linewidth]{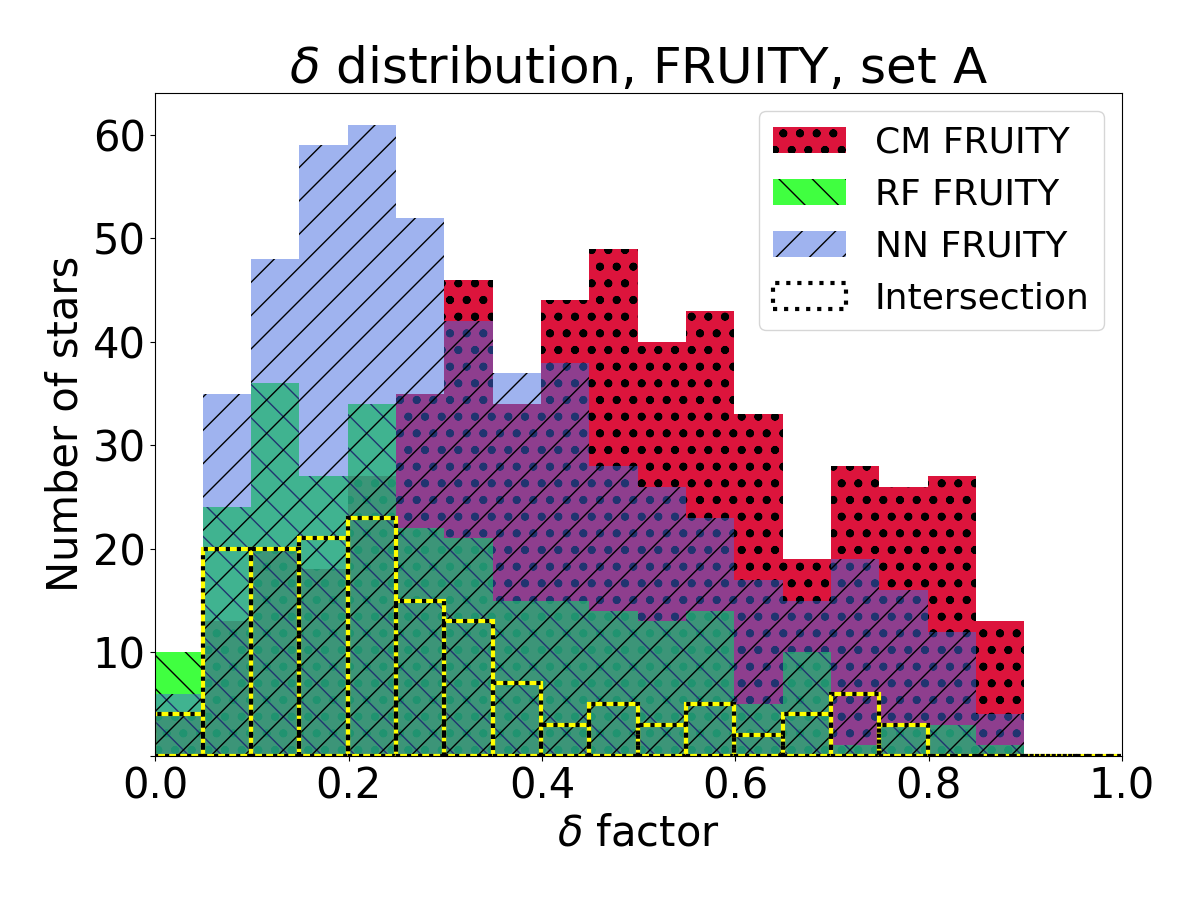}
\vspace{-.3cm}
\caption{Same as Fig. \ref{fig:distribution-dil-mon-setA}, but for FRUITY models.}
\label{fig:distribution-dil-fru-setA}
\end{figure}

The same $\delta$ distribution as in Fig. \ref{fig:distribution-dil-mon-setA} for FRUITY models can be seen in Fig. \ref{fig:distribution-dil-fru-setA}.

\vspace{-.6cm}

\newpage
The boxplots introduced in 
Sect. \ref{sec:boxplot}. give slightly different results for each classifiers (RF, CM and NN) and separately for
\vspace{.3cm}

\begin{minipage}{\textwidth}
\begin{center}
    \includegraphics[width=0.68\linewidth]{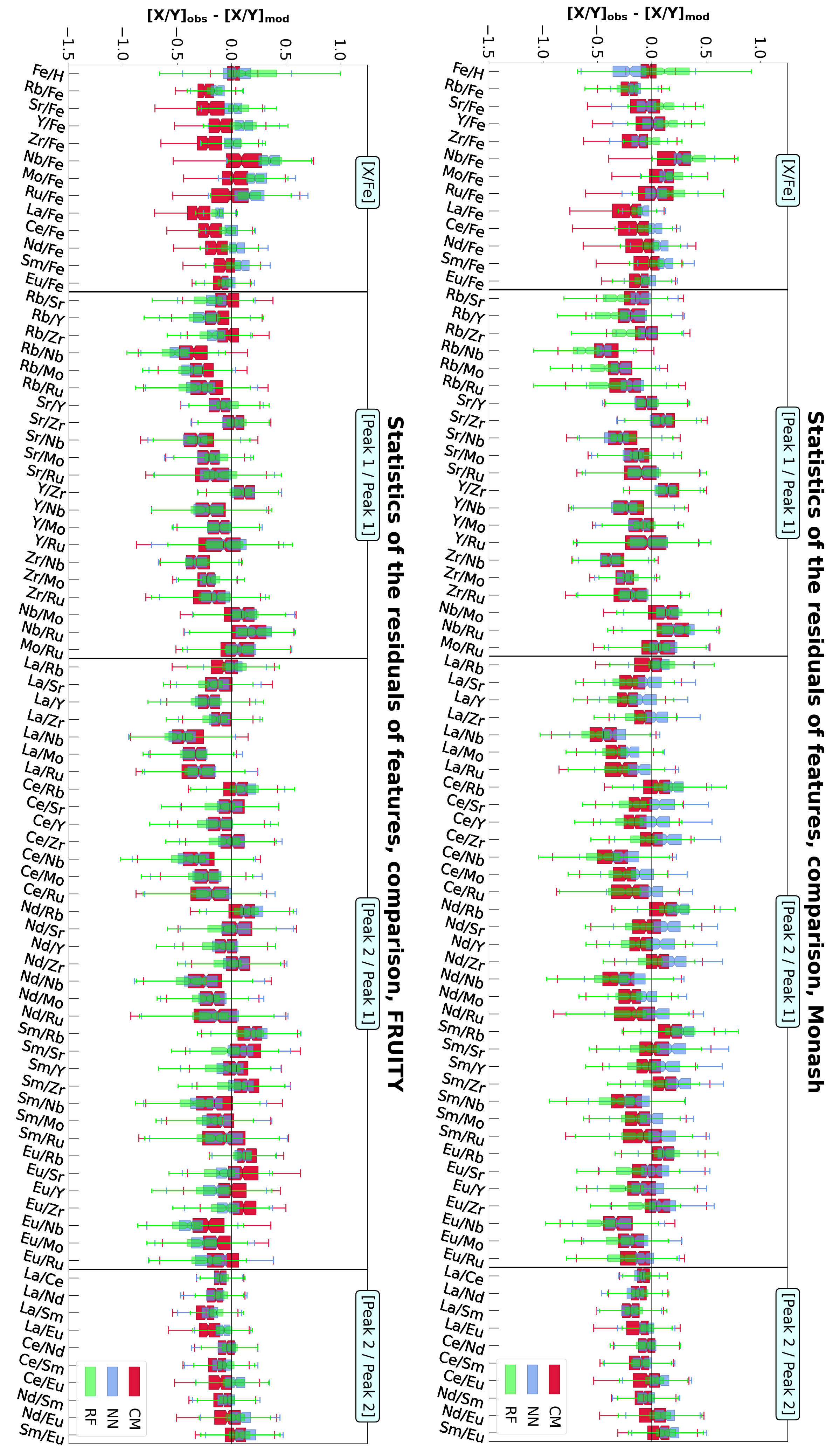}
\end{center}
  \vspace{.1cm} 
    \noindent \textbf{Fig. \ref{fig:box-big-compare}.3.} Boxplots for the different classifiers plotted over each other. 
  CM is indicated with red and the widest boxes, NN is the blue and middle-sized boxes, while the RF results are  the green and thinnest. 
  The top figure contains the results for the all three classifiers applied to Monash models, while the bottom figure shows the same for FRUITY models.
  \label{fig:box-big-compare}
\end{minipage}

\newpage
\noindent the FRUITY and Monash models. The results for the different classifiers plotted on top of each other can be seen in Fig. \ref{fig:box-big-compare}.3.

\end{appendix}

\end{document}